\documentclass[preprint,showpacs,preprintnumbers,amsmath,amssymb]{revtex4-1}
\usepackage{graphicx}
\usepackage{bm}
\usepackage{color}
\usepackage{xcolor}
\usepackage{caption}
\usepackage{ragged2e}
\usepackage{subcaption}
\usepackage{amsmath}
\usepackage{amssymb}
\usepackage{epsfig}
\usepackage{float}
\usepackage{needspace}
\usepackage{comment}
\usepackage{booktabs}   
\usepackage{multirow}   
\usepackage{float}      
\usepackage[colorlinks=true, allcolors=blue]{hyperref}

\DeclareGraphicsExtensions{.png,.pdf}
\usepackage[english]{babel}   
\usepackage{microtype}       
\usepackage{float}

\begin{document}

\title{On the Spectral Statistics of  Multiparametric Random Real Asymmetric Matrices: a hidden universality}
\author{Ratul Dutta and Pragya Shukla}
\affiliation{ Department of Physics, Indian Institute of Technology, Kharagpur-721302, West Bengal, India \\\\
Corresponding Author's E-Mail: shukla@phy.iitkgp.ac.in}
\date{\today}

\widetext

\begin{abstract}

 Theoretical analysis of a wide range of complex system e.g. those in contact with environment necessitate consideration of  the generic real-asymmetric matrix ensembles, those with varying order of matrix elements e.g. a sparse structure or a banded structure. We pursue the complexity parameter approach to analyze the spectral statistics of the multiparametric Gaussian ensembles of normal real asymmetric matrices and derive the theoretical formulations for the ensemble averaged spectral densities of the  real as well as complex eigenvalues. Considerations of the matrix elements with arbitrary choice of  mean and variances render us the freedom  to model the desired sparsity in the ensemble.  Our formulation provides a common mathematical formulation of the spectral statistics for a wide range of sparse real-asymmetric ensembles and also 
 reveals, thereby,  a deep rooted universality among them.

\end{abstract}

\maketitle

.

\section{Introduction}

With underlying complexity causing exact determination of the operators technically challenging, generic random  matrices often appear as relevant models  in various domains of complex systems.  A real physical system is typically in contact with an environment, the interactions among them often non-homogeneous and not always well-defined. This leaves a typical linear operator associated with the system  best described by a non-Hermitian system-dependent random matrix representation in many areas of physics \cite{berry, sv, tw, mss, ms, sss, ks, ab, cgh, llx, hl, fks, kr0, jnpz, r1, r2, r3, r4, r5, r6, r7, r8, r9, r10, r14, r15, r16, r17, r20, r23, fte1, fte2, fte3, fte4, cw, ben0, sc, kr1, ns, pas1, tsr, llc,  pw, mpm} (with a rapidly growing interest in the topic an exhaustive reference of all studies is impossible, compelling us to list mostly the review papers). Indeed non-Hermitian random matrices appear naturally  whenever the effective dynamics is open, dissipative, unstable, non-equilibrium, or governed by complex actions and now form an active bridge between  condensed matter, high-energy physics e.g. quantum chromodynamics (QCD) \cite{sv, tw}, string theory, gravity \cite{mss, ms, sss, ks, cgh}, statistical field theory and quantum chaos \cite{ab, hl, llx, fks, kr0} etc.  
For example, the role of non-Hermitian random matrices to model open systems in condensed matter physics has  been extensively investigated in past decades \cite{r1, r2, r3, r4, r5, r6, r7, r8, r9, r10, r14, r15, r16, r17, r20, r23, fte1, fte2, fte3, fte4, cw, ben0}.  In high energy physics, one of the most important appearances of non-Hermitian matrices is in QCD with nonzero baryon chemical potential: while  the Euclidean Dirac operator is anti-Hermitian at zero chemical potential with purely imaginary eigenvalues, it becomes non-Hermitian at finite chemical potential with eigenvalues spreading on the complex plane. The non-Hermitian random matrix (NHRM) theory  is indeed an important tool to model e.g. complex Dirac spectra, chiral symmetry breaking, QCD phase transitions, the sign problem, baryon density effects and is often used in analytical computation of many universal features of strongly interacting QCD even when lattice simulations become difficult due to the sign problem.
Another important application of NHRM theory is to open quantum systems consisting of high energy unstable particles where the effective Hamiltonians are often non-Hermitian and randomized in a physically motivated basis due to complicated many body interactions among them. The latter have been successfully used to describe the statistical properties of the resonance widths, S-matrix poles, chaotic scattering etc which have  applications in hadronic resonances, unstable nuclear states, scattering resonances, meson decay etc.  Non-Hermitian random matrices also  appear significantly in holography, black-hole information dynamics with applications in scrambling, spectral form factors and dissipative holography. Indeed recent studies indicate emerging links between non-Hermitian random matrices and near-horizon physics, SYK-like models, quantum gravity with loss/gain mechanisms. Another important realm of  NHRM models is  quantum field theories e.g non-unitary conformal field theories, effective field theories (to study gain/loss, measurement or decoherence), complex saddle-point analysis, Lee–Yang edge singularities where NHRM tools are used e.g. to analyze PT-symmetry breaking transitions, dissipative phase transitions, spectral instabilities, exceptional points and spectral universality. In case of large N matrix models central in string theory, the non-Hermitian matrices appear in connection with complex matrix models, non-unitary string theories, two dimension gravity, D-brane decay, complex saddles in path integrals etc. The spectral statistics here encode important information e.g. about topological transitions, string susceptibilities, nonperturbative effects etc. The appearance of non Hermitian random matrices is however not confined to environmental effects or many body physics only. Two important examples with huge technological applications in current context are the ensembles of synaptic matrices \cite{sc, kr1} and weight matrices \cite{pas1, tsr, llc, pw, mpm} required for  the biological and artificial  neural networks  analysis  respectively.

The choice of a matrix ensemble representing a linear operator of a complex system  is sensitive to the nature of complexity e.g. missing information about the matrix elements in a physically motivated basis due to technical complications as well as  the global constraints e.g. symmetry and conservation laws. Indeed a real complex system is in general characterized by many system conditions and, to be its appropriate representative, the ensemble must contain information about them through the  ensemble parameters. 
In past a specific ensemble of real asymmetric matrices, namely, real-Ginibre ensemble has been studied in great detail \cite{gin,  haak, meta, fh1, nh},  but it has limited applicability, only for the operator represented by a full real matrices with almost all matrix elements of the same order. A generic case in realistic applications  however may not only require matrices with varying order of entries e.g. a sparse structure or a banded structure, the randomness associated with entries may also be system specific and different for entries, with some of them even non random. In contrast to a closed system, an open system is far more sensitive to the external perturbations causing a variation of the system parameters. The latter in turn  leads to a variation of the ensemble parameters and results in evolution of the ensemble density both in matrix space as well as ensemble parameter space. Based on the nature of the perturbation, the evolution may or may not lead to an equilibrium state in matrix space. The latter  corresponds to a Ginibre ensemble only if the matrix representing the system is non Hermitian as well as normal. This motivates the present study with primary focus to statistically analyze  the multiparametric Gaussian ensembles of real asymmetric normal matrices and their evolution to Ginibre statistical equilibrium.

An important route to gain insights in the properties described by linear operators is through their formulation in terms of the eigenvalues and eigenfunctions.  In contrast to a Hermitian matrix, the diagonalization of a non-Hermitian matrix is however  far more subtle;  indeed the eigenvalues in the latter case can be both complex and real, the left and right eigenvectors are generally different and need not be orthogonal. Assuming the matrix  diagonalizable, the eigenvectors of a non-Hermitian matrix come in left-right pairs,   $w^{\dagger}_n  H = \lambda_n w^{\dagger}_n$ and $H v_n  = \lambda_n v_n$. Defining $W$ as the matrix of left eigenvectors (with $w_n$ as columns) and $V$ as the matrix of right eigenvectors (with $v_n$ as columns), their  product can in general be chosen to satisfy the biorthogonal normalization $W^{\dagger}. V =I$ if $H$ is not subjected to any additional constraint. The normal matrices, defined by the condition $H.H^{\dagger} = H^{\dagger}.H$, form  an important subclass of non-Hermitian matrices. As a diagonalizable matrix  $H$ can in general be written as $H = V \Lambda V^{-1}$, the right eigenvector matrix $V$ in case of a normal matrix is subjected to an additional unitary constraint too i.e $V^{\dagger} \cdot V =V \cdot V^{\dagger} = I$ (and similarly for the left eigenvector matrix $W^{\dagger}. W =W. W^{\dagger} = I$). The left and right eigenvectors may  then be chosen the same up to phases/permutations: $V = W. D$ with $D$ as a diagonal unitary matrix, leading to $W^{\dagger}. V =D$ (different from the above mentioned biorthogonality). Unfortunately almost no theoretical results are so far available about the spectral statistics of generic, sparse real asymmetric ensembles. This motivates us, in the present work is to attempt to bridge this information gap by a theoretical as well as numerical study of the spectral distribution of these ensembles.

The Ginibre ensembles,  representing equilibrium state of evolution, are basis invariant ensembles which in turn manifest as a spectral statistics universality, with mutually repulsive eigenvalues on the complex plane kept in equilibrium by confining potential. This also leads their successful application  to model spectral universality in chaotic/ ergodic regime of open quantum systems. A natural query arising from the above is whether the spectral universality can exist  in case of basis-dependent ensembles, also referred as  non-equilibrium regime too? The query indeed has its origin in  multiple experimental observations: as reported in past, seemingly different system conditions may  lead to analogous matrix ensembles of the linear operators, thereby indicating analogous properties for the associated observables.  The answer to query is indeed in affirmative: as discussed later in this paper, we find  the existence of a single parametric universality of the spectral statistics in non-equilibrium regime. Indeed this encourages intuitive ideas that complexity irrespective of its origin would manifest through quantitative universality in statistical properties, with universality class characterized by very few parameters dependent on system conditions e.g. symmetry, conservation laws.

The paper is organized as follows.  Section II describes the ensemble of real-asymmetric matrices $H$, represented by a probability density $\rho(H)$   (the  joint probability distribution function ({\it jpdf}) of the matrix elements of $H$)   and introduces the  definition of the ({\it jpdf}) of its eigenvalues to motivate the analysis in our next section.  in sections III,  we derive a differential equation describing  the dynamics of $\rho(H)$ both in matrix as well as in ensemble space. This section also maps the dynamics ensemble parameter space to another parameter space, referred as the complexity parameter space. The differential equation for $\rho(H)$ is used in section IV to derive the equation governing the dynamics of  the spectral {\it jpdf}.  The latter is  subsequently  used in sections V, VI  to derive the solutions for the ensemble averaged spectral densities of the real eigenvalues and complex conjugate pairs.  As the derivations  involve technically complicated steps, to avoid distraction from the main results, the details are moved to supplemental material \cite{sup}. We also note that  a given symbol with different number of  subscripts has different meaning throughout the text (e.g.  $a_{\mu \nu}$ and $a_{\mu}$ are not same).  As  the matrix space dynamics of $\rho(H)$ in the equilibrium limit approaches real-Ginibre ensemble, the same is expected for the spectral statistics too.  Section VII compares our results for generic cases with the real-Ginibre ensemble. This is followed by a numerical analysis, in section VIII,  for four different multiparametric Gaussian ensembles of real-asymmetric matrices.  The choice of these ensembles is basically motivated from their applicability to real systems as well as from the already available theoretical analysis for corresponding Hermitian ensembles. The latter  helps in the comparative studies of the spectral fluctuations under Hermitian and non-Hermitian constraint.  We conclude in section IX with a brief discussion of our main results, insights and open issues for future.

\section{Details of the theoretical model}

\subsection{Probability density in the real-asymmetric matrix space}

Consider a complex system represented by a real asymmetric matrix with known constraints only on the mean and  variances of the matrix elements and their pair wise correlations. Based on the maximum entropy hypothesis (MEH), it can be described by an ensemble of $N\times N$ real matrices $H$ defined by a multiparametric Gaussian ensemble density 
\begin{eqnarray}
\rho (H,  {\mathcal V}, \mu)  \propto  {\rm exp}\left[- (H_v-\mu)^{T} \,   {\mathcal V} \, (H_v-\mu)\right]
\label{pdf0}
\end{eqnarray}
 with $H_v$ and $\mu$ as the column vectors consisting of all matrix elements $H_{kl}$ and their mean $\mu_{kl}$, respectively,  and ${\mathcal V}^{-1}$ as the covariance matrix. (We note that eq.(\ref{pdf0}) can be rewritten as $\rho (H,  {\mathcal V}, \mu)  \propto  {\rm exp}\left[- {\mathcal H}_v^{T} \,   {\mathcal V} \, {\mathcal H}_v\right]$ with a redefinition of the matrix elements $(H_{kl} - \mu_{kl}) \to {\mathcal H}_{kl}$ and the column vector $(H_v-\mu)$  as ${\mathcal H}_v$.  The mean value $\mu_{kl}$ can therefore be set to be zero without any loss of generality.)
    
 Assuming, for technical simplification, the vanishing of pairwise correlations of type $\langle H_{kl} H_{ij} \rangle=0$ for $k,l \not= i,j$,  eq.(\ref{pdf0}) can be rewritten as  
\begin{eqnarray}
\rho (H,y,x) =  C \, {\rm exp}\left[
-  \sum_{k, l} (y_{kl} H_{kl}^2 + x_{kl} H_{kl} H_{lk})\right] 
\label{pdf}
\end{eqnarray}
 with $x_{kl}=x_{lk}$,  $C$ as a constant determined by the normalization condition $\int_{-\infty}^{\infty}  \rho \, {\rm d}H =1$:
 $C=\frac{1}{\pi^{N^2/2}}\left[\prod_{k=1}^{N}(y_{kk}+x_{kk})  \prod_{k<l}
\left\{y_{kl}y_{lk}-\frac{(x_{kl}+x_{lk})^2}{4}\right\}\right]^{1/2}$.

Here the sets  $x \equiv \{x_{kl}\}$, $y \equiv \{y_{kl}\}$ of the ensemble  parameters are chosen so as the keep the density $\rho(H, y,x)$ normalizable (e.g. requiring $y_{kl} \ge 0$) but otherwise chosen arbitrarily (permitting  considerations of the infinite variances representing  non-random entries). (Indeed the convergence of the integral imposes following conditions ensuring the positive definiteness of the quadratic form) $x_{kk}+y_{kk} >0$ for $k=1 \to N$, $y_{kl}, y_{lk} >0$  for $k,l=1 \to N$ and $ 4 y_{lk} y_{kl} > (x_{kl}+x_{lk})^2$). This permits eq.(\ref{pdf})  to represent  a large class of real asymmetric matrix ensembles e.g. varying degree of sparsity or bandedness including standard  cases. For example, for all $y_{kl} \rightarrow {N \over (1-\tau^2)}, x_{kl} \rightarrow  - {N \tau\over (1-\tau^2)}$ in eq.(\ref{pdf}), we have $\rho(H, \tau) =C \,  {\rm e}^{-{N \over 1-\tau^2} Tr(H H^T - \tau H H)}$; the latter approaches Ginibre real ensemble for $\tau \rightarrow 0$,  Gaussian orthogonal ensemble (GOE) for $\tau=1$ and the ensemble of real antisymmetric matrices (GASE real) for $\tau=-1$ and many non-stationary ones intermediate between them for $0 < |\tau | < 1$  \cite{sc, nh}.

\subsection{Spectral density on the complex plane}

Let  ${P}(z, y, x)$ be the {\it jpdf} of the $N$ eigenvalues $z_n$, $n=1 \to N$  at parameter values $y_{kl}$ and $x_{kl}$. For brevity hereafter we use the notation $z \equiv \{z_i\}$. The joint probability density function (jpdf) of the eigenvalues can then be expressed as
\begin{eqnarray}
{P} =  \int   \; f_z(z) \; \rho(H) \ dH  
\label{pdf1}
\end{eqnarray}
Where
$f_z(z) = \prod_{j=1}^{N} \delta (z_j - z_j(H))$. The eigenvalues $z_n \equiv \sum_{s=1}^2 (i)^{s-1} z_{js}$ can be either complex or real (with $z_{j2}=0$ in latter case).

A multivariable integration of the ensemble density of the Hamiltonian  is   in general technically complicated. In addition, the interactions among various subunits of a many body system are  in general sensitive to a host of system conditions, both static as well dynamic type. A variation of these conditions can lead to changes in the mutual interactions among subunits and thereby the ensemble density. can lead to serious technical issues: even if the integration is achieved by some approximations, these are often specific to system parameters and may not be applicable once the parameters are changed.  The integration route is therefore not often a viable option and  motivates a search for alternative tools which take into account arbitrary variation of the system parameters.

\section{Dynamics in the matrix and ensemble spaces}

A change in system conditions can affect the matrix elements $H_{kl}$,  resulting in evolution of $\tilde\rho$ in matrix space.  But as mentioned above, due to complexity of the interactions among subunits of the system, each $H_{kl}$ can best be determined within some uncertainty; the latter also fluctuates with changing $H_{kl}$ and changes the parameters describing its distribution. As a consequence, $\rho(H)$ undergoes dynamics both in matrix space as well as in ensemble parameter space.  Here the matrix space is a $ N^2$ dimensional  space  consisting of  $N^2$ real matrix elements,  with each point in the space corresponds to a $N \times N$ non-Hermitian matrix. The ensemble space   is a $2 N^2$ dimensional  space consisting of independent ensemble parameters $x_{kl}, y_{kl}$ with each point representing the state of the ensemble.   This leads to many important queries e.g. (i) whether the dynamics of $\rho(H)$ in the $x, y$-space has an exact equivalent in the $H$-matrix space? The information is relevant for following reason: assuming that the system under consideration is well-described by eq.(\ref{pdf}) at a given instant of time, the question is whether the system can still be described by the same ensemble with time-evolved ensemble parameters? Indeed, from eq.(\ref{pdf}), a variation of either the matrix elements or the ensemble parameters can change the ensemble density. For an evolving ensemble to continue representing the system, it is desirable that the evolution of $\rho(H)$  in $H$-matrix space is exactly mimicked by that in $(y, x)$ space. This motivates us to consider a first order variation  of the ensemble parameters $y_{kl} \rightarrow y_{kl}+\delta y_{kl}$ and $x_{kl} \rightarrow x_{kl}+\delta x_{kl}$  over time and seek (i) what type of matrix space dynamics they correspond to, and, (ii) can it lead to an equilibrium ensemble with stationary correlations? More explicitly, we consider following combination of the first order parametric derivatives of $\rho$ where $\tilde\rho = C \rho$,

\begin{eqnarray}
T \; \rho \equiv \sum_{k,l}^N   \left[ A_{kl} {\partial  \rho \over\partial y_{kl}}+ B_{kl}  {\partial  \rho\over\partial x_{kl}}\right] 
 \label{trho}
\end{eqnarray}  
with 
$A_{kl} = 2 y_{kl} \left(1 - 2 y_{kl} \right) -  4 x^2_{kl} $ and $B_{kl} = 2 x_{kl} \, \left(1 - 4 \, y_{kl}\right)$.

With help of its Gaussian form,  the multi-parametric evolution of $\rho$ can  be described exactly  in terms of a diffusion, with finite drift,  in $H$-matrix space (details of derivation given in section I of {\it supplementary material} \cite{sup}),

\begin{eqnarray}
T \rho =\sum_{k,l}{\partial \over \partial H_{kl}}\left[
 {\partial \over \partial H_{kl}}  +   H_{kl}  \right] \rho
\label{td1}
\end{eqnarray}

 As eq.(\ref{td1})  is difficult to solve for generic parametric values, this motivates us to seek whether the
 the dynamics indeed depend on individual details of all ensemble parameters or only respond to their collective influence? This query arises by ample evidence in many areas of complex systems that complexity, irrespective of its origin, is sensitive to very few details or relevant system parameters, often resulting in  universality of the physical properties.  To answer the above query, we seek whether it is possible to define a transformation from the set of $M$ parameters $\{y_{kl}, x_{kl} \}$ to another set $\{t_1,\ldots, t_M \}$,  with  $M = 2 N^2$, such that only $t_1$ varies under the evolution governed by the operator $T$ and rest of them i.e $t_2, \ldots, t_M$ remain constant: $T \rho \equiv \frac{\partial \rho}{\partial t_1}$ or $T \rho  \equiv {\partial \rho \over\partial t_1}$ 
This implies

\begin{eqnarray}
 {\partial \rho\over\partial t_1} &=&
\sum_{k,l} {\partial \over \partial H_{kl}}\left[
 {\partial \over \partial H_{kl}}  +   H_{kl}  \right] \rho
 \label{rhot1}
\end{eqnarray}
To distinguish,  the parameter-space $\{x_{kl}, y_{kl} \}$ is hereafter referred as the ensemble parameter space (or just $x-y$ space for brevity) and the $t_1, \ldots, t_M$  as the complexity parameter space (or just $t$ space for brevity).

As clear from the above, the condition $T \rho  \equiv {\partial \rho\over\partial t_1}$ can be fulfilled by three possible ways,

\begin{eqnarray}
&&  {\rm (I)} \hspace{1.in}  T t_1=1, \; T t_{\alpha}=0, \qquad \frac{\partial \rho}{\partial t_{\alpha}} = 0, \quad \forall \; \alpha >1, \nonumber \\
&&    {\rm (II)}   \hspace{1.in}  T t_1=1, \; \frac{\partial \rho}{\partial t_{\alpha}} = 0, \quad \forall \; \alpha >1,   \nonumber \\ 
&&   {\rm (III)}    \hspace{1.in}  T t_1=1, \; T t_{\alpha}=0.  \qquad \forall \; \;  {\alpha} >1,  
\label{yc3}
\end{eqnarray}

The parameters $t_1, \ldots, t_M$ for each one of the above cases can be obtained  by solving the characteristic set of equations ${{\rm d}y_{kl} \over A_{kl}}
= {{\rm d}x_{kl} \over B_{kl}}={{\rm d}t_{\alpha} \over \delta_{\alpha 1}}$ with $k,l=1 \to N$. Here $M$ corresponds to total number of ensemble parameters participating in evolution. Here $M=2 N^2$ for all of them varying, $M < 2 N^2$ in case ${\partial \rho \over \partial y_{kl}}=0$ or ${\partial \rho \over \partial x_{kl}}=0$ for some $y_{kl}$ or $x_{kl}$, with $k,l$ arbitrary. For example, for the case with $x_{kl} = 0$ and $y_{kl}$ varying $\forall k,l$, we have $M = N^2$.

A general solution of the above set of equations requires detailed calculations and is discussed  in {\it supplementary material} \cite{sup}. Here we  discuss only the case I  with $x_{kl}=0$ in eq.(\ref{pdf}),  relevant for the ensembles used later in our numerical analysis in section VIII. With  $A_{kl} = 2 y_{kl} \left(1 - 2 y_{kl} \right) $ and $B_{kl} = 0$, the characteristic set of equations in this case reduces to following form (section II of {\it supplementary material}) 
\begin{eqnarray}
\frac{d y_{11}}{y_{11;s}} = \frac{d y_{12}}{y_{12}}=\ldots =\frac{d y_{mn}}{y_{mn}} = \ldots =\frac{d t_{\alpha}}{\delta_{\alpha 1}}
\label{tk1}
\end{eqnarray}
 A particular solution of the above equation can be given as 
\begin{eqnarray}
t_1 &=&   2 \sum_{m,n} q_{mn;1}  \; {\rm log} \; {y_{mn}\over |1- 2 y_{mn}|}  + c_0,  \label{y1} \\
t_{\alpha} &=& \sum_{m,n} q_{mn;\alpha} \log y_{mn}, \hspace{1.0in} \alpha >1
\label{ta}
\end{eqnarray}
with $q_{mn; \alpha}$ as  constants subjected to condition $\sum q_{mn; \alpha}= \delta_{\alpha 1}$ and $c_0$ as an arbitrary constant of integration. (This can be checked by a direct substitution in relation  $\sum_{kl} y_{kl} {\partial t_{\alpha} \over \partial y_{kl}} = \delta_{\alpha 1}$; the latter is equivalent to eq.(\ref{tk1})). While many solutions satisfying the above condition are possible, the appropriate  solution is the one that is also applicable for initial ensemble (as the constants of evolution are constants for initial ensemble too). For example, if  $y_{mn}$ for a specific $m,n$ value remains zero throughout the evolution and does not participate in the evolution, the corresponding coefficient  $q_{mn;\alpha}$ can be set to zero. For the case in which all $y_{mn}$ remain non-zero throughout the evolution,  it is appropriate to choose $q_{mn;1}=1/M$ (for all $m,n$). The choice fulfils the condition $\sum_{m,n=1}^M  q_{mn; 1}= 1$  and leads to $t_1 =  {2\over M}  \sum_{m,n}  {\rm ln} \; {y_{mn}\over |1- 2 \, y_{mn}|}  + c_0$. 

  Eq.(\ref{rhot1}) describes a $t_1$ governed diffusion of $\rho(H)$ (eq.(\ref{pdf})) in real-asymmetric matrix space, from an arbitrary initial condition and subjected to a finite drift. The diffusion approaches an equilibrium state as 
${\partial \rho \over \partial t_1} = T \rho=0$. For the case with $x_{kl}=0$, this occurs for  $y_{kl} = N/(1-\tau^2)$ for $N$ fixed and $\tau \to 0$ which from eq.(\ref{pdf}) corresponds to Ginibre limit $\rho(H) \propto {\rm e}^{- N  Tr(H H^T)}$. Indeed the latter is also the solution of $L \rho=0$ and is therefore consistent with 
eq.(\ref{rhot1}).  Following similar steps, a single parameter governed diffusion equation can also be derived for $\rho(H)$ given by eq.(\ref{pdf0}); the solution of characteristic equations in this case are however non-trivial (see for example, \cite{psco} for correlated Gaussian ensemble of Hermitian matrices).

 As $t_1$ is a function of the ensemble parameters $y_{kl}$ and $x_{kl}$,  hereafter it will be referred as the ensemble complexity parameter.   From eq.(\ref{rhot1}),  we note that $t_1$ plays the role of time in the evolution from an arbitrary initial state (with initial value, say $t_{1,0}$ of $t_1$ determined from the initial ensemble parameters) and approaching a constant value, say  $t_{1, \infty}$  ($\approx - 2 \ln 2 + c_0$ for large $N$), in the steady state limit.   The solution of eq.(\ref{rhot1}) for $t_{1,0} < t_1 < t_{1, \infty}$ can therefore be appropriately referred as the non-stationary or non-equilibrium states of the ensemble in the $t$-space.

\subsection{Complexity parameter space}

The ensemble density in eq.(\ref{pdf}) corresponds to a point 
marked by variables $\{x_{kl}, y_{kl} \}$ in  $x-y$ space.  Ensembles with different values of $\{x_{kl}, y_{kl} \}$  therefore lie at different points in this space. 
The transformation of $\{x_{kl}, y_{kl} \}$ to complexity parameter space also implies the transformation of the ensemble density in eq.(\ref{pdf}) to $\rho(H; t_1,\ldots t_M)$. In general, the ensembles with different sets of $\{x_{kl}, y_{kl} \}$   can have different values of $t_1, \ldots, t_M$, the transformed ensemble density   is therefore represented  again by different points in the $t$-space.

Any change in system conditions with time however may change the system parameters and thereby the ensemble parameters. Under variation of the parameters $\{x_{kl}, y_{kl} \}$, different ensembles will evolve along different paths in $t$-space with $t_1$ as the evolution parameter.

While each one of the above conditions imply  $t_2, \ldots, t_M$ as the invariants along the $t_1$ governed path of the evolution of $\rho(H)$, they have different physical implications for the ensemble dynamics. In contrast to case III of eq.(\ref{yc3}),  the ensemble density for cases I, II  is however independent of $t_2, \ldots, t_M$. The latter therefore correspond to constants in complexity parameter space (also referred as "complexity constants" later) and can be related to invariant system conditions e.g. symmetry and conservation laws. For study of the evolution, therefore, we only need to determine $t_1$. The paths in $t$-space for different ensembles 
(i.e with different $x-y$ sets) satisfying the conditions I, II, therefore, move along parallel to $t_1$ axis will reach same point if the initial values of $t_1$ for them is same, irrespective of their $x-y$ details. As the spectral density is governed only by $t_1$, it is predicted to be analogous for such cases. This in turn implies a deep rooted universality underlying non-Hermitian complex systems represented by the multiparametric Gaussian ensembles; the universality class is characterized by $t_1$. As for finite $N$, the latter can vary continuously between $0 \to \infty$, the above  predicts infinite number of universality classes of the entanglement statistics among pure Gaussian states characterized by $t_1$.

 In contrast to condition I, the condition II  does not impose any constraints which can be used to determine the constants $t_2, \ldots, t_M$. A knowledge of their existence, satisfying $\frac{\partial \rho}{\partial t_{\alpha}} = 0$ is however enough, their exact knowledge not needed.  The conditions III implies that the path of the evolution of $\rho(H)$ in complexity parameter space is defined by local constraints $t_2, \ldots, t_M$. Different ensembles may evolve at the same rate governed by $t_1$ but their paths need not overlap. If however different ensembles are subjected to same set of global constraints which could be identified with $t_2, \ldots, t_M$, the ensembles are then predicted to evolve along the same path.

For case III (eq.(\ref{yc3})), while the evolution is still governed by a single parameter $t_1$, the ensemble density also depends on $t_2, \ldots, t_M$. As a consequence, the paths of evolution for different ensemble, in general, are not parallel. A relevant query in the above context is whether the constants of evolution  $t_2, \ldots, t_M$ can  be chosen same for different ensembles represented by eq.(\ref{pdf}) if they are subjected to same set of global constraints e.g symmetry and conservation laws?  Intuitively the answer is in affirmative and can be justified as follows:   a  physically motivated basis to represent  an ensemble is usually based on its global constraints.  Different ensembles with a common set of global constraints can then be represented  in same/ equivalent basis. A typical matrix in general has  many basis constants \cite{basis, basis1} which can then be chosen as the constants of evolution $t_2, \ldots, t_M$. This is later explained through three simple examples (section III.A).This in turn again  leaves the evolution characterized only by $t_1$.  At some intermediate stage of evolution, it may happen that two different ensembles (i.e., represented by different $\{x, y \}$ sets in eq.(\ref{pdf}))  may correspond to same $t_1$ value; their spectral statistics are then predicted to be equal at that point if their constants of evolution are also same. More clearly, different ensembles with same $t_1$ value  are predicted to share same spectral statistics, irrespective of the details about their $\{x, y\}$ sets and therefore form a universality class characterized by $t_1$. As for finite $N$, the latter can vary continuously between $0 \to \infty$, the above  predicts infinite number of universality classes of the statistics among pure Gaussian states characterized by $t_1$. To distinguish it from the constants $t_k$ for $k >1$,  hereafter $t_1$ will be referred as $Y$.

\section{Changing system conditions: spectral Dynamics on the complex plane}

 As mentioned in section I, a diagonalizable non-Hermitian matrix  $H$, with $\Lambda$ as the eigenvalue matrix and  $V, W$ as the  right and left eigenvector matrices, can in general be written as $H = V \Lambda V^{-1}$ with $V$ subjected to a unitary constraint if $H$ is a normal matrix (similarly for $W$). 
For notational simplification in later derivations, we redefine  $V^{-1}=V^{\dagger}=U$, equivalently, $U V = V U = I$; $H$ can now be rewritten as  $H= V \Lambda U$. (As $W$ is the left eigenvector matrix, $U$ is the inverse of right eigenvector matrix and the two are related as $W^{\dagger} = D U$,  $W$ no longer appears in our derivation).

\subsection{Response of eigenvalues and eigenfunctions for a single  normal matrix}

Prior to consideration of the response of the eigenvalues and eigenfunctions of a matrix ensemble to an external perturbation, it is imperative to understand the response at a single matrix level. 

We proceed as follows. The relation $H V = U^{\dagger} . \Lambda$ gives $ \sum_{j}  H_{ij} V_{jm}=\lambda_m U_{mi}^*$ which on differentiation with respect to $H_{kl}$ and using the relations $U.U^{\dagger} =I, V.V^{\dagger} =I$ gives   the desired responses as (detailed derivation in section III of supplementary material \cite{sup})

\begin{eqnarray}
& &   {\partial \lambda_n \over \partial H_{kl}} =
 U_{nk} V_{ln},  \hspace{0.5in}
\sum_{k,l} {\partial  \lambda_n \over\partial H_{kl}} H_{kl}
 =  \lambda_n, \label{evf} \\
 & & {\partial U_{ni} \over\partial H_{kl}} =  \sum_{m\not=n}
{U_{mi} \,  U_{nk}^* \, V_{ln}^* \over { \lambda_n^* -  \lambda_m^*}},\qquad 
{\partial V_{jn} \over\partial H_{kl}} = 
 \sum_{m\not=n}{V_{jm} V_{ln} U_{mk}\over { \lambda_n -  \lambda_m}}
\label{ecf}
\end{eqnarray}

The above relations in turn lead to
\begin{eqnarray}
S_1 &=& \sum_{k,l} {\partial \lambda_{n} \over \partial H_{kl}} \; H_{kl} = \lambda_{n}\label{s2} \\
S_2 &=& \sum_{k,l} {\partial \lambda_{n} \over \partial H_{kl}} {\partial \lambda^*_{m} \over \partial  H_{kl}} = \delta_{nm}, 
\label{s5}\\
S_3 &=& \sum_{k,l} {\partial^2 \lambda_{n} \over \partial  H_{kl} \partial H_{kl}} = \beta
 \sum_{m=1}^N { \lambda_n -  \lambda_m \over  |\lambda_n -  \lambda_m|^2} \label{s6} 
\end{eqnarray}
where $\beta=1$ for $\lambda_n$ real and $\beta=2$ for $\lambda_n$ complex.
  As clear from the above set of equations,  a degeneracy of the eigenvalues can lead to a singular rate of change of eigenvector components. With symmetry usually as the primary cause for the spectral degeneracy in the physical systems,  the latter can be avoided by representing   the matrix in a symmetry resolved basis \cite{meta}.

The above equations indicate a correlated response of the eigenvalues and eigenfunctions of a real-asymmetric matrix $H$  due to changing system conditions resulting in a change of the entries $H_{kl}$. As discussed below, the information is relevant for to derive the spectral {\it jpdf}.

\subsection{Statistical response of the eigenvalues for the ensemble}

The JPDF  $P(z,y,x)$ of the eigenvalues for an ensemble $\rho(H)$ is defined in eq.(\ref{rhot1}).
Differentiating both sides of  the former with respect to $Y \equiv t_1$, followed by a substitution of eq.(\ref{rhot1}) in the  integral 
leads to   
\begin{eqnarray}
{\partial  P\over \partial Y} =  \int f_z \; {\partial  \rho\over \partial Y}  {\rm d}H  =  I_{1}  + I_{2} 
\label{pe2}
\end{eqnarray}
where 
\begin{eqnarray}
 I_1 &=& \sum_{k,l} \int f_z \, {\partial  ( H_{kl}  \rho) \over \partial H_{kl}} \; {\rm d}H  \\
I_2  &=& \sum_{k,l} \int f_z \, {\partial^2   \rho\over \partial H_{kl} \partial  H_{kl}} \; {\rm d}H      
\label{pe4}
\end{eqnarray}
Using repeated partial integration  and the boundary condition that $\rho$ vanishes as $H_{kl} \to  \pm \infty$
$I_1$ can then  be rewritten as 
\begin{eqnarray}
 I_1  &=&  \sum_{r=1}^{\beta} \sum_{n=1}^{M} {\partial \over \partial z_{nr}}
\int f_z   \;  S_1 \; \rho   \;{\rm d}H 
\label{i1}
\end{eqnarray}
with  $S_1$ given by eq.(\ref{s2}). Similarly we have 

\begin{eqnarray}
 I_2 = \sum_{r,s=1}^{\beta} \sum_{m,n=1}^N {\partial^2 \over \partial z_{nr} \partial z_{ms}}
\int  f_z \, S_2 \; \rho  \, {\rm d}H   + \sum_{r=1}^{\beta} \sum_{n=1}^M {\partial \over \partial z_{nr}}
\int  f_z \, S_3 \, \rho \, {\rm d}H  \nonumber  \\
\label{i2}
\end{eqnarray}
Here  $S_n$, $n=2 \to 3$ are  given by eq.(\ref{s5}) and eq.(\ref{s6}).  Using  the relations in eqs.(\ref{s2}, \ref{s5}, \ref{s6}) in eqs.(\ref{i1}, \ref{i2}),  $I_1$  and $I_2$ can be rewritten as 
\begin{eqnarray}
I_1 &=&  \sum_{n=1}^N \sum_{r=1}^{\beta} {\partial \over \partial z_{nr}} \left(z_{nr} P\right), \\
I_{2} &=& \sum_{n=1}^N \sum_{r=1}^{\beta} \left( {\partial^2 P_1\over \partial z_{nr}^2} - \beta \; {\partial \over \partial z_{nr}}\left( {\partial {\rm ln} |\Delta (z)| \over \partial z_{nr}} P_1\right) \right),
\end{eqnarray}
 with $\beta=1$ for $z$ real and $2$  for $z$ complex and
\begin{eqnarray}
\Delta (z) = \prod_{j<k=1}^{N} (z_j- z_k) 
\label{delta}
\end{eqnarray}

A substitution of the above relations in eq.(\ref{pe2})  now gives (detailed derivation in section III of supplementary material \cite{sup})

\begin{eqnarray}
{\partial P\over\partial Y} &=& {\mathcal L}_z  \, P
\label{pcmp}
\end{eqnarray}
with 
 \begin{eqnarray}
{\mathcal L}_z & = & \sum_{s=1}^2 \sum_{n=1}^M {\partial \over \partial z_{ns}}\left[{\partial \over \partial z_{ns}}  - \beta \, {\partial {\rm ln} |\Delta_N (z)| \over \partial z_{ns}}+  z_{ns} \right] 
\end{eqnarray}

In the limit  ${\partial P\over\partial Y} \rightarrow 0$ or $Y\rightarrow \infty$, the  evolution described by eq.(\ref{pcmp}) reaches the equilibrium steady state of the real Ginibre ensemble of real-asymmetric matrices  with  $P= P_s = |\Delta_N(e, z)| \; 
\left[ \prod_{i=1}^{N}  {\rm e}^{-  |z_i|^2}
{\rm erfc}(z_i-{\bar z}_i)\right]^{1/2}$; the steady state limit occurs for the system conditions resulting in  almost all $y_{kl;s} \rightarrow N$ and $x_{kl;s} \rightarrow 0$. Thus eq.(\ref{pcmp}) describes a cross-over from a given initial ensemble (with $Y=Y_0$) to real-Ginibre ensemble as the equilibrium limit with $Y-Y_0$ as the crossover parameter. The non-equilibrium states of the crossover, given by non-zero finite values of $Y-Y_0$, are various ensembles of the real-asymmetric matrices.

Arising from eq.(\ref{yc3}), the $P$-dynamics in complexity parameter space is also subjected to additional constraints. Focussing on case I of eq.(\ref{yc3}), we also have ${\partial \rho\over\partial t_{\alpha}}=0$ which in turn leads to ${\partial P_1\over\partial t_{\alpha}}=0$ for $\alpha >1$. 
The spectral JPDF is thus  independent of $t_2, \ldots, t_M$ too, its behaviour  governed only by $t_1$, equivalently $Y-Y_0$. As the same value of $Y-Y_0$  can arise from  different combinations of $y_{kl}$'s and $x_{kl}$'s, the spectral JPDF of different ensembles represented by eq.(\ref{pdf}) are predicted to be analogous even in a non-equilibrium state (i.e finite $(Y-Y_0)$)  in complexity parameter space if their initial states at $Y_0$ are analogous.  

The equilibrium steady state limit  can also be achieved for 
$y_{kl} \rightarrow {N\over (1-\tau^2)}$,  $x_{kl} \rightarrow  - {N \tau \over (1-\tau^2)}$  with $\tau \rightarrow \pm 1$; while the steady state for  $\tau=1$ is a GOE, it  is a Gaussian ensemble of real anti-symmetric matrices for $\tau=-1$.    The distribution $P(z;\tau=0,\pm 1)$  in  the three steady states limits is in agreement with the expected results for the corresponding ensembles from eq.(\ref{pdf}).

\subsection{Spectral correlations}

With eigenvalues  either real or  complex conjugate pair, the spectral correlations  for a real symmetric matrix ensemble can be rich in detail as well as technically complicated. In general, three main types of correlations can be defined as follows.  

{\it (i) correlations among real eigenvalues: } The $n^{\rm th}$ order spectral density correlator,   ${ R}_{ne}$ among real eigenvalues $e_1,\ldots, e_n$ can be defined as
 \begin{eqnarray}
 {R}_{ne}(e_1,\ldots, e_n) ={N! \over (N-n)!} \int  \prod_{k=1}^n \delta(e_k - \lambda_k) \; {P}(\lambda; Y) \, {\rm  D}\Omega
\label{rn1}
\end{eqnarray}
with ${\rm D} \Omega \equiv  \prod_{k=1 }^N \; {\rm d}^{\beta} \lambda_k $ where  ${\rm d}^{\beta} \lambda \equiv {\rm d} \lambda$ for $\lambda$ real and  ${\rm d}^{\beta} \lambda \equiv {\rm d}\lambda \; {\rm d} \lambda^*$ for $\lambda$ complex.

{\it (ii) correlations among complex eigenvalues:} The $n^{\rm th}$ order spectral density correlator,  
${ R}_{nz}$ among complex eigenvalues $z_1,\ldots, z_n$ can be defined as
 \begin{eqnarray}
  { R}_{nz}(z_1,\ldots, z_n) ={ N! \over (N-n)!}  \int   \prod_{k=1}^n \delta^2(z_k - \lambda_k)   \; {P}(\lambda; Y) \; {\rm D} \Omega
\label{rn2}
\end{eqnarray}
with $\delta^{2}(x) = \delta(x)  \delta(x^*)$.

 {\it (iii) mixed correlations among real and complex conjugate pairs:}
The $n^{\rm th}$ order spectral density correlator,  ${\tilde R}_{ntez}$ among $s$ real eigenvalues $e_1,\ldots, e_s$ and $t$ complex conjugate pairs  $z_1,\ldots, z_t$, with $n=s+t$ can be defined as
 \begin{eqnarray}
 {R}_{nez}(e_1, \ldots, e_s, z_1,\ldots, z_t)  = { N! \over (N-n)!}  \int  \left(\prod_{k=1}^s \delta(e_k - \lambda_k) \right) \left( \prod_{l=1}^t \delta^2(z_l - \lambda_l) \right) \; {P}(\lambda; Y) \; {\rm D} \Omega 
 \nonumber \\
\label{rn3}
\end{eqnarray}

In principle, any spectral fluctuation measure can  derived from ${ R}_{ne}, { R}_{nz}$ or ${ R}_{ntez}$.  But a determination of the latter  measures for $n, t >1$ requires, in addition to a prior knowledge of the {\it jpdf} ${ P}$,  a multi-dimensional integration thus making it  technically complicated and requires a detailed analysis. This compels us to consider only $n=1$ case in the present work.

\section{Spectral density on real axis }

The spectral density $\rho(e)$ at a  point $e$ on the real spectral line is defined as $\rho_e(e)  ={1\over N} \; \sum_{k=1}^N \delta(e-e_k)$. Following from the definition in eq.(\ref{rn1}),  the ensemble averaged spectral density $\langle \rho_e(e) \rangle$ then   corresponds to the case $n=1$:  $ {R}_{1e} = N \langle \rho_e(e) \rangle$, 

\begin{eqnarray}
 R_{1e}(e; Y)  = N \; \int \delta(e-\lambda_1) P_1 (\lambda; Y) \, {\rm  D}\Omega
\label{rr1}
\end{eqnarray}
The normalization condition   $\int {P}(\lambda, Y) \, {\rm  D}\Omega =1$  gives  $\int {\rm d}e \, {R}_{1e}(e) = N$ and thereby  $\langle \rho_e(e) \rangle=1$. 


\subsection{Evolution of the Probability Density for Real Eigenvalues}

 Differentiating eq.(\ref{rr1}) with respect to $Y$ and subsequently using eq.(\ref{pcmp}), the $Y$- governed 
evolution of $R_{1e}$ can be described as (derived in section IV of {\it supplemental material} \cite{sup})

\begin{eqnarray}
\frac{\partial R_{1e}}{\partial Y} = 
     \frac{\partial^2 R_{1e}}{\partial e^2} 
    + \frac{\partial}{\partial e} ( e R_{1e}) 
     -   \frac{\partial  }{\partial e} \bigg({I_{ee}}+{I_{ez}} \bigg) 
\label{rr2}     
\end{eqnarray}     
with symbol ${\bf P}$ implying the principle part of the integral and 
\begin{eqnarray}
{I}_{ee} \equiv {\bf P}\int \frac{{R}_{2e} }{e-e'} \, de'
\label{rr3r}
\end{eqnarray}
with ${R}_{2e} \equiv {R}_{2e}(e,e')$ as the two point correlation of the eigenvalue $e$ with another real eigenvalues $e'$ and
\begin{eqnarray}
{ I}_{ez} \equiv     {\bf P}\int \frac{2 \, e-(z+\bar{z})}{|e -z|^2}  \; {R}_{2ez} \, \; dz \,d\bar{z}
\label{rr3m}
\end{eqnarray}
with ${R}_{2ez}$ as the two point correlation of $e$ with a complex conjugate pair $z, \bar{z}$. 

We note that, except for the term $\frac{\partial {I}_{ez} }{\partial e}$,  eq.(\ref{rr2}) is similar to the one that governs the evolution of the  spectral density of the Gaussian Hermitian ensembles and is known as the Dyson-Pastur equation. The presence  of complex eigenvalues pairs therefore affects  the real eigenvalue density  only through a repulsion term. With latter expected to weaken at large distances, only the complex eigenvalues close to real axis may have any influence on  $R_{1e}$.

\subsection{Solution of eq.(\ref{rr2}) for arbitrary $Y$}

Using polar coordinates $z=r e^{i\theta}$ and  $\bar{z}=r e^{-i\theta}$, the integral  in eq.(\ref{rr3m}) can be rewritten as 

\begin{eqnarray}
{I}_{ez}(e) \equiv  2 \int_{r=0}^{\infty}\int_{\theta=0}^{2\pi} \,  C_e(e,r,\theta) \, {R}_{2ez} \; d\theta dr
\label{pi2}
\end{eqnarray}
with $C_e(e,r,\theta) = \frac{r \, (e-r \, cos\theta)}{e^2+r^2-2e r cos\theta}$. While $e$ can be both positive and negative, it is sufficient here to consider the case $e \ge 0$. (This can be explained as follow: for  case $e < 0$, we can write $C_e(e,r,\theta) = \frac{r \, (|e|-r \, cos\phi)}{e^2+r^2-2e r cos\phi}$ with $\phi=\pi-\theta$. Now as ${R}_{2ez}(e, r, \theta) = {R}_{2ez} (e, r, \pi-\theta)$, the integral for $e < 0$ is same as from $e \ge 0$).

For generic cases,  the local spectral correlations are expected to be negligible for eigenvalues separations $|e-r|$ larger than unit mean level spacing $(R_{1e})^{-1}$.The above permits the approximation $R_{2ez}(e,r,\theta) \approx R_{1e}(e) \, R_{1z}(r,\theta)$ where $R_{1z}$ is the ensemble averaged spectral density of complex eigenvalues at a given point $z$ on the complex plane (defined in eq.(\ref{rc1})). As this is not valid for $r \sim e$ (where $C_e \sim {1\over 2 e}$), we divide the above integral in two regions $0 \le r < e$ and $r \ge e$. This in turn permits following approximation (details in sections IV, V of {\it supplemental material} \cite{sup})

\begin{eqnarray}
{ I}_{ez}(e) =  {s_1 \, R_{1e}(e) \over e}   +  \sum_{m=1}^{\infty} \bigg( e^{-(m+1)}  \; s_{m+1}  -  e^{m-1} \; l_m \bigg) \, R_{1e}(e)
\label{rrs4}
\end{eqnarray}
with
 $s_{m+1}(e)= 2 \int_{0}^{e} {\rm d}r \, r^{m+1} \int_{\theta=0}^{2\pi} {\rm d}\theta \, \cos(m\theta) \, {R}_{1z}(r,\theta) $ and $l_m(e) = 2 \int_e^{\infty} {\rm d}r \int_{\theta=0}^{2\pi} {\rm d}\theta \frac{\cos(m\theta)  }{r^{m-1}} \,  R_{1z}(r,\theta)$. Based on insights given by our numerical analysis for three prototypical cases of the ensemble (\ref{pdf}) (discussed later in section VII),  $R_{1z}(r,\theta)$ does not vary significantly with $\theta$. This along with $\cos(m\theta)$ present in both $s_m$ and $l_m$ implies a faster oscillation of integrand  as $m$ increases. With $\cos(m\theta)$ rapidly changing sign on integration over entire $\theta$-range from $0 \to 2 \pi$,  the net contribution from the integrand becomes negligible even for $m=1$ and it is sufficient to consider only the first term in eq.(\ref{rrs4}). Approximating   ${I}_{ez}(e) \approx  {s_1 \, R_{1e}(e) \over e} $ and its substitution in eq.(\ref{rr2}) then leads to 
 
 \begin{eqnarray}
 	\frac{\partial R_{1e}}{\partial Y} = 
 	\frac{\partial^2 R_{1e}}{\partial e^2} 
 	+  \frac{\partial}{\partial e} \bigg(e - {s_1\over e}-{\bf P}\int \frac{{R}_{1e}(e') }{e-e'} \, de' \bigg) R_{1e} 
\label{rr3}     
 \end{eqnarray}  
with $s_1(e, Y)$ as the cumulative spectral density on the complex spectral plane at radial distance $e$, 
\begin{eqnarray}
s_1(e, Y) \equiv s_{1e} = \int_0^e  dr'  \, r' \,  \langle { R}_{1z}(r', \theta', Y) \rangle_{\theta}
\label{m0} 
\end{eqnarray}
where $\langle R_{1z}(r, \theta)\rangle_{\theta} = \int_0^{2 \pi} {\rm d}\theta \, R_{1z}(r, \theta)$. From eq.(\ref{rn2}), we have the normalization condition $\int_{0}^{\infty} \int_0^{2\pi} \, {\tilde R}_{1z}(r,\theta) \, r \, dr \, d\theta =N$, equivalently, $\int_{0}^{\infty} \int_0^{2\pi} \, { R}_{1z}(r,\theta) \, r \, dr \, d\theta =c_e$. The latter implies, from eq.(\ref{m0}), $s_1(e)=2 c_e - p_1$ with $p_{1}(e)= 2 \int_{e}^{\infty} \, dr \, r \, \langle { R}_{1z}(r,\theta) \rangle_{\theta}$. Based on the $e$-range, eq.(\ref{rr3}) can further be simplified as follows.


\noindent {\bf Case $ e^2 \gg s_1(e)$:} 
Under the condition, the term ${s_1 \over e}$  in eq.(\ref{rr3}) can be neglected with respect to the term $e$.  This reduces eq.(\ref{rr3}) to well-known Dyson-Pastur equation \cite{pastur1, haak}

\begin{eqnarray}
\frac{\partial R_{1e}}{\partial Y} = 
     \frac{\partial^2 R_{1e}}{\partial e^2} 
    +  \frac{\partial}{\partial e} \bigg( e -{\bf P}\int \frac{{R}_{1r}(e') }{e-e'} \, de' \bigg) R_{1e} 
\label{rr3r}     
\end{eqnarray}  

The general solution of the above equation for an arbitrary 
 initial condition $R_1(e;Y_0)$ can be given as $R_1(e; Y)=-{1\over \pi} \; \lim_{\varepsilon \to 0} \;{\rm Im} \, G(e-i  \varepsilon; Y)$  where
 
\begin{eqnarray}
G(z; Y) = G(z- (Y-Y_0) \, G(z; Y); Y_0)
\label{g1}
\end{eqnarray}

As an example, we determine $R_1(e, Y)$ from eq.(\ref{g1}) for a Gaussian initial level-density $R_1(e; Y_0) =  { c_e \over\sqrt{2 \pi \sigma^2}} \; {\rm e}^{-{e^2 \over 2 \sigma^2}}$; (this is the case analyzed later in our numerical analysis too discussed in section VI). The initial condition on $G(z;Y)$ with $z=e-i \varepsilon$ then becomes 
\begin{eqnarray}
G(z; Y_0) =   - {i  \; c_e \, \sqrt{ \pi \over 2 \sigma^2}}  \; {\rm e}^{-{z^2 \over 2 \sigma^2}} \; {\rm erfc}\bigg( -{i z \over \sqrt{2 \sigma^2}}\bigg)
\label{gg5}
\end{eqnarray}

Using the above in eq.(\ref{g1}) now gives
\begin{eqnarray}
G(z; Y) =   -{ i  \, c_e \; \sqrt{ \pi \over 2 \sigma^2}}  \; {\rm e}^{-{(z- (Y-Y_0) \; G)^2 \over 2 \sigma^2}} \; 
{\rm erfc}\bigg( -{i (z- (Y-Y_0) \; G \over \sqrt{2 \sigma^2}}\bigg)
\label{g5}
\end{eqnarray}

To solve the above equation for small $Y-Y_0$, we replace $G$ in the right side of the equation by $G_0$. A substitution of eq.(\ref{gg5}) on the right side and subsequent limit ${\varepsilon \to 0}$ then leads to  
(derived in section VI of {\it supplemental material} \cite{sup})

\begin{eqnarray}
R_{1e}(e; Y-Y_0) &=& {c_e \over \sqrt{2 \pi \sigma^2}} \; 
{\rm e}^{-{(e-\eta (Y-Y_0))^2\over 2 \sigma^2} - {\alpha^2 (Y-Y_0)^2 \over 2 \sigma^2}} \; 
\bigg(\delta \cos\phi - \gamma \, \sin\phi \bigg)
\label{gr5}
\end{eqnarray}
 with $\eta = (\lim_{\varepsilon \to 0}  {\rm Re} \,  G_0(z))$, 
 $\alpha = (\lim_{\varepsilon \to 0}  {\rm Im}  \, G_0(z)) =- \pi R_{1e}(e; Y_0)$ and
\begin{eqnarray}
\phi &=& { \alpha \over \sigma^2}  (Y-Y_0) (e-\eta  (Y-Y_0)), \nonumber \\
\gamma + i\delta &=& {\rm erfc}\!\left(-\frac{i e}{\sqrt{2\sigma^2}}\right)
+\frac{i(Y-Y_0) G_0}{\sqrt{2\sigma^2}}\,
{\rm erfc}'\!\left(-\frac{i e}{\sqrt{2\sigma^2}}\right) 
\label{gd1}
\end{eqnarray} 
with ${\rm erfc}'(x)= \frac{\rm d}{{\rm d}x} {\rm erfc}(x)$. 
The latter gives
$\gamma=1 + \frac{2\, \alpha (Y-Y_0)}{\sqrt{2\pi\sigma^2}}
\exp\!\left(\frac{e^{2}}{2\sigma^{2}}\right) $ and $\delta
={\rm erfi}\left(\frac{e}{\sqrt{2\sigma^2}}\right)-\frac{2\,(Y-Y_0)\,\eta}{\sqrt{2\pi\sigma^2}} \exp\!\left(\frac{e^{2}}{2\sigma^2}\right)$ (with ${\rm erfc}$ and ${\rm erfi}$ as error functions).   As discussed later in section VI, a Gaussian form of $R_1(e)$ for small $Y-Y_0$ is also indicated by our numerical analysis of  three ensembles defined by eq.(\ref{pdf}) with different $x, y$-sets.

Alternatively  another $G$-form i.e  $G(z; Y_0) =    {c_e \sqrt{2 \pi \sigma^2} \over \varepsilon \; e}  \; {\rm e}^{-{z^2 \over 2 \sigma^2}}$ also leads to the Gaussian initial condition $R_1(e; Y_0) =  {c_e \over\sqrt{2 \pi \sigma^2}} \; {\rm e}^{-{e^2 \over 2 \sigma^2}}$. To determine $R_{1e}(e; Y-Y_0)$ for large $Y-Y_0$,  it is easier to use the second form; this gives $G \approx {1\over \sigma^2} \bigg(e \pm i c_e \sqrt{2 (\sigma^2-e^2)} \bigg)$. This in turn leads to a semicircle form for the spectral density: $R_{1e}(e) \approx {c_e \; \sqrt{2 (\sigma^2-e^2)} \over \pi \sigma^2} $.  A previous study \cite{sc} on Ginibre ensembles of real asymmetric matrices  predicted a semicircle form for the spectral density projected on a real plane too. While the above result is derived for the range $ e^2 \gg s_1(e)$, our numerics discussed in section VIII indicates its validity for almost all $e$-ranges.

The form of the eq.(\ref{rr3}) is similar to that for a Hermitian  ensemble \cite{haak}. The solutions for the two cases are then expected to be analogous if the initial as well as boundary conditions are analogous for the two cases. The real spectral density for a non-Hermitian ensemble represented by eq.(\ref{pdf}) is therefore expected to approach that of the Hermitian ensemble  (represented by  the ensemble density $\rho_{hm}(H) = \rho(H) \; \prod_{k,l=1}^N \delta(H_{kl}-H_{lk})$ with $\rho(H)$ given by eq.(\ref{pdf})) for large $e$ ranges if both are subjected to same initial and boundary conditions.

\vspace{0.1in}

\noindent {\bf Case $ e^2 \ll s_1(e)$:} 
The term ${s_1\over e}$ in eq.(\ref{rr3}) now dominates over   $e$ as well as  the term $ {\bf P}\int \frac{{R}_{1r}(e') }{e-e'} \, de'$; (this can also be seen by  using the identity ${1 \over e} =\int \frac{\delta(e') }{e-e'}\, de' $ to rewrite the second term of eq.(\ref{rr3}) as $\frac{\partial}{\partial e} \bigg(e - {\bf P}\int \frac{({R}_{1r}(e') - s_1 \, \delta(e') }{e-e'} \, de' \bigg) R_{1e}$. For $e^2 \ll s_1(e)$, the evolution equation can now be approximated as 

 \begin{eqnarray}
 	\frac{\partial R_{1e}}{\partial Y} =\frac{\partial^2 R_{1e}}{\partial e^2} -\frac{\partial}{\partial e} \bigg(\frac{s_1 }{e}\bigg) R_{1e} 
\label{rr4} 	
 \end{eqnarray} 
with $s_1(e)$ defined in eq.(\ref{m0}). The latter  gives $\frac{\partial s_1}{\partial e} = e \;  \langle R_1(e, \theta) \rangle_{\theta}$ and thereby permits a  simplification of the above equation by noting that $e \frac{\partial s_1}{\partial e} = e^2 \,  \langle R_1(e, \theta) \rangle_{\theta} \ll s_1(e)$.

An explicit closed form solution of the above equation requires a prior information about  the $e$-dependence of $s_1(e)$.   Assuming specific  $e$-dependence of the latter and using the separation of variables, some important solvable cases  of eq.(\ref{rr4}) for an arbitrary initial condition can however be obtained. For example, 
from eq.(\ref{m0}),  $s_1(e)$ is approximately a constant for large $e$ ranges ($s_1(e) \approx 2 c_e$ for cases $2 c_e  \gg p_{1}(e)$).  The solution for this case can be given as 

\begin{eqnarray}
 R_{1e}(e) &=& e^{c_e+(1/2)} \, \left[ c_1\, J_{\nu} \left({ e \sqrt{E}} \right) + c_2 \, N_{\nu} \left(e \sqrt{E} \right) \right] \; {\rm exp}[-E(Y-Y_0)]
 \label{us2}
\end{eqnarray}
 with $\nu= \bigg|c_e -(1/2) \bigg|$ and  $J_{\nu}, N_{\nu}$ as the Bessel's functions,  $c_1(E), c_2(E)$ as arbitrary constants, determined by the initial condition on $R_{1e}(e)$. 
As the arbitrary constant $E$ is not subjected to any additional known constraints except semi-positive definite one, a valid particular solution can occur for continuous range of $E$ from $0 \to \infty$.  The  general solution of eq.(\ref{rr4}) for arbitrary initial condition (but finite at $e=0$) can now be given as

\begin{eqnarray}
& & R_{1e}(e; Y) =   e^{c_e+(1/2)} \, \int_0^{\infty}  {\rm d}E \; \left[ c_1\, J_{\nu} \left({ e \sqrt{E}} \right) + c_2 \, N_{\nu} \left(e \sqrt{E} \right) \right] \, {\rm e}^{ - E\, (Y-Y_0)} 
\label{dsol}
\end{eqnarray}

\vspace{0.1in}

\noindent{\bf Arbitrary $e$: } We note that   eq.(\ref{rr3}) is  valid for arbitrary $e$ if $R_1(r, \theta)$ has a  weak dependence on $\theta$. For cases where the integrals in $s_m(e)$ and $l_m(e)$ contribute significantly, we need to include higher order terms too. The series can however be simplified again for cases for which $s_m  \ll e^m $ and $l_m \gg e^{m-1}$.

\section{Spectral density on the complex plane}

The spectral density $\rho_z(z)$ of the complex eigenvalues at a  point $z$ on the  complex spectral plane is defined as $\rho_z(z)  ={1\over N} \sum_{k=1}^N \delta^2(z-z_k)$ where $\delta^2(z-z_k) \equiv \delta(z-z_k) \delta({\bar z}-{\bar z}_k)$ with $\bar z$ as complex conjugate of $z$. Its average over the ensemble in eq.(\ref{pdf})  at a given point $z$ on the spectral plane can then be expressed as   
 \begin{eqnarray}
{ R}_{1z}(z, {\bar z}) = N  \langle \rho(z) \rangle = N \int \delta^2 (z-z_1)  P_1(E,Z,Y) \, {\rm  D}\Omega
\label{rc1}
\end{eqnarray}
subjected to the normalization condition $\int {\rm d}z \, {\rm d}{\bar z} \, R_{1z}(z, {\bar z}) = N$ (equivalently $\int \rho_z(z)  \, {\rm d}z \, {\rm d}{\bar z} =1$). 


\subsection{Evolution of the probability density for complex eigenvalues}

Again differentiating eq.(\ref{rc1}) with respect to $Y$ and subsequently using eq.(\ref{pcmp}), the evolution equation for $R_{1z}$ can  be given as (derivation in section VIII of {\it supplementary material} \cite{sup})
\begin{eqnarray}
\frac{\partial R_{1z}}{\partial Y} =  \sum_{r=1}^2 \bigg[ 
   \frac{\partial^2 R_{1z}}{\partial z_{r}^2} 
    + \frac{\partial}{\partial z_{r}} ( z_{r} R_{1z}) - \frac{\partial}{\partial z_{r}} \bigg({ I}_{zz} + {\tilde I}_{ez} \bigg)\bigg]
\label{rc3}   
\end{eqnarray} 
where $z_r$ in the above equation now refers to real or imaginary parts of the eigenvalue $z$. 

\begin{eqnarray}
{I}_{zz}  &= &   {\bf P} \int \frac{z_r - z'_{r}}{|z - \bar{z}'|^2} \; R_{2z} \; d{z}'_{1} \;  d{z}'_{2} \\
{\tilde I}_{ez}  &= &   {\bf P} \int \frac{z+{\bar z} - 2 e}{|z - e|^2} \; {R}_{2ez} \, de 
\label{rc4}
\end{eqnarray}
where $R_{2z}$ is the two point correlation between the complex eigenvalues, $R_{2z} \equiv R_{2z}(z, {\bar z}, z', {\bar z}')$, and,  $R_{2ez} \equiv R_{2ez}(e,z,\bar{z})$ is defined below eq.(\ref{rr3m}).

To solve eq.(\ref{rc3}), we  transform it to the polar coordinates $r, \theta$; this  leads to (derived in section IX of {\it supplementary material} \cite{sup}),

\begin{eqnarray}
{\partial R_{1z}\over \partial Y} =  \Bigg( \frac{\partial^2}{\partial r^2} + \frac{1}{r^2}\frac{\partial^2}{\partial \theta^2} + 2 + \bigg( r + \frac{2}{r} \bigg)\frac{\partial}{\partial r}   \bigg ){R}_{1z} +  I_2 -I_3
\label{r1rt1}    
\end{eqnarray}
with $I_n \equiv I_n(r, \theta; Y) $ where
\begin{eqnarray}
I_2 &=&   -2 \, \int_{0}^{\infty}\int_{0}^{2\pi}  \; \left[ \left({1\over r} \, {\partial (r \, C_z) \over \partial r} + {1\over r} \,  {\partial D_z \over \partial \theta} \right) \; { R}_{2z} + \left( C_z \, {\partial R_{2z} \over \partial r} + {D_z\over r} \,  {\partial R_{2z} \over \partial \theta} \right) \right]d\Omega 
\label{i2ia} \\
I_3 &=& -2 \, \int_{-\infty}^{\infty}  \; \left[ \left({1\over r} \, {\partial (r \, C_e) \over \partial r} + {1\over r} \,  {\partial D_e \over \partial \theta} \right)\; { R}_{2ez} + \left( C_z \, {\partial R_{2ez} \over \partial r} + {D_z\over r} \,  {\partial R_{2ez} \over \partial \theta} \right) \right] \, de
\label{i3ia}
\end{eqnarray}
with ${ R}_{2z} \equiv { R}_{2z}(r,\theta,r',\theta') $ and ${R}_{2ez} \equiv {R}_{2z}(r,\theta,e) $.
with ${\rm d}\Omega = r' \, d\theta' \, dr' $, $C_z(r,\theta,r',\theta')=\frac{(r-r'cos(\theta -\theta'))}{r^2+r'^2-2rr'cos(\theta-\theta')}$ and  $D_z(r,\theta,r',\theta')= \frac{r' sin(\theta-\theta')}{r^2+r'^2-2rr'cos(\theta-\theta')}$, and $C_e(e,r,\theta) = \frac{(e-r \, \cos\theta)}{e^2+r^2-2e r \cos\theta}$
and $D_e(e,r,\theta)= \frac{r \, \sin\theta}{e^2+r^2-2e r \cos\theta}$.

Using the exact relations (i) $\frac{1-p \,\cos x}{1+p^2 -2 p \cos x} = \sum_{k=0}^{\infty} p^k \, \cos(k x)$, (ii) $\frac{p \, \sin x}{1+p^2 -2 p \cos x} = \sum_{k=0}^{\infty} p^k \, \sin(k x)$ and (iii) $\frac{1-p^2}{1+p^2 -2 p \cos x} = 1 + 2 \sum_{k=1}^{\infty} p^k \, \cos(k x)$  (for  $|p| <1$) \cite{grad}, eq.(\ref{i2ia}) and eq.(\ref{i3ia}) can further be simplified. Using
$\int_0^{\infty} dr [..] = \int_0^{r}  dr [..] + \int_r^{\infty}  dr [..]$  in eq.(\ref{i2ia}), we have

\begin{eqnarray}
{I}_{2} &=&
-{2\over r} \int_0^r d\Omega \,
{\partial {R}_{2z} \over \partial r}
+ 2 \sum_{k=1}^{\infty}
\left[
{1\over r^{k+1}}
\, {\partial C^{0r}_{kk} \over \partial r}
+ {1\over r^{k+2}}
\, {\partial S^{0r}_{kk} \over \partial \theta}
- {k \, C^{0r}_{kk} \over r^{k+2}}
\right.
\nonumber \\
&&
\left.
- r^{k-1}
\, {\partial C^{r \infty}_{k,2-k} \over \partial r}
- r^{k-2}
\, {\partial S^{r \infty}_{k,1-k} \over \partial \theta}
+ k r^{k-2}
\, C^{r \infty}_{k,1-k}
\right]
\label{i2ib}
\end{eqnarray}

where
$C^{ab}_{kn}(r,\theta)
=
\int_a^b dr' \, r'^n
\int_0^{2\pi} d\theta'
\, \cos\!\left(k(\theta-\theta')\right)
\, {R}_{2z}$,

$S^{ab}_{kn}(r,\theta)
=
\int_a^b dr' \, r'^n
\int_0^{2\pi} d\theta'
\, \sin\!\left(k(\theta-\theta')\right)
\, {R}_{2z}$.
Proceeding similarly, eq.(\ref{i3ia}) can be rewritten as 
\begin{eqnarray}
 { I}_{3} &=&  -2 \int_r^{\infty} \,{de\over e} \, {\partial { R}_{2ez} \over \partial r}  +
 2 \sum_{k=1}^{\infty} 
\left[{c_k \over r^{k}} \, {\partial W^{0r}_{k-1} \over \partial r}  + {s_k\over r^{k}}  \, {\partial W^{0r}_{k-1} \over \partial \theta}  -  r^{k} \, c_k \, {\partial W^{r \infty}_{-(1+k)} \over \partial r}  - r^{k} \,  s_k \,  {\partial W^{r \infty}_{-(k+1)} \over \partial \theta}  \right]  \nonumber \\
\label{i3ib}
\end{eqnarray}
with $W^{ab}_{n}  =\int_a^b \, e^n \, { R}_{2ez} \, {\rm d}e$ and 
$c_k = \cos(k \theta)+\cos(k(\pi-\theta))= 2 \cos(k \pi/2) \cos(k \pi/2 - kx)$ and $s_k= \sin(k \theta)-\sin(k(\pi-\theta)) = - 2 \cos(k \pi/2) \sin(k \pi/2 - kx)$.

Eq.(\ref{r1rt1}), with $ { I}_{2}$ and $ { I}_{3}$ given by eq.(\ref{i2ib}) and eq.(\ref{i3ib}), is exact. To solve it, however, it is imperative to consider following approximation:
\begin{eqnarray}
R_{2z}(r, \theta, r', \theta'; Y) \approx R_{1z}(r, \theta; Y) \, R_{1z}(r', \theta'; Y).
\label{2to1}
\end{eqnarray}
 With the spectral density correlations relatively weaker, the approximation is valid  at length scales $|r-r'| > \Delta(r)$ (for $\theta \sim \theta'$). Here $\Delta(r)$ is the local mean level spacing at $r$ and it usually decays as a power law for large $N$).

Using eq.(\ref{2to1}), we rewrite   $C^{ab}_{kn}(r, \theta)  = \tau^{ab}_{kn}  \; R_{1z}(r, \theta)$,  $S^{ab}_{kn}(r, \theta)  = \sigma^{ab}_{kn}  \; R_{1z}(r, \theta)$ where $\tau^{ab}_{kn} =  \langle r'^n \, {\cos}(k (\theta-\theta') \rangle_{ab}$ and $ \sigma^{ab}_{kn}  = \langle r'^n \, {\sin}(k (\theta-\theta') \rangle_{ab}$
 with $\langle r'^n \, f(k (\theta-\theta') \rangle_{ab} =\int_a^b \, dr'  r'^n \,\int_0^{2 \pi}  d \theta' f(k (\theta-\theta')) R_{1z}(r', \theta')$.  But as $0 \le R_{1z} \le 1$, and, $\cos(k (\theta-\theta')$ oscillate rapidly taking both positive and negative values as $\theta'$ varies between $0$ and $2 \pi$, (similarly for $\sin(k (\theta-\theta')$),  the higher $k$ order terms become very small in magnitude: $\tau^{ab}_{kn} \sim 0, \sigma^{ab}_{kn} \sim 0$ for all $ k \ge 1$. 
Based on the above reasoning, the dominant contribution to eq.(\ref{i2ib}) comes from the first term and relatively negligible contribution from the  series terms. Retaining only first term of eq.(\ref{i2ib}), we thus have 
\begin{eqnarray}
 { I}_{2} &\approx & -{1 \over r}  {\partial (s_1(r) \,  { R}_{1z}) \over \partial r}  
 \label{i2ic} 
\end{eqnarray}
The above also follows from a numerically observed weak dependence of
$R_1(r', \theta')$ on $\theta'$. We note that $s_1(r)/2$ gives, from
eq.(\ref{m0}), the number of eigenvalues within a radial distance $r$
from the origin and thereby implies
$\lim_{r \to \infty} s_1(r) \approx 2N$
(with $N$ as the total number of eigenvalues). Additionally,
eq.(\ref{m0}) also gives
$\partial_r s_1(r) \equiv {\partial s_1 \over \partial r}
= 2r \langle R_{1,z}(r,\theta) \rangle_{\theta}$,
and, assuming a constant spectral density within a small $r$, i.e.
$R_{1,z}(r,\theta,Y) \approx \alpha_0(Y)$, we have
$s_1(r) \approx 2\pi r^2 \alpha_0$
and
$\partial_r s_1 \approx 4\pi r \alpha_0$.

Similar to eq.(\ref{2to1}), we now approximate
\begin{eqnarray}
 { R}_{2ez}(r, \theta, e) \approx { R}_{1z}(r, \theta) \, { R}_{1e}(e)
\label{e2to1}
\end{eqnarray}
valid  at length scales $|r-e| $ bigger than a mean level spacing). This in turn  leads to $W^{ab}_{n}(r, \theta) = \omega^{ab}_{n}(r) \, { R}_{1z}(r, \theta)$ where $\omega^{ab}_{n}(r)$ are partial moments of the real spectral density: $\omega^{ab}_{n}(r)  = \langle e^n \rangle_{ab} = \int_a^b \, e^n \, { R}_{1e} \, {\rm d}e$. The latter can be determined from the ${ R}_{1e}(e)$ formulation derived in previous section, suggesting a Gaussian form for small $Y-Y_0$ and a semicircle for large $Y-Y_0$. For the Gaussian $R_{1e}(e)={c_e\over \sqrt{2 \pi \sigma^2}} \, {\rm e}^{-{e^2 \over 2 \sigma^2}}$, we have 
\begin{eqnarray}
 \omega_{n}^{r \infty}(r) &=&  
 {c_e\over 2 \sqrt{2 \pi} } \; r^{n+1} \; E\left({1-n \over 2}, {r^2 \over 2 \sigma^2} \right) \nonumber \\
 \omega_{n}^{0 r }(r) &=& {c_e\over \sqrt{2 \pi} }  \; 2^{{n-1}/2} \; \sigma^n 
 \left[\Gamma\left({n+1\over 2}\right) - \Gamma\left({n+1\over 2}, {r^2 \over 2 \sigma^2}\right)\right]
 \label{gau1}
 \end{eqnarray} 
 with $E(n,x)$ as the exponential integral function: $E(n,x)=\int_1^{\infty} {{\rm e}^{-x t} \over t^n} dt$. 
 Similarly,  for the semicircle case $R_1(e)={c_e \over \pi \sigma^2} \sqrt{2(\sigma^2 -e^2)}$,  we have, 
  for $r < \sigma$,
  
\begin{eqnarray}
 \omega_{n}^{r \infty}(r) &=&  {c_e\over 2 \sqrt{2} \pi \sigma}  \Gamma\left({n+1\over 2}\right)
 \left[ {\sqrt{\pi} \sigma^{n+1} \over \Gamma\left(2+{n\over 2}\right)} 
 - 2 \, r^{n+1}  \;  _2 F_{1,reg}\left[-{1\over 2}, {n+1\over 2}, {n+3\over 2}, {r^2 \over \sigma^2}\right]
    \right], \nonumber \\
 \omega_{n}^{0 r }(r) &=& 
 {\sqrt{2} \, c_e \,  r^{n+1} \over (n+1) \pi \sigma} 
 \left[  _2 F_{1,reg}\left[-{1\over 2}, {n+1\over 2}, {n+3\over 2}, {r^2 \over \sigma^2}\right]
    \right].  \label{sc1}
\end{eqnarray}

Substitution of the above in eq.(\ref{i3ib}), again  leads to  ${ I}_{3}$ as a sum over rapidly oscillatory terms for large and even $k$ (with odd $k$ terms exactly zero due to $\cos(k \pi/2) =0)$. Again retaining only the lowest $k^{th}$ order term of the series i.e  with $k=2$, we have, with $c_2=\cos(2\theta), s_2=-\sin(2\theta)$, 
\begin{eqnarray}
 { I}_{3} &\approx &  -2  \left( {\partial  \over \partial r}   \omega_{-1}^{r \infty}-{c_2 \over r^2} \, {\partial \over \partial r}  \omega^{0r}_{1}  + {s_2\over r^{2}}  \, {\partial \over \partial \theta} \omega^{0r}_{1}  -  r^2 \, c_2 \, {\partial \over \partial r}   \omega^{r \infty}_{-3} - r^2 \,  s_2 \,  {\partial \over \partial \theta} \omega^{r \infty}_{-3}  \right)   { R}_{1z}
 \label{i3ic}
 \end{eqnarray}

With   $\omega_{-1}^{r, \infty}, \omega_{-3}^{r, \infty}, \omega_{1}^{0 r}$ given by eq.(\ref{gau1}) and eq.(\ref{sc1}), these depend on the form of real spectral density i.e Gaussian (for small $Y-Y_0$) and semicircle (large $Y-Y_0$).  This in turn leads to different expressions for $I_3$ from eq.(\ref{i3ic}) and thereby different reductions of eq.(\ref{r1rt1}). The detailed steps are as follows.

\vspace{0.1in}

\subsubsection{\bf Small $Y-Y_0$}

As discussed in previous section, $R_{1e}(e)$ in this range behaves as a Gaussian and, consequently,   $\omega_{-1}^{r, \infty}, \omega_{-3}^{r, \infty}, \omega_{1}^{0 r}$ are given by eq.(\ref{gau1}): $ \omega_{1}^{0 r }= {0.564 \, c_e \, \sigma}   \left[ 1 - {\rm e}^{-{r^2\over 2\sigma^2}}  \right] $, 
$ \omega_{-1}^{r \infty}= {0.2} \, c_e \, \Gamma\left(0, {r^2 \over 2 \sigma^2} \right)$,  
  $ \omega_{-3}^{r \infty}={0.2 \, c_e\over  r^2} \left[{\rm e}^{-{r^2\over 2\sigma^2}} - 0.5 \, {r^2 \over \sigma^2} \Gamma\left(0, {r^2 \over 2 \sigma^2} \right) \right]$. A substitution of the above in  $I_2-I_3$  lead to   terms containing both positive and negative powers of $r$.  To simplify, we consider  small and large $r$ ranges of $I_2-I_3$ separately.

{\center{\it Large $\mathbf{r}$ range ($\mathbf{r^2 > 2 \sigma^2}$) } } 
From eq.(\ref{i2ic}) and eq.(\ref{i3ic}), we have in this case

\begin{eqnarray} 
I_2 -  I_3 & \approx &  -2 \left({s_{1r} \over 2 r}  + {0.564 \, c_e \, c_2 \over r^2}  \right)  {\partial  { R}_{1z} \over \partial r}   +  {1.128 \, c_e \, s_2 \over r^2}   {\partial  { R}_{1z} \over \partial \theta}  - {1 \over r} {\partial  s_{1r} \over \partial r} { R}_{1z} 
\label{i3id}
\end{eqnarray}
with  $s_{1r} \equiv s_{1}(r)$, $c_2=\cos(2\theta), s_2=-\sin(2\theta)$.
For  the cases in which  ${ R}_{1z}$ decays with large $r$, the upper integration limit in eq.(\ref{m0}) can be replaced by $\infty$, leading to the approximation  $s_{1}(r) \approx s_{1\infty} =2 \int_0^{\infty} dr' d\theta' r' \,  { R}_{1z}(r', \theta') = 2 c_e$ (from eq.(\ref{m0})). Substituting the above in eq.(\ref{r1rt1}), we now have

\begin{eqnarray}
{\partial R_{1z}\over \partial Y} =  \Bigg( \frac{\partial^2}{\partial r^2}  + 2 + \left( r + \frac{2(1- c_e)}{r} \right)\frac{\partial}{\partial r}   \bigg ){R}_{1z} + \frac{1}{r^2} \left( \frac{\partial^2}{\partial \theta^2} + 1.128 \, c_e \, s_2 \frac{\partial}{\partial \theta} \right) {R}_{1z}
\label{r1rt2}    
\end{eqnarray}

As discussed below eq.(\ref{i2ic}), for $Y-Y_0 < Y_c-Y_0$, the  right side is approximately independent of $Y$ and the above equation can then be solved by a separation of variables approach.  Substituting the relation 
\begin{eqnarray}
R_{1z}(r,\theta,Y-Y_0)= A(r) \; T(\theta) \; e^{-E (Y-Y_0)},
\label{r1zy}
\end{eqnarray}
 with $E$ as an arbitrary positive constant, reduces it in two separate equations for $r, \theta$,
\begin{eqnarray}
 r^2 \frac{\partial^2 A(r)}{\partial r^2}  +  r \, (r^2 + 2(1-c_e))  \frac{\partial A(r)}{\partial r}   + ((E+2) r^2 -E_t) \, A(r) =0 \label{rrc1} \\
\frac{\partial^2 T}{\partial \theta^2} - 1.128 \, c_e \, \sin(2\theta) \, \frac{\partial T}{\partial \theta} + E_t \, T =0 
\label{rtc1}
 \end{eqnarray}
with  $E_t$ as another arbitrary constant. The latter is determined by the boundary conditions that  ${\mathcal R}_{1z}(r, \theta)$ vanishes at large radial distances and is single valued in $\theta$, equivalently, $\lim_{r\to \infty} \, A(r) \to 0$ and $T(\theta; E_t) = T(2 \pi +\theta; E_t)$.   

To solve the above pair of equations, we redefine  $E, E_t$ as $E_t= { (\mu^2-1) \over 4}\,  (2 c_e -1)^2$ and $E=4  \nu  \, c_e$ (without loss of generality as both constants being arbitrary) with $\mu, \nu $ as arbitrary constants  to be determined by boundary conditions; the conditions that $E,  E_t \ge 0$ however also requires $\mu \ge 1$ and $\nu \ge 0$.  A general solution of eq.(\ref{rrc1}) is (discussed in supplementary material \cite{sup}),
\begin{eqnarray}
A_{\mu \nu}  &=&  c_{\mu \nu} \, \left({ r^2 \over 2}\right)^{-\frac{(2 c_e -1)(\mu-1)}{4}} \,   F_1\left(\frac{(4 \nu-\mu+1) c_e}{2}, \mu c_e, -\frac{ r^2}{2}\right)  + \nonumber \\
& & d_{\mu \nu} \, \left({ r^2 \over 2}\right)^{\frac{(2 c_e -1)(\mu+1)}{4}} \,   F_1\left(\frac{(4 \nu+\mu+1) c_e}{2},  \mu c_e, -\frac{ r^2}{2}\right)
\label{umnn}
\end{eqnarray}

Eq.(\ref{rtc1}) is a linear second-order ordinary differential equation with variable coefficients and has  no
elementary closed-form solution. The general solution can be given as 
\begin{eqnarray}
T_{\mu}(\theta) ={\rm e}^{-0.28 \, c_e \, \cos (2\theta)} \; v(\theta)
\label{tmn1}
\end{eqnarray}
 where $v(\theta)$ satisfies 
\begin{eqnarray}
{{\rm d}^2 v\over {\rm d}\theta^2} + (E_t + 1.128 \, c_e \, \cos (2\theta)) v =0
\label{tmnv}
\end{eqnarray}

Eq.(\ref{tmnv}) is a form of the Mathieu equation with its general solution as 
\begin{eqnarray}
v(\theta) = c_{\mu} \; {\mathcal C}(E_t, -0.564 \, c_e, \theta) + d_{\mu} \; {\mathcal S}(E_t, -0.564 c_e, \theta)
\label{vv}
\end{eqnarray}
with $ {\mathcal C}(E_t, -0.564 c_e, \theta)$  and $ {\mathcal S}(E_t, -0.564 c_e, \theta)$ as the even and odd Mathieu functions, respectively. With $v(\theta)$ subject to $2 \pi$-periodicity, the solution exist only for special discrete values of $E_t$ and is either even periodic ($v(\theta) = v(-\theta)$) or odd periodic ones ($v(\theta) = - v(-\theta)$).

From eq.(\ref{r1zy}), the general solution for $R_{1z}(r,\theta)$ in  large $N$ limit, for arbitrary initial conditions and in the neighborhood of $\theta \sim \theta_0$ can now be given as
\begin{eqnarray}
R_{1z}(r, \theta; Y) \approx   \sum_{\mu, \nu}  \, A_{\mu \nu}(r)  \; T_{\mu} \, {\rm exp}\left[- 4 \, \nu \, c_e   (Y-Y_0)\right] 
\label{ri} 
\end{eqnarray} 
with  the arbitrary constants $b_{\mu}, \phi_{\mu}, c_{\mu \nu}$ and  $d_{\mu \nu} $ determined from the initial solution ${ R}_{1z}(r, \theta; Y_0)$.  In large $r$-limit,  the first term  in eq.(\ref{umnn}) for $\nu \not=0$,  is  negligible in comparison to the second term but  for $\nu=0$ it approaches a constant value.  To satisfy the condition $\lim_{r \to \infty}{\mathcal R}_{1z}(r, \theta) \to 0$ we choose $c_{\mu 0}=0$.  

As eq.(\ref{ri}) indicates, $\nu$ appears in the exponent with $Y-Y_0$ and also in its $\theta$-independent part;   the summation over $\nu$  replaces them  by a $Y$-dependent function. The $\theta$-dependent part of the equation is however independent of $\nu$ and therefore  remains independent of $Y-Y_0$.  This implies an analogous  angular dependence of  $R_{1z}(r, \theta; Y)$ for all $Y-Y_0$ values and arbitrary initial conditions.

\vspace{0.2in}

{\center{\it Small $\mathbf{r}$ range ($\mathbf{r^2 < 2 \sigma^2}$:) } }  From eq.(\ref{i2ic}) and eq.(\ref{i3ic}), we have for Gaussian $R_{1e}(e)$ form,
\begin{align} 
I_2 -  I_3  \approx   - \left({s_{1r} \over r}  + {1.128 \, c_e \over \sigma}\left(0.6+ 2 c_2 + { \log r^2 \over 2 \sigma^2} \right)  \right)  {\partial  { R}_{1z} \over \partial r}  - {c_e \over r} \left({1.128 \over \sigma} + {0.2 \, r^2\over \sigma^5} - {0.4 \over \sigma^3 r} + {1\over c_e} {\partial  s_{1r} \over \partial r} \right){ R}_{1z}  \nonumber \\
\label{i3id}
\end{align}
For $\sigma \sim N^{\alpha}$, with $\alpha >0$ and system dependent, the above expression can further be approximated as 
\begin{eqnarray} 
I_2 -  I_3 & \approx &  -{s_{1r} \over r}  {\partial  { R}_{1z} \over \partial r}  -  \left({\partial  s_{1r} \over \partial r} \right) \, {{ R}_{1z}  \over r}
\label{i3if}
\end{eqnarray}
The above on substitution in eq.(\ref{r1rt1}) leads to 
\begin{eqnarray}
{\partial R_{1z}\over \partial Y} =  \Bigg( \frac{\partial^2}{\partial r^2}  + \frac{1}{r^2}  \frac{\partial^2}{\partial \theta^2} + 2 + \left( r + \frac{(2-s_{1r})}{r} \right)\frac{\partial}{\partial r}   \bigg ){R}_{1z}  - \left( {\partial  s_{1r} \over \partial r} \right) { { R}_{1z}  \over r}
\label{r1rt3}    
\end{eqnarray}
The solution of the above equation depends on $s_{1r} \equiv s_1(r)$ defined in eq.(\ref{m0}).
Here again, neglecting $Y$-dependence of $s_{1r}$, it can be solved by the separation of variable approach.  A substitution of  eq.(\ref{r1zy}) now gives 
\begin{eqnarray}
  r^2 \frac{\partial^2 A(r)}{\partial r^2}  +  r \, (r^2 + 2 - s_{1r})  \frac{\partial A(r)}{\partial r}  + ((E+2) r^2 -E_t - r \; (\partial_r \, s_{1r}) ) \, A(r) =0 \label{rrc2s} \\
\frac{\partial^2 T}{\partial \theta^2} + E_t \, T =0 
\label{rtc2s}
 \end{eqnarray}
 with $E$ and $E_t$ as  arbitrary constants, determined by the boundary conditions that  $A(r)$ vanishes at large radial distances and $T(\theta)$ is single valued in $\theta$.

To solve eq.(\ref{rrc2s}), a prior information about the $r$-dependence of $s_{1r}$ is needed. 
As an example, we again consider  the case of complex eigenvalues within small $r$ with homogeneous density $\alpha_0$.  With $s_{1r}=2 \pi r^2 \alpha_0$, the general solution for the above equation pair turns out to be

\begin{eqnarray}
R_{1z}(r, \theta; Y) &\approx &   \sum_{\mu, \nu} \,  b_{\mu} \, A_{\mu \nu}(r) \;  \cos\left(\frac{1}{2} \, \sqrt{\mu^2-1} \,  \, \theta \right)  \, {\rm e}^{- E  (Y-Y_0) }  
\label{rij} 
\end{eqnarray}
with  $E_1= {1\over 4} (\mu^2-1)$, $E=2 \chi \nu$ and 

\begin{eqnarray}
A_{\mu \nu}(r) &=& \left({ \chi \, r^2 \over 2}\right)^{-\frac{1}{4}}
\left[ {\overline p}_{\mu \nu} \, \left({ \chi \, r^2 \over 2}\right)^{-\frac{ \mu}{4}} \; F_1\left(\frac{(4 \nu + 3 -\mu)}{4}, \frac{2-\mu}{2},   -\frac{\chi r^2}{2}\right)  \right.\,  + \nonumber \\
&+ & {\overline q}_{\mu \nu} \,  \left({ \chi \, r^2 \over 2}\right)^{\frac{\mu}{4}} \, \left. F_1\left(\frac{(4 \nu+3+\mu)}{4}, \frac{2+\mu}{2} ,-\frac{\chi r^2}{2}\right)
\right]
\label{uja1}
\end{eqnarray}
with $\chi= 1- \pi \alpha_0$ and ${\overline p},  {\overline q}$ as arbitrary constants determined from the initial solution ${ R}_1(r, \theta; Y_0)$.  With above solution applicable for small $Y-Y_0$,  $E$ can take finite negative values too.

With $\nu$  in eq.(\ref{rij}) appearing only through $E$ and $A_{\mu \nu}$,  $\sum_{\nu} A_{\mu \nu}(r) {\rm e}^{- E  (Y-Y_0) }$ transforms  to a $Y$-dependent function, leaving $\theta$-dependent part independent of $Y$.   The latter implies  $\theta$-dependence of $R_1(r, \theta; Y)$ does not vary significantly with $Y$.  As discussed in section VIII, this is confirmed by our numerical analysis too which in turn supports our separability conjecture  too (recalling that eq.(\ref{rij}) is  based on the conjecture).

\vspace{0.2in}

\subsubsection{{\bf Large $Y-Y_0$}}

As discussed in section V, $R_{1e}$ is described by a semicircle in large $Y-Y_0$ limit
and, with $R_1(e)={c_e \over \pi \sigma^2} \sqrt{2(\sigma^2 -e^2)}$, 
$\omega_n^{ab}$ is defined only for $r < \sigma$. From eq.(\ref{sc1}), we now have
$\omega_{1}^{0 r }=
{c_e \sqrt{2} \over 3 \pi \sigma^2}
\left[
\sigma^3
-
\sqrt{(\sigma^2-r^2)^3}
\right]$,
$\omega_{-1}^{r \infty}=
{c_e \over \pi \sigma^2 \sqrt{2}}
\left[
\sigma \log\left(
{\sigma \over r}
+
{\sqrt{\sigma^2-r^2} \over r}
\right)
-
\sqrt{\sigma^2-r^2}
\right]$.
 $ \omega_{-3}^{r \infty}= {c_e \over \pi \sigma^3 \sqrt{2} }  
 \left[ \log\left({r \over \sigma -\sqrt{\sigma^2-r^2}}\right) + {\sigma \, \sqrt{\sigma^2-r^2}\over r^2}   \right] $.
  From eq.(\ref{i2ic}) and eq.(\ref{i3ic}) along with the above relations, we now have, for  $r < \sigma$ range,
\begin{align} 
I_2 -  I_3 & \approx &  - \left({s_{1r} \over r}  - {2\sqrt{2} c_e (1 +  c_2) \over \pi \sigma} +{c_e \, c_2 \, r^2 \over  \pi \sigma^3 \sqrt{2} } \right)  {\partial  { R}_{1z} \over \partial r}   -  {c_e \, s_2 \, r^2 \over 2 \, \sigma^2}   {\partial  { R}_{1z} \over \partial \theta}  -  \left({\partial  s_{1r} \over \partial r}\right) \, \frac{ R_{1z}}{r}
\label{i3ie}
\end{align}
With $\sigma \sim o(N)$, the $2^{nd}$ and $3^{rd}$ terms inside the circular brackets and the $\theta$-derivative term with a prefactor ${r^2 \over \sigma^2}$  can be neglected.  The above expression then again reduces to the same form as in Gaussian case i.e eq.(\ref{i3if}). The evolution equation for $R_{1z}$ is then given by eq.(\ref{rrc2s}) and its general solution for an arbitrary initial condition by eq.(\ref{rij}).

Following from  eq.(\ref{rij}),    $R_1(r, \theta; \infty)$ here again corresponds to  $E=0$ (equivalently  $\nu=0$)  and can be given as 
\begin{eqnarray}
R_{1z}(r,\theta; \infty) \approx    \sum_{\mu} \,  b_{\mu} \,  A_{\mu 0} \;
  \cos\left(\frac{\theta }{2} \, \sqrt{\mu^2-1} \right) 
\label{rinf3}
\end{eqnarray}
with  $A_{\mu 0}$  given by eq.(\ref{uja1}).  For a finite $R_1(r,\theta; \infty) $ at $r=0$,  we then need  $\overline{p}_{\mu0}=0$ and $\mu \ge 1$. This leads to
\begin{eqnarray}
R_{1z}(r,\theta; \infty) \approx
\sum_{\mu} b_{\mu}
\left(\frac{\chi r^2}{2}\right)^{\frac{\mu-1}{4}}
\, {}_1F_1\!\left(
\frac{3+\mu}{4},
\frac{2+\mu}{2},
-\frac{\chi r^2}{2}
\right)
\cos\left(
\frac{\theta}{2}\sqrt{\mu^2-1}
\right)
\label{rinf3}
\end{eqnarray}
where $b_{\mu} {\overline q}_{\mu 0}$   (both arbitrary) is redefined as $b_{\mu}$.

\subsection{Calculation of $\langle R_1(r,  \theta, Y) \rangle_{\theta}$ and $\langle R_1(r,  \theta, Y) \rangle_{r}$}

Using the definition  $\langle R_1 \rangle_{\theta}=\int_0^{2\pi} R_1 \, {\rm d}\theta$ and  $\langle R_1\rangle_{r}=\int_0^{\infty} R_1 \, r \, {\rm d}r$, the radial and angular dependence of $R_1$  can separately be analyzed.

{\bf Radial Dependence:}  Different formulation  for $R_1(r,  \theta, Y)$ for different ranges of $r$ and  $Y-Y_0$   leads to three different results for $\langle R_1(r, \theta,Y) \rangle_{\theta}$. For small $Y-Y_0$ and  small $r$,  $R_1(r,  \theta, Y)$ is described by eq.(\ref{rij}). A $\theta$-integration from $0 \to 2\pi$ of the equation gives 

\begin{eqnarray}
\langle R_1(r, \theta,Y) \rangle_{\theta}  &=&   \sum_{\nu}    \, A_{1\nu}  \, {\rm e}^{- \nu \, \phi }   
 \label{rima}
\end{eqnarray}
with  $A_{1\nu}$ given by eq.(\ref{uja1}) and $\phi= 2 \chi (Y-Y_0)$.   For small $Y-Y_0$,  approximating ${\rm e}^{- \nu \phi }  \approx 1 - \nu \phi + O(\phi^2)$ leads to

\begin{eqnarray}
\langle R_1(r,\theta,Y) \rangle_{\theta}  
\approx  \langle R_1(r,\theta,Y_0) \rangle_{\theta} - 2 \chi (Y-Y_0)  \sum_{ \nu=1}^{\infty}  \nu \; q_{1 \nu} \; F_1\left(\nu+1,  {3 \over 2},  -\frac{\chi r^2}{2}\right)   
\label{rio3a} 
\end{eqnarray}
with unknown constants $q_{1 \nu}$ determined by the initial spectral density. 

For   small $Y-Y_0$ and large $r$, $R_1(r,  \theta, Y)$ is given by eq.(\ref{ri}) with $A_{1\nu}$ now given by eq.(\ref{umnn}). Here again a $\theta$ integration of eq.(\ref{ri}) leads to eq.(\ref{rima}) and thereby eq.(\ref{rio3a}) but with $\phi=4 c_e (Y-Y_0)$.  This leads to 
\begin{eqnarray}
\langle R_1(r,\theta,Y) \rangle_{\theta}  
\approx  \langle R_1(r,\theta,Y_0) \rangle_{\theta} - 4 c_e (Y-Y_0) \;  \sum_{ \nu=1}^{\infty}  \nu \; d_{1\nu} \left({ r^2 \over 2}\right)^{2 c_e-1\over 2} \,   F_1\left(\frac{(2 \nu+1)c_e}{2},  c_e, -\frac{ r^2}{2}\right)  \nonumber \\
\label{rio3b} 
\end{eqnarray}

For large $Y-Y_0$ and with  $r$ less than the real spectral range,  $R_1(r,  \theta, Y)$ is again described by eq.(\ref{rij}) which leads to 
$\langle R_1(r,\theta,Y) \rangle_{\theta}$ is given by
eq.(\ref{rima}), with $A_{1\nu}$ given by eq.(\ref{uja1})
and the same $\phi = 2\chi (Y-Y_0)$.
With only the $\nu=0$ term contributing in the limit
$Y-Y_0 \to \infty$, we then have
$\langle R_1(r,\theta; \infty) \rangle_{\theta}
= A_{10}
= {\tilde q}_{10}
\, {}_1F_1\left(
1,
{3\over 2},
-\frac{\chi r^2}{2}
\right)$.

Eq.(\ref{rima}) for large $Y-Y_0$ can then be rewritten as
\begin{eqnarray}
\langle R_1(r,\theta,Y) \rangle_{\theta}
&=&
\langle R_1(r,\theta; \infty) \rangle_{\theta}
+ \sum_{\nu=1}^{\infty}
A_{1\nu}
\, {\rm e}^{-\nu \phi}
\label{rlargeY}
\end{eqnarray}
or
\begin{eqnarray}
\langle R_1(r, \theta,Y) \rangle_{\theta} =  \langle R_1(r,  \theta; \infty) \rangle_{\theta}  +  \sum_{\nu=1}^{\infty}    \,  q_{1 \nu} \; F_1\left(\nu+1,  {3 \over 2},  -\frac{\chi r^2}{2}\right)  \, {\rm e}^{- 2 \nu \chi (Y-Y_0) }   
 \label{rio3c}
\end{eqnarray}
Due to large $Y-Y_0$, however, the terms with lower $\nu$ dominate the sum $\sum_{\nu}$ and the latter can be approximated only by first few terms. The density in this case approaches uniformity on the complex plane within a circle of radius $\sqrt{N}$ with increasing $Y-Y_0$.

{\bf Angular Dependence:}  
To calculate $\langle R_1(r, \theta,Y) \rangle_{r}  = \int_0^{\infty}  {\rm d}r \, r \, R_1(r, \theta,Y)  $, we need in principle a functional dependence of $R_1(r, \theta,Y)$ for entire $r$-range. But, with $R_1(r, \theta,Y) $  rapidly decaying away from $r=0$,  its small $r$-behavior dominates  the  integral. As eq.(\ref{rij})  describes the $R_1(r, \theta,Y)$-behavior for small $r$,  for  both small as well as large $Y-Y_0$, its substitution in the above integral     leads to
\begin{eqnarray}
\langle R_1(r, \theta,Y) \rangle_{r}  \approx   \sum_{\mu=1}^{\infty} \cos\left((1/2) \sqrt{\mu^2-1} \, \alpha_1 \, \theta \right) \; X_{\mu}(Y) 
\label{rio3d} 
\end{eqnarray}
where $X_{\mu}(Y)$ is independent of $\theta$: $X_{\mu}(Y) = \sum_{\nu} \,   b_{\mu } \, H_{\mu \nu}   \, {\rm e}^{- 2 \nu \chi \, (Y-Y_0) } $  with $ H_{\mu\nu}=\int_0^{\infty} A_{\mu \nu}  \; r \;   {\rm d}r $ and $A_{\mu \nu}$ given by eq.(\ref{uja1}) and unknown constants $b_{\mu}$ determined from the initial condition.   As clear from the above,   the dominant contribution to eq.(\ref{rio3d}) comes from $\mu=1$ term, thereby suggesting 
\begin{eqnarray}
\langle R_1(r, \theta,Y) \rangle_{r}  \approx    X_1(Y).
\label{rio3e} 
\end{eqnarray}
The above result, indicating $\langle R_1(r, \theta; Y) \rangle_{r}$   almost constant  in $\theta$,
 is consistent with our numerical results  discussed in section VIII.

\section{Approach to real Ginibre ensemble}

As $y_{kl} \rightarrow {N\over (1-\tau^2)}$, $x_{kl} \rightarrow -{N \tau\over (1-\tau^2)}$, the ensemble density in eq.(\ref{pdf}) approaches three different limits, namely, a real Ginibre ensemble of real asymmetric matrices for $\tau=0$, a Gaussian orthogonal ensemble of real-symmetric matrix for  $\tau=1$ and a Gaussian ensemble of real anti-symmetric matrices for $\tau=-1$. From eq.(\ref{y1}), the ensemble parameters $y_{kl}, x_{kl}$ for each of these cases correspond to limit $Y \to \infty$. A substitution of ${\partial P_1 \over \partial Y}=0$ in eq.(\ref{pcmp}) and solving the resulting equation also leads to the expected solution  $P(z;\tau=0,\pm 1)$ for each of the three steady states.

The study \cite{sc} analyzed the average spectral density $R_1(e,z)$ for $N \times N$ real asymmetric matrices $\rho(H) \propto {\rm exp} \bigg[-{N\over 2(1-\tau^2)} {\rm Tr}(H H^T -\tau H H) \bigg]$ in the limit $N \to \infty$ and found that $R_1$ is uniform inside an ellipse, in the complex plane, whose real and imaginary axes are $1+ \tau$ and $1-\tau$, respectively: $R_1(e, z) = (\pi (1-\tau^2))^{-1}$ for 
$\bigg({z_r\over 1+\tau}\bigg)^2 + \bigg({z_i\over 1+\tau}\bigg)^2 \le 1$ and $0$ otherwise (with $z=z_r+iz_i$). The projection of $R_1(e,z,z^*)$ on the real/ imaginary axis leads to a generalized semi circle  $\int R_1(e,z_r,z_i) \, {\rm d}z_i ={2\over \pi (1+\tau)^2} \bigg((1+\tau)^2-z_r^2 \bigg)^{1\over 2}$ for $z_r \le 1+\tau$ and $\int R_1(e,z_r,z_i) \, {\rm d}z_r ={2\over \pi (1-\tau)^2} \bigg((1-\tau)^2-z_r^2 \bigg)^{1\over 2}$ for $z_i \le 1-\tau$. For finite $N$ however the study found significant deviation from the uniformity of the density of states near the real axis;  the observed density of states on the real axis was found to be higher than the average density, whereas the density slightly above and below the real axis is less than the average. This excess density originates from the reduced degree of  the level repulsion of eigenstates near the real axis in comparison to the average level repulsion. To verify that the non-uniformity is a finite-size effect, the study also  measured the average number of real eigenvalues for different  $N$ and showed that the excess density of real eigenvalues vanishes in $N \to \infty$ limit.

With present study focussed on real-asymmetric ensembles, here we confine the discussion of the steady state limit to the real-Ginibre ensemble ($\tau=0$): $R_{1, \text{Ginibre}} = \frac{1}{\pi}$. It is relevant to compare this result with the asymptotic behavior of our findings as $Y \to \infty$.  Based on our theoretical analysis of complex eigenvalues  (details given in supplementary material \cite{sup}), we find that \( R_1(r, \theta, \infty) \) becomes a constant within the region \( r \sim \sqrt{N} \) and exhibits exponential decay beyond this range. Furthermore, it is invariant with respect to \( \theta \). Our results  are thus in agreement with previous results in  \cite{haak} for large \( N \).

Based on the analysis discussed in previous section, we find that \( R_1(r, \theta, \infty) \) becomes a constant within the region \( r \sim \sqrt{N} \) and exhibits exponential decay beyond this range. Furthermore, it is invariant with respect to \( \theta \). Our results  are thus in agreement with previous results in  \cite{haak} for large \( N \).

\section{Numerical Analysis}

As mentioned in section II,  the multiparametric ensemble density described  eq.(\ref{pdf})  can in general represent a wide range of non-Hermitian complex systems. But based on our theoretical analysis discussed in previous section, the ensemble averaged spectral density is insensitive to their individual details and is governed by the complexity parameter $Y-Y_0$ only with $Y_0$ as the $Y$-value for the initial ensemble. Recalling that quantitatively a same $Y-Y_0$ can results from different combinations of the ensemble parameters, this in turn implies that  $R_1$ behaviour  for two different ensembles with same $Y-Y_0$ is expected to be analogous if both evolve from same initial condition $R_1(Y_0)$ at $Y=Y_0$. In addition, with constants $t_2, \ldots, t_M$ marking the path of the evolution in $Y$-space,  different ensembles are expected to evolve along the same path in $Y$-space as their ensemble parameters are varied if they share same set of complexity constants ({\it appendix B}). The robustness of our prediction makes it necessary to pursue a detailed numerical investigation.


 To pursue the above objective, we  numerically analyse four prototypical Gaussian  ensembles of real asymmetric nomal random matrices. 
 Due to computational time limitations as well as other technical issues, we choose three of them as univariate ensembles i.e with independent, Gaussian-distributed elements with zero mean but varying functional dependencies for their variances. Due to independence of the elements,  the ensemble density $\rho(H)$ is given by eq.(\ref{pdf}) with $x_{kl} = 0 $ for all $k,l$ but $y_{kl}$ varying among elements, with $Y \equiv t_1$ given by eq.(\ref{y1}).  The necessary $y_{k}$ details for these cases, referred jointly as univariate cases, are as follows (with details of $Y$-calculation in each case given in {\it supplementary material}). The fourth case is a multivariate case, with $\rho(H)$ given by eq.(\ref{pdf0}), and will be described later.

{\bf (i) Ensemble with a constant ratio for off-diagonal to diagonal variances:} The ensemble density in this case corresponds to  eq.(\ref{pdf}) with $y_{kk}=\frac{1}{2}$, $y_{kl} = \frac{1}{2} (1+\frac{N}{b})^2 $ for $k \not=l$.  Substitution of the above parameters in eq.(\ref{y1}) gives 
\begin{eqnarray}
Y-Y_0 &=& 2\ln\left[\frac{(b+N)^2(2b_0+N)}{(b_0+N)^2(2b+N)}\right].
\label{ybe}
\end{eqnarray}
with $Y_0$ as $Y$-value for the initial ensemble chosen with $b=b_0$. For brevity this case is later referred as the Brownian ensemble (BE).

{\bf (ii) Ensemble with power law decay for off-diagonal to diagonal variances:}  
 Referred later as the power law ensemble (PE), this case corresponds to  eq.(\ref{pdf}) with $y_{kk}=\frac{1}{2}$, 
$y_{kl}=\frac{1}{2}(1+\frac{|k-l|^2}{b^2})^2 $ for $k \not=l$. Taking here again the initial ensemble with  $b=b_0$, we then have 
\begin{equation}
Y - Y_0 ={1\over{N^2}} \sum_{r=1}^{N} (N-r)\\ln \!\left[
\frac{\left( b^2 + r^2 \right)^2\left( 2b_0^2 + r^2\right)}{
\left( b_0^2 + r^2 \right)^2\left( 2b^2 + r^2 \right) } \right]  
\label{ype}
\end{equation}

{\bf (iii) Ensemble with exponential decay of the off-diagonal to diagonal variances:}
 Referred later as the  exponential ensemble (EE), this case corresponds to  eq.(\ref{pdf}) with 
 $y_{kk}=\frac{1}{2}$, $y_{k,l}=\frac{1}{2}e^{\frac{(|k-l|}{b})^2} $ for $k \not=l$.
 Taking here again $b=b_0$ for initial state, eq.(\ref{y1}) now gives
\begin{equation}
Y - Y_0 = {1\over{N^2}} \sum_{r=1}^{N} (N-r) \Big[ \ln| F(r,b)| - \ln|F(r,b_0)| \Big] 
\label{yee}
\end{equation}
with $F(r,b) = \frac{1}{2} \frac{e^{r^2/b^2}}{|1 - e^{r^2/b^2}|}$

As clear from the above, while each ensemble is influenced by two free parameters i.e band width $b$ and the matrix size $N$, their functional dependence on $b$ and $N$ is  quite different: it varies from a constant variance for the off-diagonals  to a power-law decay (for PE) and an exponential decay (for EE) away from the diagonals. In addition, the off-diagonal variances in PE and EE depend on the distances in basis space and thereby  basis parameters too. Indeed the latter also affect the constants $t_2, \ldots, t_M$ for BE, PE and EE; a solution of eq.(\ref{tk1}) to determine these constants is given in  {\it appendix A}.

Following from eq.(\ref{pdf}) along with  the variance details given above, the real-Ginibre limit for BE, PE and EE is expected  in  $b\to \infty$ limit when almost all off-diagonal elements of a typical matrix in the ensemble become of the same order of magnitude.  Indeed the  $b \to \infty$ limit  for each ensemble also corresponds to $Y \to \infty$ limit (see secrion IX of supplementary material). As the solution of eq.(\ref{rhot1}) also  approaches  a Ginibre statistics in $Y \to \infty$, the prediction based on complexity parameter formulation is consistent with the ensemble density implied by eq.(\ref{pdf}) in $b\to \infty$ limit. 
Further, as the functional dependence of variances indicate, the matrices in each of these ensemble become diagonal for $b=0$. With independent Gaussian distributed diagonals and zero off-diagonals, the  eigenfunctions in the limit $b=0$ are localized on the basis states and the ensemble approaches Poisson limit of the statistics. For our numerical analysis, we choose however the initial condition $b_0 =1/N $ for each ensemble. Indeed the choice is arbitrary and $b_0 = 0$ could have been chosen just as well.  It must be noted here that BE, PE and EE  are expected to approach exact Poisson statistics only in $N \to \infty$ limit, the case $b=1/N$ does not exactly correspond to a Poisson limit.

\begin{figure}[ht!]
\centering

\vspace{-1.0in}

\includegraphics[width=18cm,height=25cm]{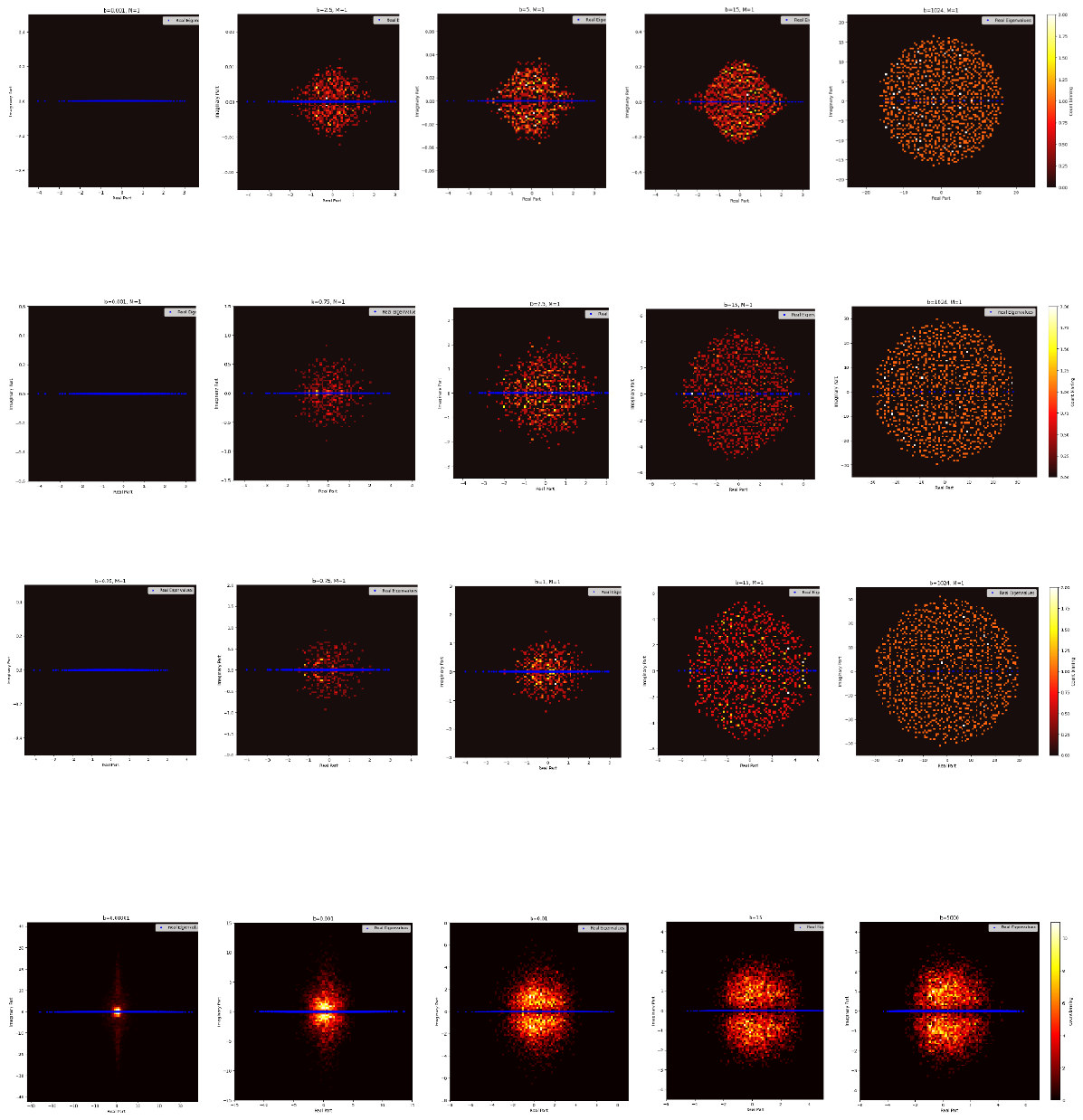}

\vspace{-1.5in}
\caption{\justifying{\bf Spectral density on complex plane case for many fixed $N=1024$:} distribution of both real and complex eigenvalues of a single matrix on the complex plane for five $b$ values
 (i) {\it  first row}: BE case (for $b=0.001, 2.5, 5, 15, N$)
(ii)  {\it second row}: PE case (for $b=0.001, 0.75, 2.5, 15, N$
(iii) {\it third column}:  EE case (for $b=0.001, 0.75, 1, 15, N$)
(iv)  {\it fourth column}:  ME case  (for $b=10^{-5}, 10^{-3}, 10^{-2}, 15, 5000$)}
\label{den-all}
\end{figure}

\begin{figure}[h]
\centering

\includegraphics[width=17cm,height=25cm]{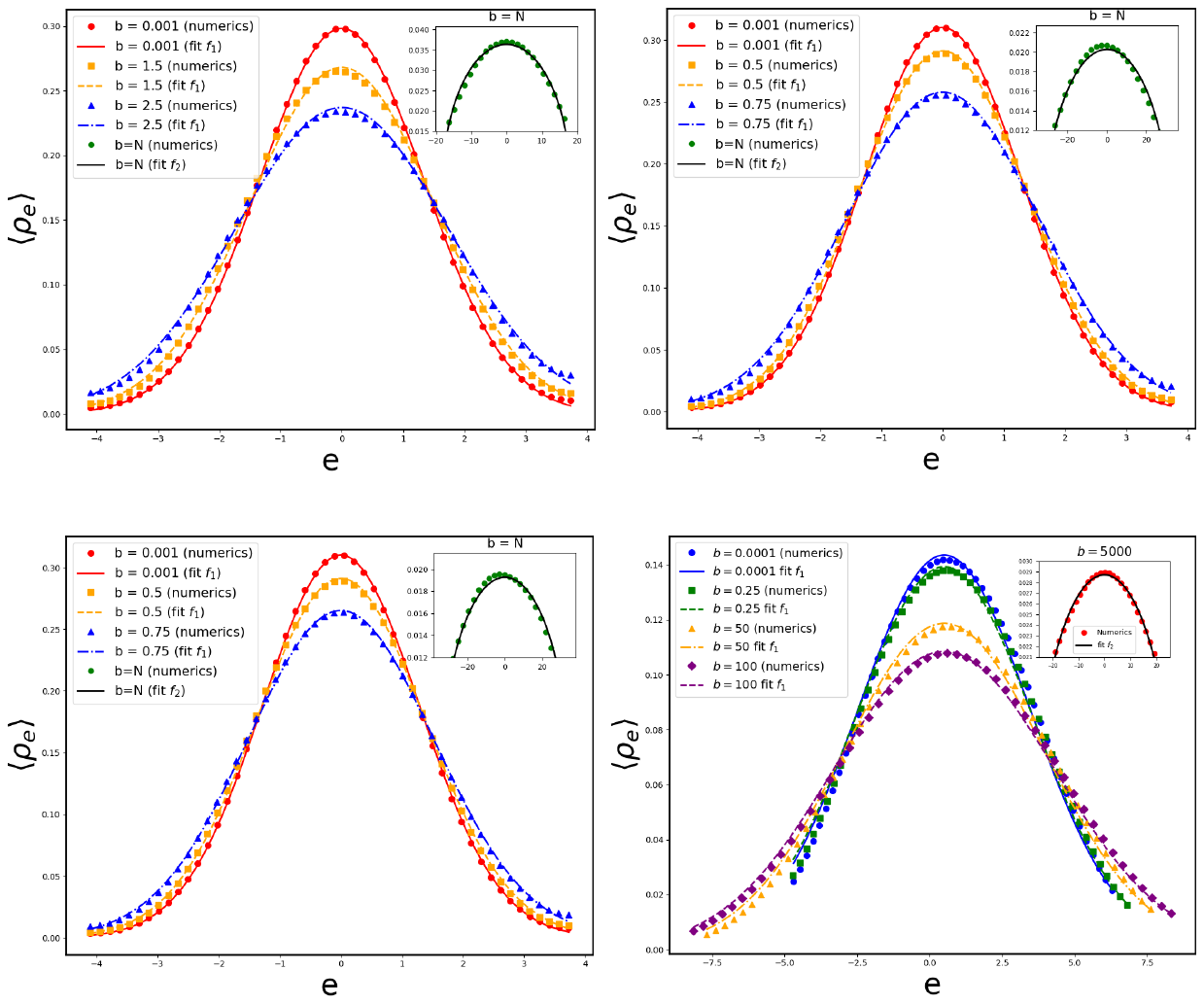}

\vspace{-3.0in}

\caption{\justifying{\bf Ensemble averaged spectral density on the real line :}  The figure displays  the  behavior of the average spectral density $\langle \rho_e(e) \rangle$ (normalized to $1$) of the real eigenvalues for four matrix ensembles for a fixed matrix size $N=1024$ and ensemble size ${\mathcal M}=10$: BE (top left), PE (top right), EE (bottom left), ME (bottom right). 
The theoretical fits here are based on the discussion in section V and fitting parameters for all cases are given in table I. A good agreement with theoretical prediction for four different ensembles, each considered for four different $b$ values, provides a strong support to our theory.  }
\label{real-all}
\end{figure}

 As mentioned in section II, while our theoretical analysis is based on the ensemble described by eq.(\ref{pdf}), its generalization  to the one in eq.(\ref{pdf0}) is possible although technically difficult. This motivates us to numerically analyze a fourth ensemble of type described by eq.(\ref{pdf0}) that not only corresponds to $x_{kl}\not=0$  but also correlations between other elements. The ensemble in this case consists of $10^4$ matrices $H$  with matrix size $N=5$. The corresponding covariance matrix ${\mathcal V}^{-1}$ is a sparse  $25 \times 25$ real-symmetric matrix with its entries ${\mathcal V}_{\eta \nu}$ describing correlation between elements $H_{\eta} \equiv H_{kl}$ and $H_{\nu} \equiv H_{ij}$ where $\eta, \nu$ are column indicies for $H_{kl}$ and $ H_{ij}$ determined by following rule $\eta= (k-1)N+l, \nu=(i-1)N +j$. All diagonal elements $V_{kk}$ are set to $1$, representing unit variances. We choose off-diagonal entries $V_{\eta \nu }=(1+b)^{-2}$ for $k=16,l=17$ or $k=17,l=16$; this implies all $H_{kl}$ and $H_{ij}$ satisfying the rules $5(k-1)+l=16, 5(i-1)+j=17$ will be correlated. In addition all  $V_{\eta \nu}=0$ for all other $\eta, \nu$-pairs. Further the column vector $\mu$ is of dimension $25 \times 1$ representing the mean values of matrix elements of $H$, with $\mu_{\eta}=0.5$. 
{  While the theoretical determination of $Y$ parameter in this case is well-defined by a set of  characteristic equations, their solution is not available in a closed form. (Indeed the $Y$ parameter in this case is a generalization of corresponding Hermitian case discussed in \cite{psco} (eqs.(1, 15, 17) therein) and as it does not provide any new physics insights, we do not elaborate it here any further).  To distinguish it from the three univariate ensembles mentioned above, this case will hereafter be referred as the multivariate ensemble (ME).

To explore the parametric dependence of $R_1$ for finite, non-zero $b$, we perform exact diagonalization of each ensemble for multiple $b$ values and for a fixed $N=1024$. 
As mentioned above, $b=0$ and $b \to \infty$ limits corresponds, from eq.(\ref{pdf}),  to Poisson and Ginibre limits of each of the three ensembles.   We also note, from eqs.(\ref{ybe}, \ref{ype}, \ref{yee}), that the limit $b\to N$,  with $b_0=1/N$,  corresponds to the limit $Y-Y_0 \to \infty$. This is  consistent with the complexity parameter formulation, predicting a Ginibre statistics in $Y-Y_0 \to \infty$. A variation of $b$ from $1/N \to N$ is therefore expected to result  in a crossover from  the Poisson $\to$ Ginibre statistics. This motivates a natural  query whether the crossover in each case occurs at same or different rates, at a single matrix level?  To gain a qualitative insight into the spectral density behaviour on the complex plane,  figure \ref{den-all} depict the distribution of  eigenvalues for a single matrix taken from BE, PE, EE and ME respectively, for five $b$ values. (While we have analyzed the spectrum for many $b$ values ranging between $b=1/N$ to $b=N$, here  only those cases are included where some visible change is occurring).  As figure \ref{den-all} indicates, the eigenvalues of a single matrix taken from  BE, PE and EE for $b=1/N$ are real, with  complex eigenvalues  beginning to appear for $b=b_c > 1/N$. As intuitively expected,  $b_c$  varies from one ensemble to another. Indeed the complex conjugate pairs of eigenvalues  cluster near the center  for very small $b$ values and spread out  with $b$ increasing,  indicating increasing level repulsion giving rise to a tendency towards uniform  distribution. For the system represented by the matrix $H$, this indicates an approach to an almost ergodic state.   As visible from the figure, the Poisson $\to$ Ginibre crossover occurs  at different rates for different ensembles. Noting that $Y$ arises from a consideration of an ensemble, it is not expected to play any role at a single matrix level.

Our next query  is whether $Y$ governs the crossover if we consider some quantitative statistical measures e.g analyze ensemble averaged spectral density? To seek answers, we numerically analyze the four ensembles mentioned above with number ${\mathcal M}$ of matrices in each ensemble optimized to result in a smooth statistics; we find that a choice of ${\mathcal M}=10$ is sufficient for current purpose. Figure  \ref{real-all} depicts the  $R_{1e}(e)$ for five $b$ values  of BE, PE, EE and ME cases along with the numerically fitted functions; the latter in each case agrees well with our theoretical prediction i.e a Gaussian for small $Y-Y_0$ and a semicircle for large $Y-Y_0$. The specific $b$ values considered for each case and corresponding $(Y-Y_0)$ along with the fitted functions are given in table I.  (Consistent with our theory,  here we have  $(Y-Y_0)$   increasing from $0 \to \infty$ for each case as $b$ varies from $1/N \to N$). As the theoretical formulation for $R_{1e}$  (the real eigenvalue density) is derived from $Y$ governed diffusion equation for the spectral JPDF $P_1$, this corroborates our claim regarding  the spectral statistics crossover governed by $Y$ and not by functional forms/ sparsity details of the ensembles. A detailed analysis of the higher order spectral  measures, both theoretically and numerically is  however needed for a complete verification of our claim.

The spectral density on the complex plane has both  radial and angular dependence. To seek the role of $Y$ in crossover on the complex plane, we analyze the ensemble averaged spectral density of the complex eigenvalues for each ensembles for many $b$ values.
As our theoretical solution $R_{1z}(r,\theta,Y)=R'_{1z}(r,\theta) \, e^{-E(Y-Y_0)}$, with $R'_{1z}(r,\theta)=A(r) \, T(\theta)$, indicates a  separability of $r$ and $\theta$ variables,  the numerically obtained $\langle R_{1z} \rangle_{\theta}$ and $\langle R_{1z} \rangle_r $  are thus expected to agree with our theoretical predictions given in section VI. B if the separability assumption holds true.
To verify the above, we numerically determine  $\langle R_{1z} \rangle_{\theta}$, for a fixed $b$,   by counting the number of eigenvalues in an annular disk at a distance $r$, of width ${\rm d}r$ and centred at $r=0$ and then rescaling the number  by the area $2 \pi r {\rm d}r$. Similarly, $\langle R_{1z} \rangle_{r}$ is numerically obtained, for a fixed $b$,  by counting the eigenvalues lying between sector $\theta$ and $\theta+{\rm d}{\theta}$ on the complex plane  and then rescaling the number  by the area $(1/2) r^2 {\rm d}{\theta}$ of the sector. Figures 3 and 4  illustrate the radial and angular dependence   of $R_{1z}(r, \theta)$ (i.e $\langle{R_{1z}\rangle}_{\theta}$  and $\langle{R_{1z}}\rangle_r$),  respectively, for many $b$ values  of the four ensembles. As theoretically predicted (see eq.(\ref{rio3d})), $\langle R_1\rangle_r$ is almost constant for entire $\theta$-range as well as for different $b$ values, but $\langle R_1 \rangle_{\theta}$ depends on both $r$ and $Y-Y_0$. A similar behaviour was also observed for the statistics of complex matrix ensembles too \cite{psgs1}. The latter analogy suggests that the presence of real eigenvalues in the spectrum has very weak or almost no impact on the distribution of complex ones. For near Poisson limit (small $b$), the clustering of complex eigenvalues for small-$r$ ranges indicates almost missing repulsion. But the repulsion rapidly builds in with increasing $b$. We also note a smooth transition from Poisson to real-Ginibre behavior as $Y-Y_0$  increases. However, in contrast to BE,   the approach to a constant density in PE and EE case is relatively rapid for both radial and angular variables, thereby indicating a rapid transition from the initial state to Ginibre limit.

Figure \ref{rad-all} also displays a comparison with eq.(\ref{rio3a}, \ref{rio3b}, \ref{rio3c}); this however requires a prior knowledge of the initial density $\langle R_{1z}(r, \theta; Y_0) \rangle_{\theta}$. Taking the  ensemble at  $b=b_0$  as the initial state, we determine the initial density $R_1(r, \theta; Y_0)$ numerically; a fit to numerical data gives $R_{1z}(r, \theta; Y_0) \equiv R_1(r; Y_0) = A \; r^{-1/2} \, J_{1/2}(B r)  \; {\rm e}^{-C r^2}$ with $A, B, C$  as constants (the values $b_0, A, B, C$ for each ensemble given in table II). We note that here  the initial state $\langle R_{1z}(r, \theta; Y_0) \rangle_{\theta}$, required for comparison,  is obtained by fitting the case $b=b_0=1.5$ for BE and $b=b_0=0.75$ for EE and PE; this is because $b_0=1/N$ case for each ensemble has only real eigenvalues.  
As indicated by  figure \ref{rad-all},  $\langle R_1(r, \theta; Y) \rangle_{\theta}$ varies slowly from its initial state $b=b_0$  and can be well fitted by eq.(\ref{rio3a})   for two intermediate values of $b$. Eq.(\ref{rio3b}) however seems to be a better fit (again keeping only first few terms of $\sum_{\nu}$ referred by $fit_{bn}$); this is expected due to slow change in $(Y-Y_0)$ for the chosen $b$-values for BE.
 A possible explanation of this deviation could be attributed to numerical stability.  In contrast to BE, PE and EE case indicates a rapid variation of $\langle R_1(r, \theta; Y) \rangle_{\theta}$ with $b$ and is well-fitted by eq.(\ref{rio3a}) for higher $(Y-Y_0)$ values; (as mentioned above, $b$ dependence of $(Y-Y_0)$ for this case is almost negligible).} This indeed reconfirms the sensitivity of the  spectral density to $(Y-Y_0)$ instead of $b$.   As expected on theoretical grounds, the behavior for the remaining case i.e $b=N$ for each ensemble  is  almost constant and is fitted by eq.(\ref{rio3c}).  The display in figure \ref{ang-all} for $\langle R_{1z}(r, \theta; Y_0) \rangle_{r}$ is also consistent with eq.(\ref{rio3d}) that predicts an almost constant behavior. An agreement of our theoretical predictions  with fitted functional form for both $\langle R_1(r, \theta; Y) \rangle_{\theta }$ and $\langle R_1(r, \theta; Y) \rangle_{r}$ also lends credence to the separability ansatz made in section IV.

While a lack of complexity parameter formulation for the ensemble (\ref{pdf0}) handicaps us from a comparison of ME case with equations in section VI.B, the corresponding $\langle R_1(r, \theta; Y) \rangle_{\theta}$ behavior  is  fitted well by Eq.(\ref{rio3b}); the fitted parameter values are given in table II (with no theoretical prediction available for  $Y-Y_0$ values for this case, the corresponding column is left blank). 
The behavior for  $\langle R_1(r, \theta; Y) \rangle_{r}$  for this case too turns out to be almost constant.  Indeed while our theoretical derivations presented here  is based on eq.(\ref{pdf}), the above numerical agreement suggests their validity extending to the ensembles described by eq.(\ref{pdf0}) too. 

  The complexity parameter formulation of spectral density discussed above predicts a same mathematical form for a wide range of Gaussian ensembles  e.g. with varying degree of the sparsity if they  share the same $(Y-Y_0)$-value  and share the same complexity constants $t_2, \ldots, t_M$.  
As discussed in {\it appendix A}, the complexity constants can be chosen same for the three univariate ensembles. Figure \ref{same-y} displays a comparison of two spectral measures for them at a  $Y-Y_0$-value chosen arbitrarily. The collapse of statistical behaviour to a single curve for three different ensembles for both measures reinforces our theoretical prediction. We note the analogy is not limited to a single static point on the evolutionary path; if a random  perturbation subjects the ensemble to evolve, it  continues lying along the same evolutionary path  in the ensemble  space (constrained by  fixed values of $t_2, \ldots, t_M$ based on  global  constraints).

While a real matrix of size $N \times N$ has $N$ total number of eigenvalues, the numbers $L$ of its real eigenvalues  and and $M$ of its complex conjugate pairs depends on the details of the matrix structure (with $N=L+2M$). As the later details appear through the ensemble parameters (for the ensemble to be suitable representation of the system), it is relevant to seek as to how the relative number of real eigenvalues varies with $b$.
As mentioned in previous section, the ratio approach zero limit for real-Ginibre ensemble as $N \to \infty$. 
 The insight can be obtained by analyzing the ratio, {\bf say $S(b)=L/N$}, of the $L$ real eigenvalues to $M$ complex conjugate pairs as $b$ varies. Figure \ref{ratio} displays, for the three ensembles, the variation of ensemble averaged ratio $\langle S(b) \rangle$ as $b$ varies. As the figure indicates, the variations for the three ensembles collapse onto same curve as a function of $Y \sim log b$ notwithstanding their different sparsity structures. This is consistent our theoretical claim that the spectral dynamics on the complex plane is governed by the complexity parameter and not by the ensemble details. (It is worth re-emphasizing: although each of the three ensembles considered have only one free parameter i.e $b$, the functional dependence of the variances for each ensemble is different. This in turn give rise to many different variance parameters for each ensemble, the ratio $S(b)$ is however not governed by their details but only by $b$ or $Y$).

  Another interesting insight given by figure \ref{ratio} is the potential existence of a critical $b$-range, analogous for each of the three ensembles where the ratio $S(b)$ rapidly changes from $1$ to $0$ as $b$ increases from $1/N$ to $N$ for $N=1024$. This implies very few real eigenvalues survive in the limit $b=N$ and almost all eigenvalues form complex-conjugate pairs. In addition, this also suggest,  in limit $N \to \infty$,  a possible phase transition from a real eigenvalue "phase"  to "complex conjugate" pair phase.    A detailed theoretical as well as numerical analysis of this aspect however is beyond the scope of current study and will be pursued elsewhere.

\section{Conclusion}

To conclude, we summarize our key insights and findings along with some unresolved questions.
Based on the representation by a multiparametric Gaussian ensemble of real assymetric  normal matrices, we show that the spectral distribution  of a non-Hermitian complex system is influenced by system-specific details solely through the complexity parameter—a single functional of the system conditions. While changes in system conditions can drive the spectral distribution to evolve both in spectral and ensemble space, the global constraints, such as conservation laws and symmetries, act as constants of the dynamics. Our quest for an evolution in the ensemble parameter space that mirrors the one on the spectral plane, leads us to complexity parameter space wherein lies a single parametric "universal"  path for evolution of different ensembles determined by a set of global constraints e.g. basis constants shared by them. The significance of a universal path in complexity parameter space can be emphasized as follows: when system parameters change, the ensembles representing different systems  are constrained to evolve along different paths determined by a set of complexity constants. The latter and thereby the path can be chosen same for the systems sharing the same set of global constraints e.g. symmetries and conservation laws.  This describes a universal dynamic, in terms of the complexity parameter \( Y \), for the eigenvalues of real assymetric matrices (with Gaussian randomness) from any initial state subjected to multiparametric perturbations. The dynamics approach the steady state of the real Ginibre ensemble for large \( Y \) limits if the matrices in the ensemble are normal. More explicitly, since different real normal matrix ensembles at the same point on this path (i.e., those sharing the same value of \( Y \) and similar initial statistics) share the same spectral distribution, this reveals a hidden universality even in non-equilibrium regimes, extending beyond the real Ginibre ensemble. The results obtained here have  huge potential applications to many other areas e.g. condensed matter and quantum information, biological and artificial neural networks etc.

One of the key outcomes of our analysis is the solution of the evolution equation for the average spectral density as the ensemble parameters vary. Despite the technical intricacies of the differential equation requiring us to assume weak second-order spectral correlations on the complex plane, the assumption stems from the very nature of local spectral correlations which tend to diminish beyond a single local mean level spacing. Thus, it remains valid for calculating the average spectral density. Moreover, since our analysis allows for arbitrary choices of the ensemble parameters, the solution extends to a broad spectrum of ensembles, including various sparse matrix structures that characterize the statistical behavior of real normal non-Hermitian operators in complex systems. Our claim is substantiated by the numerical simulations of four different ensembles with varying degrees of off-diagonal sparsity; the numerically derived average spectral densities closely match our theoretical predictions based on the complexity parameter and the separability conjecture.
We note that our current numerical analysis is based on an initial condition corresponding to Poisson spectral statistics on a complex plane, chosen due to the intense interest in the transition from Poisson to Ginibre spectral statistics. This transition, with intermediate states described by various sparse random matrix ensembles, represents many real non-Hermitian systems. However, since our evolution equations for the joint probability density function (jpdf) of eigenvalues, and consequently for the spectral density, are valid for any arbitrary initial condition, it would be highly desirable to pursue the numerical verification for other initial conditions as well.

The present study  still leaves many issues unresolved e.g. can complexity parameter formulation help to seek existence of the phase transitions in the spectral distributions as a function of the complexity parameter? Such transition can occur for example due to  breaking of certain global symmetries e.g. time-reversal symmetry or due to varying local constraints e.g. disorder. This information is relevant for fundamental as well as technological applications in quantum systems and quantum information theory. The understanding of local density fluctuations at the spectral edge as well as bulk, for sparse real non-Hermitian matrices in large size limit, (for both normal as well as non-normal matrices) is another important issue yet to be addressed.
One of the most challenging issues however is to derive mathematical formulations for the spectral statistics under system condition which permit  exceptional points i.e the non-Hermitian degeneracies where eigenvalues merge and eigenvectors coalesce; these points are believed to play an important role in many areas of many body physics. We hope to answer some of these questions in near future.

\acknowledgments

One of the authors (P.S.)  is grateful to Anusandhan National Research Foundation (ANRF), India for the financial support provided for the  research under Matrics grant scheme.

\vspace{0.1in}


 
 {\bf Declarations}

{\bf Data Availability} Data sets generated during the current study are available from the corresponding author on reasonable request. 
 
{\bf Conflict of interest} The authors declare no conflicts of interest.




\begin{table}
\caption{\justifying{\bf Details of the fits used in figure \ref{real-all} on the real line:} 
While the lower $Y-Y_0$ values can be well-fitted by the function 
 $f_1= \frac{c_e}{\sqrt{2\pi\sigma^2}} \exp\!\left(-\frac{(e - \eta (Y-Y_0))^2}{2\sigma^2}
-\frac{(\alpha (Y-Y_0))^2}{2\sigma^2} \right) (\delta \cos\phi - \gamma \sin\phi)$, the
large $Y-Y_0$ behavior is better described by the function $f_2= \frac{c_e\sqrt{2(\sigma^2 - e^2)}}{\pi\sigma^2}$. The parameter $\gamma, \delta, \phi$ are defined in eq.(\ref{gd1}). 
The parameters  $\eta$ and $\alpha$ are determined  from the initial spectral density on the real line (i.e by fitting $f_1$ to the case $b=b_0$). 
Here we have $\eta=\alpha=6678.494$ (For BE),
$\eta=\alpha=101.571$ (For PE),
$\eta=\alpha=162.961$ (For EE), 
$\eta=-0.0409$, $\alpha=-2.133$ (For ME) and
$\eta=\alpha=-48189.205$ (For figure 6).
We note that $Y-Y_{0}$ for the ME case given here is a fitted value; This is due to  inaccessibility of a closed form solution for $Y-Y_{0}$ in this case. We recall that $\langle \rho_e(e) \rangle$ is normalized to $1$ and thereby $c_e=1$ here.}


\begin{tabular}{|c|c|c|p{10cm}|}
\hline
\textbf{Case} & \textbf{$b$} & \textbf{$Y-Y_0$} & \textbf{Fitted curve and the parameters} \\
\hline
\multirow{5}{*}{BE} 
& $\frac{1}{N}( \equiv b_0)$ & 0.00 &  $f_1$, $\sigma= 1.341$, $\delta=1.002$\\
& 1.5 & $4.279\times 10^{-6}$ & $f_1$,  $\sigma= 1.477$, $\delta=1.023$\\
& 2.5 & $1.186\times 10^{-5}$ & $f_1$,  $\sigma= 1.746$, $\delta= 1.071$\\
& N & 0.575 &  $f_2$, $\sigma=18.155$ \\
\hline
\multirow{5}{*}{PE} 
& $\frac{1}{N}(\equiv b_0)$ & 0.00 & $f_1$, $\sigma= 1.341$, $\delta=1.002$\\
& 0.26 &  $4.279\times 10^{-6}$ & $f_1$,  $\sigma= 1.477$, $\delta=1.023$\\
& 0.5 & $4.436\times 10^{-5}$ & $f_1$,  $\sigma=  1.482$, $\delta=1.024$\\
& 0.75 &  $2\times 10^{-4}$ & $f_1$,  $\sigma= 1.539$, $\delta=1.026$\\
& N &  1.249 & fit$f_2$,  $\sigma= 32.898$ \\
\hline
\multirow{5}{*}{EE} 
& $\frac{1}{N}(\equiv b_0)$ & 0.00 & $f_1$,  $\sigma= 1.341$, $\delta=1.002$\\
& 0.43 & $4.279\times 10^{-6}$ & $f_1$,  $\sigma= 1.477$, $\delta=1.023$\\
& 0.5 & $1.803\times10^{-5}$ & $f_1$, $\sigma= 1.482$, $\delta=1.027$\\
& 0.75 & $1.811\times10^{-4}$ & $f_1$,  $\sigma=1.504$, $\delta= 1.034$ \\
& N & $1.532$ & $f_2$,  $\sigma=34.831$ \\
\hline
\multirow{5}{*}{ME} 
&  0.0001($\equiv b_0$) & 0.036 & $f_1$, $\sigma = 3.270$, $\delta=1.178$ \\
&  0.25 & 0.040 & $f_1$, $\sigma=3.338$, $\delta=1.236$ \\
&  50 & 0.450 & $f_1$, $\sigma=3.649$, $\delta=1.140$ \\
& 100 & 0.561 & $f_1$,  $\sigma=3.784$, $\delta=1.096$\\
& 5000 & $>>1$ &  $f_2$, $\sigma=28.271$\\
\hline
\end{tabular}
\end{table}


\begin{figure}[h]
\centering

\includegraphics[width=17cm,height=25cm]{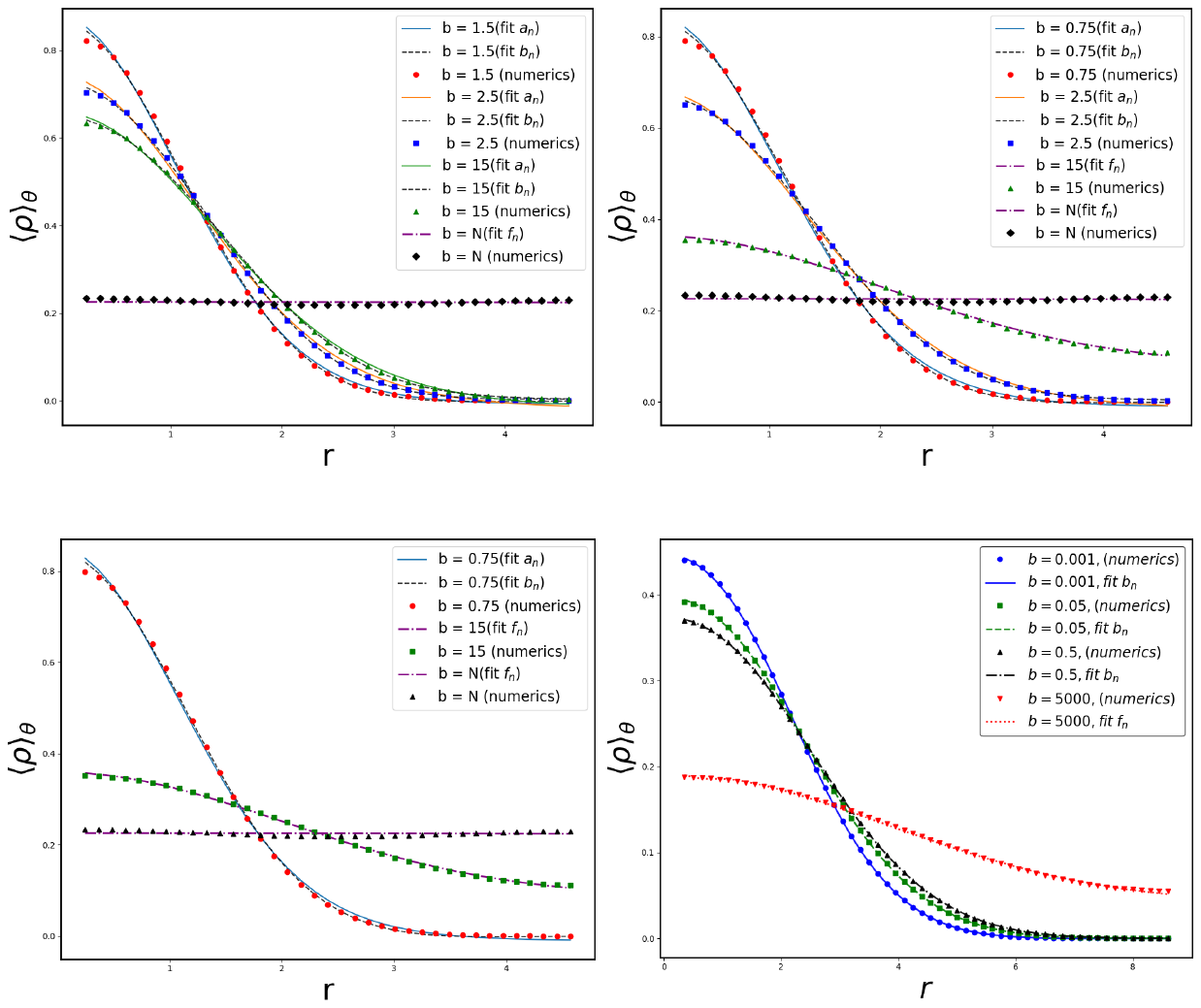}

\vspace{-2.95in}

\caption{\justifying{\bf Ensemble averaged spectral density of the radial parts of the complex eigenvalues :}  The figure displays  the  behavior of the average spectral density of the radial part of the complex conjugate  eigenvalue-pairs  for the four matrix ensembles for a fixed matrix size $N=1024$ and ensemble size ${\mathcal M}=10$: BE (top left), PE (top right), EE (bottom left), ME (bottom right). The theoretical fits here are based on the discussion in section VI.B and fitting parameters for all cases are given in table II. Here again a  good agreement with numerics for many cases  confirms our theoretical prediction.}
\label{rad-all}
\end{figure}

\begin{figure}[ht]
\centering
\vspace{-0.9in}
\includegraphics[width=19cm,height=30cm]{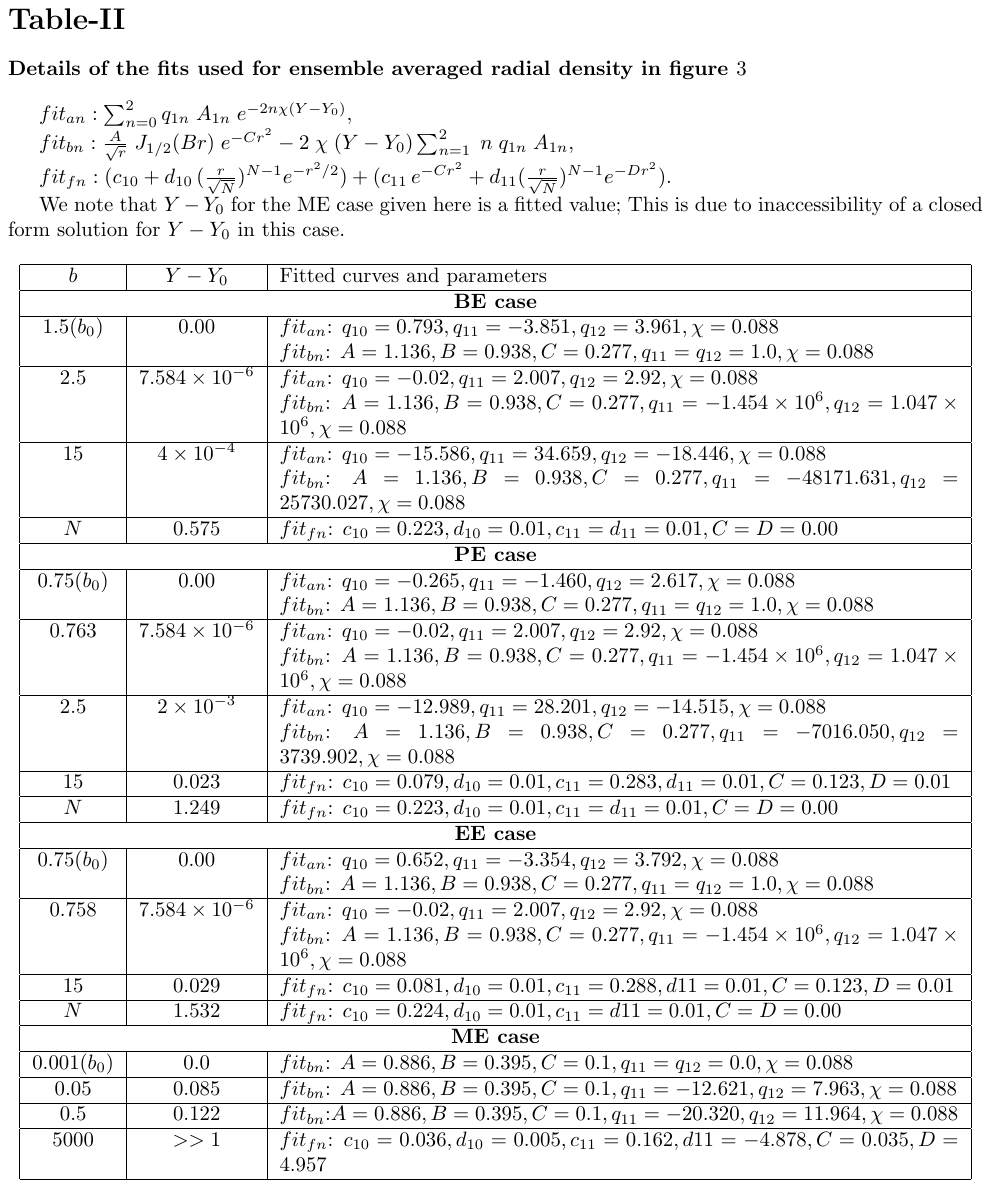}
\vspace{-1.5in}
\label{tcomp}
\end{figure}

\begin{figure}[h]
\centering

\includegraphics[width=17cm,height=25cm]{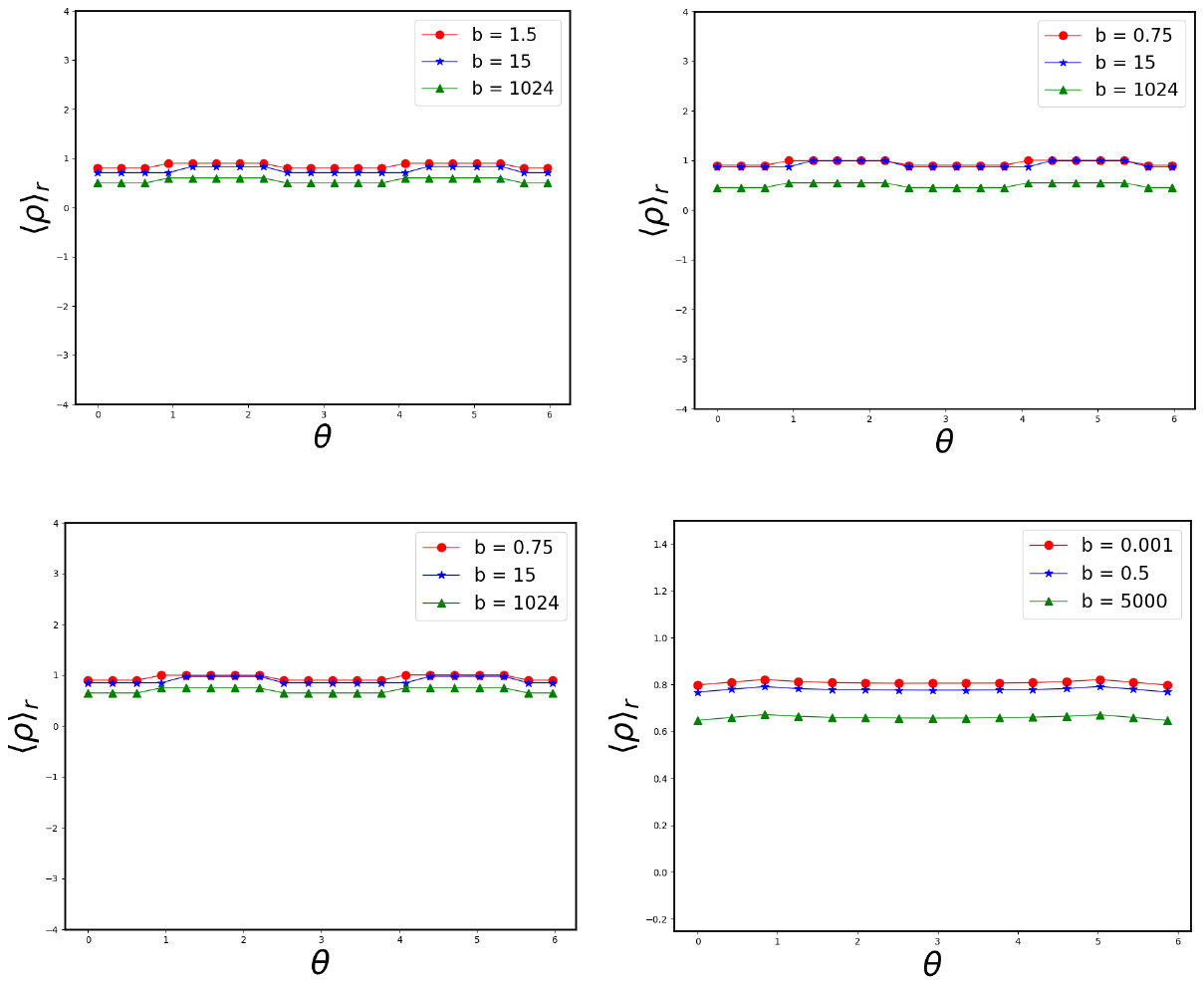}

\vspace{-3.0in}

\caption{\justifying{\bf Ensemble averaged Spectral density of the angular parts of the complex eigenvalues :}  The figure displays  the  behavior of the average spectral density of the angular part of the complex conjugate  eigenvalue-pairs  for the four matrix ensembles for a fixed matrix size $N=1024$ and ensemble size ${\mathcal M}=10$: BE (top left), PE (top right), EE (bottom left), ME (bottom right). The behavior for each ensemble is displayed here for three $b$ values. Consistent with our theory (see eq.(\ref{rio3e}), $\langle R_1(r, \theta; Y) \rangle_{r}$   is almost constant  in $\theta$ and has a weak dependence on $Y-Y_0$.}
\label{ang-all}
\end{figure}


\begin{figure}[ht]
\centering
\includegraphics[width=17cm,height=26cm]{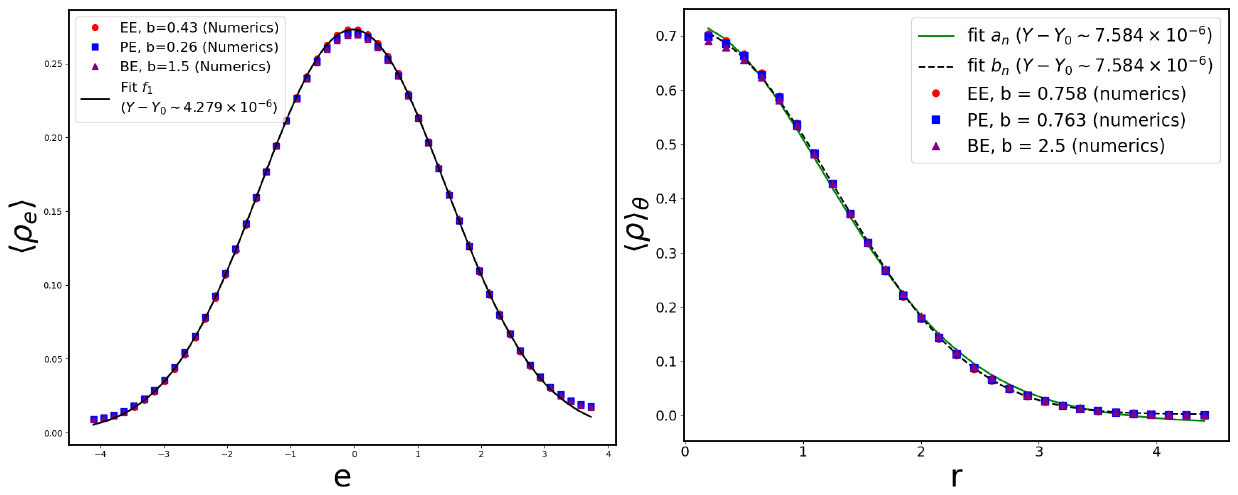}

\vspace{-6.2in}

\caption{\justifying{\bf Ensemble average spectral density  for three different ensembles at a same $Y-Y_0$ value:} The figure displays a comparison of the ensemble-averaged real eigenvalue density (left panel, for $Y-Y_0=4.279\times 10^{-6}$), and the ensemble-averaged radial part of the complex eigenvalues (right panel, for $Y-Y_0=7.584\times 10^{-6}$) for BE, PE and EE. The collapse of the curves for  all three ensembles for both measures supports our theoretical prediction.}
\label{same-y}
\end{figure} 



\begin{figure}[ht]
\centering

\includegraphics[width=16cm,height=12cm]{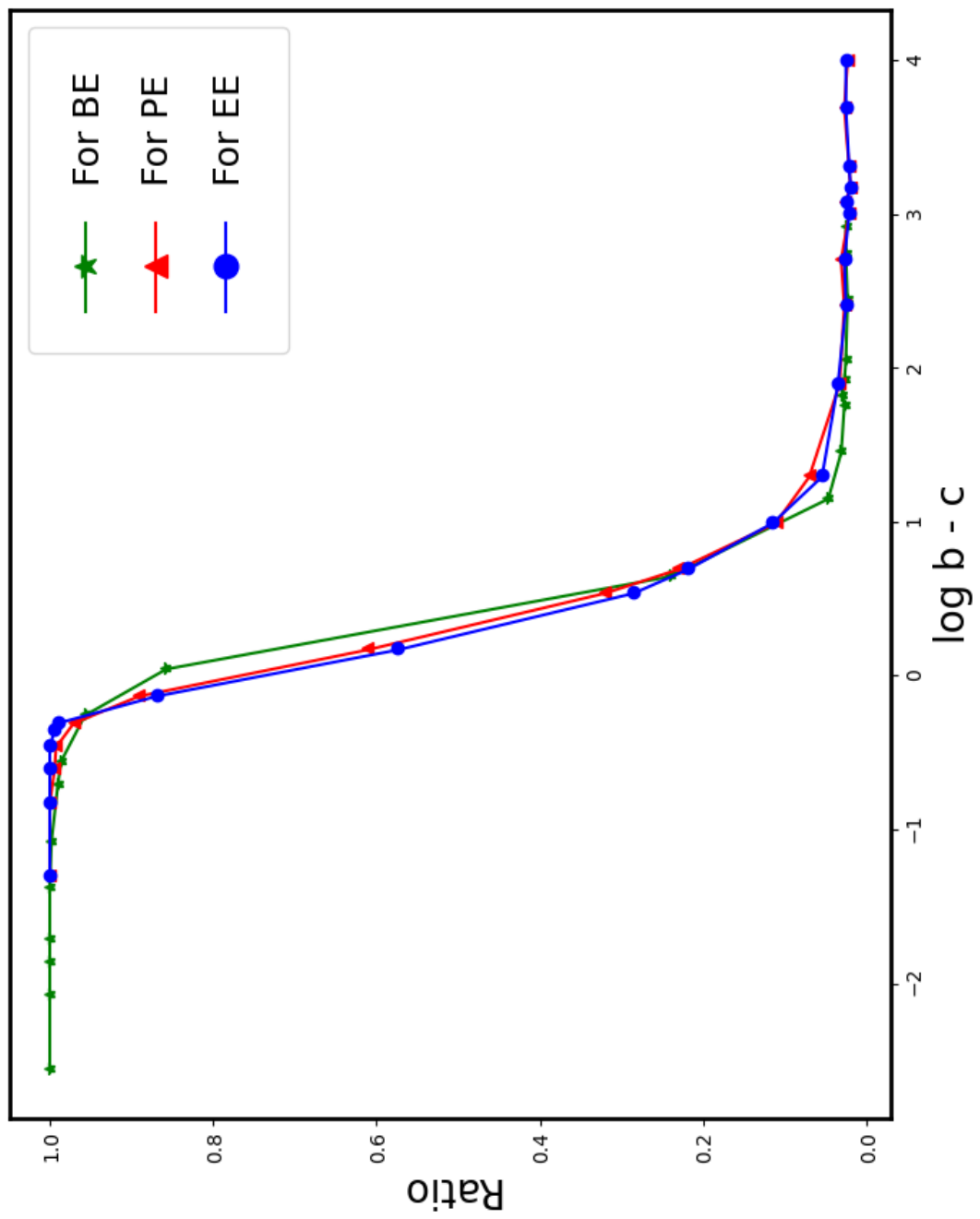}

\caption{\justifying{\bf Ratio of the real eigenvalues and complex conjugate pairs:} 
 The figure displays  the  ratio ${L \over N}$ of $L$ real eigenvalues and $N=L+2M$ for many $b$ values and fixed $N=1024$ for three ensembles with $M$ as number of complex conjugate pairs. As clear from the figure, the number of real eigenvalues decrease as the free parameter $b$ increases. The rate of change however is same for the three ensembles irrespective of the difference of their variance matrices. }
\label{ratio}
\end{figure}

\clearpage

\appendix

\section{Determination of constants $t_2, \ldots, t_M$ for BE, PE and EE}

As mentioned in section II, $t_1$ and the constants of evolution are given by the solution of following set of characteristics equations

\begin{eqnarray}
 {{\rm d}y_{11;1} \over A_{11}}
= {{\rm d}x_{11} \over B_{11}}
= \ldots 
= {{\rm d}y_{kl} \over A_{kl}}
= {{\rm d}x_{kl} \over B_{kl}}
={{\rm d}t_{\alpha} \over \delta_{\alpha 1}},
\label{tka}
\end{eqnarray}
where $\alpha=2\to M$ and 
$A_{kl} = 2 y_{kl} \left(1 - 2 y_{kl} \right) - 4 x^2_{kl} $ and $B_{kl} = 2 x_{kl} \, \left(1 - 4 \, y_{kl}\right)$.


For case  $x_{kl}=0$ used in our numerics, eq.(\ref{tka})  reduces to following form
\begin{eqnarray}
\frac{d y_{11}}{y_{11}} = \frac{d y_{12}}{y_{12}}=\ldots =\frac{d y_{mn}}{y_{mn}} = \ldots =\frac{d t_{\alpha}}{\delta_{\alpha 1}}
\label{tk1}
\end{eqnarray}
A particular solution of the above equation for $\alpha >1$ can be given as 
\begin{eqnarray}
t_{\alpha} = \sum_{m,n} q_{mn;\alpha} \log y_{mn}
\label{ta}
\end{eqnarray}
with $q_{mn; \alpha}$ as  constants subjected to condition $\sum q_{mn; \alpha}=0$ for $\alpha >1$. (This can be checked by a direct substitution in relation  $\sum_{kl} y_{kl} {\partial t_{\alpha} \over \partial y_{kl}} = \delta_{\alpha 1}$; the latter is equivalent to eq.(\ref{tk1})). 
While many solutions satisfying the above condition are possible, the appropriate  solution is the one that is also applicable for initial ensemble (as the constants of evolution are constants for initial ensemble too).

For BE, we choose $t_{\alpha}=\log y_{mn}-\log y_{ij}$ (for  $1 < \alpha \le M$ and any two pairs of indices $m,n$ and $i,j$ with $1 \le m,n,i,j \le N$ where $m \not=n$ if $i \not=j$,  $m=n$ if $i=j$). The same variance of all off-diagonals (diagonals) then gives $t_{\alpha}=0$. The choice  corresponds to solution (\ref{ta}) with pair $q_{mn;\alpha} =- q_{ij;\alpha}=1$; $q_{kl;\alpha}=0$ if $(k,l) \not= (m,n)$ or $(i,j)$. As total number of such $q$ pairs can be more than $ N^2$, different pairs can be chosen for each $\alpha$. 

For PE and EE, we choose  $t_{\alpha} = \log\left(\frac{y_{m m+r}}{y_{i i+r}}\right)$. Noting that $y_{m m+r} = y_{ii+r}$ (for $1 \le m, i \le N$, $1 \le r < N$), this again gives  $t_{\alpha}=0$. The choice  corresponds to solution (\ref{ta}) with pair $q_{m m+r;\alpha} =- q_{ii+r;\alpha}=1$; $q_{kl;\alpha}=0$ if $(k,l) \not= (m,m+r)$ or $(i,i+r)$.  

As the above indicates, it is possible to choose same set of constants $t_{\alpha}$ for BE, PE and EE, namely $t_{\alpha}=0$ for $\alpha >1$. Further as the latter results by choosing specific pairs of variances,  irrespective of $b$-value, these are valid for initial condition $b=1/N$ too.  The three ensembles can thus evolve along the same path in the $t$-space from same initial condition i.e Poisson spectral statistics. As described above, although $(Y-Y_0)$-forms for them are in general different, they may take same value at some point of evolution. Our theory predicts same spectral statistics for the three ensembles at such points.

\section{Determination of coefficients in eq.(\ref{rio3a}), eq.(\ref{rio3b})}

The initial condition used for our numerical analysis correponds to $b=b_0$ with $b_0$  for each of the three ensembles mentioned in the table II. 
As displayed in figures 3 for cases $b=b_0$ for BE, PE and EE,   the spectral density in this case is as follows:
\begin{eqnarray}
R_1(r, \theta; Y_0) = A \, r^{-1/2} \; J_{1/2}(B r) \, {\rm e}^{-C r^2}.   \label{rina}
\end{eqnarray}

Eq.(\ref{rima}) gives   $\langle R_1(r, \theta,Y_0) \rangle_{\theta}  =  \sum_{\nu}    \, A_{1\nu}  $; the unknown constants  $p_{1\nu}$ and $q_{1\nu}$ of $A_{1\nu}$ can then be determined from the initial condition.
For example,  for initial density given by  eq.(\ref{rina}), we have (using eq.(\ref{uja1}))
\begin{eqnarray}
{A \over \sqrt{r}} \; J_{1\over 2}(B r) \, {\rm e}^{-C r^2} =  \sum_{ \nu=0}^{\infty}  \; \left[ p_{1 \nu} \, \left({r^2 \over 2}\right)^{-\frac{1}{2}} \; F_1\left(\frac{1}{2}+\nu \chi,  \frac{1}{2}, \,  -\frac{ r^2}{2}\right)  + q_{1 \nu} \, \, F_1\left( \nu \chi +1,  \frac{3}{2},-\frac{ r^2}{2}\right)  \right] \nonumber \\
 \label{rinaa3a}
\end{eqnarray}

Using $J_{1/2}(B r) = {\sin(B r)\over \sqrt{\pi B r}}$ and expanding both sides in a power series near $r=0$ and  comparing various powers of $r$, we have  $p_{1 \nu}=0$ and
\begin{eqnarray}
z_n  =   \sum_{ \nu=0}^{\infty}  \;  q_{1\nu}  \, (\nu \chi +1)_n
 \label{rina4a}
\end{eqnarray}
where $z_{n} =\frac{A}{\sqrt{\pi}} \; \left(\frac{2}{\gamma}\right)^n  \left(\frac{3}{2}\right)_n \, \sum_{k=0}^{n}  {n!  B^{2k+1/2} C^{n-k} \over (n-k)!  (2k+1)!}$.   Here $(a)_n$ is a Pochhammer symbol i.e $(a)_n = a(a+1)...(a+n)$.

Eq.(\ref{rina4a}) can further be written in form of  a matrix equation $Z =  W Q$   with $Z$ and  $Q$  as the column vectors,  $Z \equiv [z_n]$, $Q_{\nu} \equiv [q_{1 \nu}]$ (both $n, \nu=1 \to \infty$)
and $W$  as an infinite dimensional  matrix with entries $W_{\nu n} =(\nu \chi +1)_n$.  Inverting the above  gives  $Q = W^{-1} \; Z$ and thereby $q_{1 \nu}$ for each $\nu$.   In  practice however inversion is not easy due to infinite dimansionality of the matrices involved, leaving the determination of the constants $q_{1\nu}$ only by numerical fitting.


\newpage

{\Large {\bf Supplementary Material}}

\newpage 


{\bf The text below contains detailed derivation  of some of the mathematical results in the main text. }

\section{Complexity parameter space formulation of ensemble density, eq.(6) of main text}

We consider an ensemble of $N\times N$ real-asymmetric matrices $H$ defined by a multiparametric Gaussian ensemble density ${\rho}(H, y, x)$ defined as
\begin{eqnarray}
{\rho}(H, y, x) =  C \, {\rm exp}\left[
-  \sum_{k, l} (y_{kl} H_{kl}^2 + x_{kl} H_{kl} H_{lk})\right] 
\label{pdf}
\end{eqnarray}
 with $C$ as a constant determined by the normalization condition $\int  \rho \, {\rm d}H =1$. Here the sets $y$ and $x$, consisting of  ensemble  parameters $y_{kl}, x_{kl}$ are chosen arbitrarily (also including infinite values for non-random entries). 


A change in ensemble parameters result in evolution of $\rho$ in ensemble space. We consider following combination of the first order parametric derivatives of $\rho$ where $\rho = C \rho$,
\begin{eqnarray}
T \; \rho \equiv \sum_{k,l}^N   \left[ A_{kl} {\partial  \rho \over\partial y_{kl}}+ B_{kl}  {\partial  \rho\over\partial x_{kl}}\right] 
 \label{trho}
\end{eqnarray}  
with
\begin{eqnarray}
A_{kl} &=& 2 y_{kl} \left(1 - 2 y_{kl} \right) - 4x^2_{kl}  
 \nonumber  \\
B_{kl} &=& 2 x_{kl} \, \left(1 - 4 \, y_{kl}\right)
\label{dab}
\end{eqnarray} 
%
%

From eq.(\ref{pdf}) we have ${\partial \rho \over\partial y_{kl}} = - H_{kl}^2 \rho$,   ${\partial \rho \over\partial x_{kl}} =- H_{kl}  H_{lk} \rho$ and ${\partial \rho \over\partial H_{kl}} = 
- 2 ( y_{kl} H_{kl} + x_{kl} H_{lk})\rho$. The above relations lead to
\begin{eqnarray}
T \rho 
=\sum_{k,l}
{\partial \over \partial H_{kl}}\left[
 {\partial \over \partial H_{kl}}  + H_{kl}  \right] \rho
\label{td1}
\end{eqnarray}

As eq.(\ref{td1})  is difficult to solve for generic parametric values,  we seek a transformation from the set of $M$ parameters $(y,x) \equiv \{y_{kl}, x_{kl} \}$ to another set $t \equiv \{t_1,\ldots, t_M \}$,  with  $M \le 2 N^2$, 
such that only $t_1$ varies under the evolution governed by the operator $T$ and rest of them i.e $t_2, \ldots, t_M$ remain constant: 

\begin{eqnarray}
 T \rho  \equiv {\partial \rho\over\partial t_1}
 \label{tt0}
\end{eqnarray}

This implies

\begin{eqnarray}
 {\partial \rho\over\partial t_1} &=&
\sum_{k,l} {\partial \over \partial H_{kl}}\left[{\partial \over \partial H_{kl}}  + H_{kl}  \right] \rho 
 \label{rhot1}
\end{eqnarray}

To determine the conditions under which the relation in eq.(\ref{tt0}) is valid, we proceed as follows:
as the transformation maps $\rho(y,x) \to \rho(t)$, the chain rule leads to  

\begin{eqnarray}
 {\partial \rho\over\partial f_{kl}}  = \sum_{\alpha=1}^M  {\partial \rho\over\partial t_{\alpha}}  \; {\partial  t_{\alpha}\over\partial  f_{kl}}  
 \label{tt1}
\end{eqnarray}
with $f_{kl}= x_{kl}, y_{kl}$. Replacing $\rho$ by $\rho$ in eq.(\ref{trho}) and substituting the above derivatives  in its right side, we have
\begin{eqnarray}
T \; \rho   = \sum_{\alpha=1}^M  (Tt_{\alpha} ) \; {\partial \rho\over\partial t_{\alpha}} 
 \label{tt2}
\end{eqnarray} 
where 
\begin{eqnarray}
T \; t_{\alpha} \equiv \sum_{k,l}^N   \left[ A_{kl} {\partial  t_{\alpha}\over\partial  y_{kl}}  
 + B_{kl}  {\partial  t_{\alpha}\over\partial  x_{kl}} \right]  
 \label{tt3}
\end{eqnarray}


As clear from the above, the condition $T \rho  \equiv {\partial \rho\over\partial t_1}$ can be fulfilled by three possible ways,

\begin{eqnarray}
(i) \qquad &&  T t_1=1, \; T t_{\alpha}=0, \quad \frac{\partial \rho}{\partial t_{\alpha}} = 0, \quad \forall \; \alpha >1, 
\label{yc1}  \\
(ii) \qquad  && T t_1=1, \; \frac{\partial \rho}{\partial t_{\alpha}} = 0, \qquad \forall \; \alpha >1, \label{yc2} \\
(iii) \qquad  && T t_1=1, \; T t_{\alpha}=0.  \qquad \forall \; \;  \alpha >1,  
\label{yc3}
\end{eqnarray}

\section{Determination of $t_1, t_2, \ldots, t_M$}


For the condition $ T \, \rho = {\partial \rho\over\partial t_1}$ to be satisfied, we
 require $t_{\alpha}$  to satisfy following condition 
\begin{eqnarray}
  \sum_{k,l}
 \left[ A_{kl} {\partial t_{\alpha} \over\partial y_{kl}}
 +  B_{kl}  {\partial t_{\alpha} \over\partial x_{kl}}\right] = \delta_{\alpha 1}.
\label{ttk} 
\end{eqnarray}
with  $A_{kl}, B_{kl}$ defined in eq.(\ref{dab}). 


The above equation can be solved by standard method of characteristics; the latter leads to a set of $M$ relations
\begin{eqnarray}
 {{\rm d}y_{11} \over A_{11}}
= {{\rm d}x_{11} \over B_{11}}
= \ldots 
= {{\rm d}y_{kl} \over A_{kl}}
= {{\rm d}x_{kl} \over B_{kl}}
={{\rm d}t_{\alpha} \over \delta_{\alpha 1}}, 
\label{t1}
\end{eqnarray}


{\bf Solution for $t_1$:} For clarity, we first consider the solution of the above equation for $t_1$.
The above is a first-order linear partial differential equation and can be solved using the standard method of characteristics. The characteristic equations are
\begin{eqnarray}
\frac{dy_{kl}}{A_{kl}}
=
\frac{dx_{kl}}{B_{kl}}
=
\frac{dt_\alpha}{\delta_{\alpha1}}.
\label{char1}
\end{eqnarray}
Substituting $A_{kl}$ and $B_{kl}$ from eq.(\ref{dab}) and rearranging eq.(\ref{t1}), with $x_{kl}=x_{lk}$ and for $\alpha=1$, this gives;
\begin{eqnarray}
\frac{dy_{kl}}
{2y_{kl}(1-2y_{kl})-4x_{kl}^2}
=
\frac{dx_{kl}}
{2x_{kl}(1-4y_{kl})}
=
dt_1.
\label{char2}
\end{eqnarray}

From the first equality in Eq.(\ref{char2}), one obtains
\begin{eqnarray}
\frac{dx_{kl}}{dy_{kl}}
=
\frac{x_{kl}(1-4y_{kl})}
{y_{kl}(1-2y_{kl})-2x_{kl}^2}.
\label{char3}
\end{eqnarray}

To simplify the above nonlinear equation, we define
\begin{eqnarray}
u_{kl}=x_{kl}^2.
\label{u_def}
\end{eqnarray}

Differentiating Eq.(\ref{u_def}) with respect to $y_{kl}$ gives
\begin{eqnarray}
\frac{du_{kl}}{dy_{kl}}
=
2x_{kl}\frac{dx_{kl}}{dy_{kl}}.
\label{u_diff}
\end{eqnarray}

Substituting Eq.(\ref{char3}) into Eq.(\ref{u_diff}), we obtain
\begin{eqnarray}
\frac{du_{kl}}{dy_{kl}}
=
\frac{2u_{kl}(1-4y_{kl})}
{y_{kl}(1-2y_{kl})-2u_{kl}}.
\label{u_eq}
\end{eqnarray}
To solve Eq.~(\ref{u_eq}), let us define
\begin{eqnarray}
v_{kl}
=
\frac{u_{kl}}{y_{kl}(1-2y_{kl})}.
\label{v_def}
\end{eqnarray}
Thus,
\begin{eqnarray}
u_{kl}
=
v_{kl}\,y_{kl}(1-2y_{kl}).
\label{u_sub}
\end{eqnarray}
Differentiating Eq.~(\ref{u_sub}) with respect to $y_{kl}$, we obtain
\begin{eqnarray}
\frac{du_{kl}}{dy_{kl}}
=
v_{kl}(1-4y_{kl})
+
y_{kl}(1-2y_{kl})\frac{dv_{kl}}{dy_{kl}}.
\label{du}
\end{eqnarray}
Substituting Eqs.~(\ref{u_sub}) and (\ref{du}) into Eq.~(\ref{u_eq}) gives
\begin{eqnarray}
v_{kl}(1-4y_{kl})
+
y_{kl}(1-2y_{kl})\frac{dv_{kl}}{dy_{kl}}
=
\frac{2v_{kl}(1-4y_{kl})}{1-2v_{kl}}.
\end{eqnarray}
Rearranging,
\begin{eqnarray}
y_{kl}(1-2y_{kl})\frac{dv_{kl}}{dy_{kl}}
=
(1-4y_{kl})
\left(
\frac{2v_{kl}}{1-2v_{kl}}-v_{kl}
\right).
\end{eqnarray}
so that
\begin{eqnarray}
y_{kl}(1-2y_{kl})\frac{dv_{kl}}{dy_{kl}}
=
(1-4y_{kl})
\frac{v_{kl}(1+2v_{kl})}{1-2v_{kl}}.
\label{v_eq}
\end{eqnarray}
Separating variables,
\begin{eqnarray}
\frac{1-2v_{kl}}{v_{kl}(1+2v_{kl})}\,dv_{kl}
=
\frac{1-4y_{kl}}{y_{kl}(1-2y_{kl})}\,dy_{kl}.
\label{sep}
\end{eqnarray}
Now integrating both side of eq. (\ref{sep}) we obtain;
\begin{eqnarray}
\int
\left(
\frac{1}{v_{kl}}
-
\frac{4}{1+2v_{kl}}
\right)dv_{kl}
=
\int
\left(
\frac{1}{y_{kl}}
-
\frac{2}{1-2y_{kl}}
\right)dy_{kl}+ C_{kl}
\end{eqnarray}
So,
\begin{eqnarray}
\ln v_{kl}
-
2\ln(1+2v_{kl})
=
\ln y_{kl}
+
\ln(1-2y_{kl})
+
C_{kl}.
\end{eqnarray}
Exponentiating both sides,
\begin{eqnarray}
\frac{v_{kl}}{(1+2v_{kl})^2}
=
c_{kl}\,y_{kl}(1-2y_{kl}),
\label{implicit_v}
\end{eqnarray}
where $c_{kl}=e^{C_{kl}}$ is the integration constant.
Substituting from Eq.~(\ref{v_def}), we obtain the exact implicit solution
\begin{eqnarray}
\frac{u_{kl}}
{\left(y_{kl}(1-2y_{kl})+2u_{kl}\right)^2}
=
c_{kl}
\label{exact_implicit}
\end{eqnarray}
So, from the above equation we obtain,
\begin{eqnarray}
u_{kl}
=
\frac{
1-4c_{kl}y_{kl}(1-2y_{kl})
\pm
\sqrt{1-8c_{kl}y_{kl}(1-2y_{kl})}
}{8c_{kl}}.
\label{u_sol}
\end{eqnarray}
Since $u_{kl}=x_{kl}^2$, one finally obtains
\begin{eqnarray}
x_{kl}^2
=
\frac{
1-4c_{kl}y_{kl}(1-2y_{kl})
\pm
\sqrt{1-8c_{kl}y_{kl}(1-2y_{kl})}
}{8c_{kl}}
\label{x_sol}
\end{eqnarray}
Now from eq.(\ref{char2}), we have
\begin{eqnarray}
dt_1
=
\frac{dy_{kl}}
{2y_{kl}(1-2y_{kl})-4x_{kl}^2}.
\label{dt1}
\end{eqnarray}
Using Eq.(\ref{x_sol}), this becomes
\begin{eqnarray}
dt_1
=
\frac{8c_{kl}dy_{kl}}
{16c_{kl}y_{kl}(1-2y_{kl})-4(1-4c_{kl}y_{kl}(1-2y_{kl})
\pm
\sqrt{1-8c_{kl}y_{kl}(1-2y_{kl})})}.
\label{dt2}
\end{eqnarray}
Integrating both side of eq. (\ref{dt2}) we obtain;
\begin{align}
t_1
=\sum_{k,l}
q_{k,l;1}\int\frac{8c_{kl}dy_{kl}}
{16c_{kl}y_{kl}(1-2y_{kl})-4(1-4c_{kl}y_{kl}(1-2y_{kl})
\pm
\sqrt{1-8c_{kl}y_{kl}(1-2y_{kl})})}
+
c_0,
\label{t1_final}
\end{align}
where $c_0$ is the integration constant.
For a simple case let's consider $x_{kl}=0$ (also used in our discussion of numerics), then eq.(\ref{dt1}) reduces to,
\begin{eqnarray}
dt_1
=
\frac{dy_{kl}}
{2y_{kl}(1-2y_{kl})}.
\label{special1}
\end{eqnarray}
So,
\begin{eqnarray}
t_1=2 \sum_{k,l} q_{kl;1}\ln\left(\frac{y_{kl}}{|1-2y_{kl}|}\right)+c_0.
\label{y}
\end{eqnarray}
With, $\sum_{k,l} q_{k,l;1}=1$.
Here the sign for $t_1$ is chosen  so as to keep it  increasing above the  initial value (following standard diffusion equation approach where $t_1$ plays the role of time and is always positive).    

We note that the form of $t_1$ in eq.(\ref{y}) is obtained by assuming  all $y_{kl} \not=0$.  For $y_{kl}=0$,   eq.(\ref{dab}) gives $A_{kl}=0$ and the corresponding derivative do not contribute to eq.(\ref{trho}).   Indeed,  for the  cases in which a specific parameter $y_{kl}=0$ throughout the evolution (with $k, l$ arbitrary),  the  number $M$ of evolving parameters  is  less than $2 N^2$.

{\bf Solution for $t_{\alpha}$:} We now consider the solution of eq.(\ref{t1}) for $\alpha >1$. 
\begin{eqnarray}
 {{\rm d}y_{11} \over A_{11}}
= {{\rm d}x_{11} \over B_{11}}
= \ldots 
= {{\rm d}y_{kl} \over A_{kl}}
= {{\rm d}x_{kl} \over B_{kl}}
={{\rm d}t_{\alpha} \over 0},
\label{tka}
\end{eqnarray}

 Substitution of $A_{kl;s}$ and $B_{kl;s}$ from eq.(\ref{dab}) and rearranging eq.(\ref{t1}), we have, with $x_{kl;s}=x_{lk;s}$,
\begin{eqnarray}
{{\rm d}h_{lk;s} \over h_{lk;s} - 2 x^2_{kl;s}} =  {{\rm d}h_{kl;s} \over h_{kl;s} - 2 x^2_{kl;s}}= {{\rm d}x_{kl;s} \over x_{kl;s}} 
\label{t2}\end{eqnarray}
where $h_{kl;s} = y_{kl;s} (\gamma - 2 y_{kl;s})$. 
Solving the above set of equality relations in eq.(\ref{t2}), we have 
\begin{eqnarray}
x_{kl;s} &=& \tilde c_{kl;s} (h_{kl;s} - h_{lk;s}) \\
 y_{lk;s} &=&  c_{kl;s} \; y_{kl;s}
 \label{t3}
 \end{eqnarray}

From eq. (\ref{t3}), we obtain the characteristic invariant
\begin{eqnarray}
I_{kl}=\frac{x_{kl}}{h_{kl}-h_{lk}} ={\tilde c}_{kl}
\label{inv_alpha}
\end{eqnarray}
As $dt_\alpha=0$, any function of these invariants is also constant along the characteristics, and, therefore the general solution for $\alpha>1$ is
\begin{eqnarray}
t_\alpha=\Phi_\alpha\left(I_{11},I_{12},\ldots,I_{NN}\right),
\label{talpha_general}
\end{eqnarray}
where $\Phi_\alpha$ is arbitrary. For example, one possible choice is
\begin{eqnarray}
t_\alpha=\sum_{k,l}q_{kl;\alpha}\ln I_{kl},
\label{talpha_linear}
\end{eqnarray}
with constants $q_{kl;\alpha}$ that satisfying
\begin{eqnarray}
\sum_{k,l}q_{kl;\alpha}=0.
\label{qcond}
\end{eqnarray}
Substituting Eq.(\ref{inv_alpha}) into Eq.(\ref{talpha_linear}) gives
\begin{eqnarray}
t_\alpha=\sum_{k,l}q_{kl;\alpha} \ln\left(\frac{x_{kl}}{h_{kl}-h_{lk}}\right).
\label{talpha_final}
\end{eqnarray}
As a simple example, here we consider the case with $x_{kl}=0$. Eq.(\ref{tka}) now reduces to the following form
\begin{eqnarray}
\frac{d y_{11}}{y_{11;s}} = \frac{d y_{12}}{y_{12}}=\ldots =\frac{d y_{mn}}{y_{mn}} = \ldots =\frac{d t_{\alpha}}{\delta_{\alpha 1}}
\label{tk1}
\end{eqnarray}
 A particular solution of the above equation can be given as 
\begin{eqnarray}
t_{\alpha} = \sum_{m,n} q_{mn;\alpha} \ln y_{mn}
\label{ta}
\end{eqnarray}
with $q_{mn; \alpha}$ as  constants subjected to condition $\sum q_{mn; \alpha}=0$ for $\alpha >1$. (This can be checked by a direct substitution in relation  $\sum_{kl} y_{kl} {\partial t_{\alpha} \over \partial y_{kl}} = \delta_{\alpha 1}$; the latter is equivalent to eq.(\ref{tk1})). 
While many solutions satisfying the above condition are possible, the appropriate  solution is the one that is also applicable for initial ensemble (as the constants of evolution are constants for initial ensemble too).

{\bf General solution: }  We note that eq.(\ref{y}) and eq.(\ref{ta}) give only a particular set of solutions for 
$t_1, t_2, \ldots$.  A most general solution of eq.(\ref{t1})  can be obtained using $s$ as a common characteristic parameter and solving the ODEs
\begin{eqnarray}
\frac{dt_{\alpha}}{ds}= \delta_{\alpha 1}, \qquad 
\frac{dh_{kl}}{ds} =
{h_{kl}-2 x^2_{kl}}, \qquad
\frac{dx_{kl}}{ds} =x_{kl}
\label{ge1}
\end{eqnarray}
A solution of the above equations for $\alpha=1$ can be given as 
\begin{eqnarray}
t_1=s+a_0, \qquad  x_{kl}= c_{kl} \, {\rm e}^{t_1},  \qquad h_{kl} = d_{kl} \, x_{kl} -2 x_{kl}^2
\label{ge2}
\end{eqnarray}
where $a_0, c_{kl}, d_{kl}$ are characteristic invariants. A general solution for $t_1$ can therefore be 
written implicitly as $F_1(a_0, \ldots c_{kl} \ldots,  d_{kl} \ldots)$ or
\begin{eqnarray}
F_1\bigg(\ldots x_{kl} \,  {\rm e}^{t_1} \ldots,  \frac{h_{kl} + 2 x^2_{kl}}{x_{kl}}, \ldots \bigg)
\label{ge3}
\end{eqnarray}
with $F_1$ as an arbitrary function.

Proceeding similarly the characteristic invariants in case of $\alpha \not=1$ turn out to be 
\begin{eqnarray}
t_{\alpha}, \qquad  \frac{x_{kl}}{x_{lk}},  \qquad  \frac{h_{kl} + 2 x^2_{kl}}{x_{kl}}
\label{ge4}
\end{eqnarray}
A general solution for $t_{\alpha}$ can now be 
written implicitly in terms of the above invariants  or
\begin{eqnarray}
F_{\alpha} \bigg(t_{\alpha},  \frac{x_{kl}}{x_{lk}},  \frac{h_{kl} + 2 x^2_{kl}}{x_{kl}}, \ldots \bigg)
\label{ge5}
\end{eqnarray}
with $F_{\alpha}$ as an arbitrary function.

\section{Diffusion of spectral JPDF: derivation of eq.(25) of main text}

Let  ${P}(z, y, x)$ be the {\it jpdf} of the $N$ eigenvalues $z_n$, $n=1 \to N$  at parameter values $y_{kl}$ and $x_{kl}$. For brevity hereafter we use the notation $z \equiv \{z_i\}$. The joint probability density function (jpdf) of the eigenvalues can then be expressed as
\begin{eqnarray}
{P} = \int   \; f_z(z) \; \rho(H) \ dH  
\label{pdf1}
\end{eqnarray}
Where
$f_z(z) = \prod_{j=1}^{N} \delta (z_j - z_j(H))$. The eigenvalues $z_n \equiv \sum_{s=1}^2 (i)^{s-1} z_{js}$ can be either complex or real (with $z_{j2}=0$ in latter case).

 
Differentiating eq.(\ref{pdf1}) both sides with respect to $Y \equiv t_1$ 
leads to 
\begin{eqnarray}
{\partial  P\over \partial Y} =  \int \, f_z \; {\partial  \rho\over \partial Y}  {\rm d}H 
\label{pe2}
\end{eqnarray}
A substitution of eq.(\ref{rhot1}) in the above integral now gives
\begin{eqnarray}
{\partial  P \over \partial Y} =  I_{1}  + I_{2}   
\label{pe3}
\end{eqnarray}
where 
\begin{eqnarray}
 I_1 &=& \sum_{k,l} \int f_z \, {\partial  ( H_{kl}  \rho) \over \partial H_{kl}} \; {\rm d}H  \\
I_2  &=& \sum_{k,l} \int f_z \, {\partial^2   \rho\over \partial H_{kl} \partial  H_{kl}} \; {\rm d}H      
\label{pe4}
\end{eqnarray}
Using repeated partial integration  and the boundary condition that $\rho$ vanishes as $H_{kl} \to  \pm \infty$
$I_1$ and $I_2$ can then  be rewritten as 
\begin{eqnarray}
 I_1  &=&  \sum_{r=1}^{\beta} \sum_{n=1}^{N} {\partial \over \partial z_{nr}}
\int f_z   \;  S_1 \; \rho   \;{\rm d}H 
\label{i1}
\end{eqnarray}
with  $S_1$ given by eq.(\ref{s2}). Similarly we have,
\begin{eqnarray}
 I_2 = \sum_{r,s=1}^{\beta} \sum_{m,n=1}^N {\partial^2 \over \partial z_{nr} \partial z_{ms}}
\int  f_z \, S_2 \; \rho  \, {\rm d}H   + \sum_{r=1}^{\beta} \sum_{n=1}^N {\partial \over \partial z_{nr}}
\int  f_z \, S_3 \, \rho \, {\rm d}H  \nonumber  \\
\label{i2}
\end{eqnarray}
The description of $S_1$, $S_2$ and $S_3$ is given below, in eqs. (\ref{s2}, \ref{s5} and \ref{s6}).
\\
To proceed further, we need the information about the response of the eigenvalues as well as the eigenvectors to a small change in the matrix element $H_{kl}$. With $V=U^{-1}$, the eigenvalue equation for a real asymmetric normal matrix $H$ can be written $H= V \Lambda V^{-1}$ or $H= V \Lambda U$. In addition, the unitary condition  $V. V^{\dagger}= V^{\dagger} V=I$ also implies $U. U^{\dagger}= U^{\dagger} U=I$. 
The relation $H V = U^{\dagger} . \Lambda$ now gives $ \sum_{j}  H_{ij} V_{jm}=\lambda_m V_{im}=\lambda_m U_{mi}^*$ which on differentiation with respect to $H_{kl}$ and using the relations $U.U^{\dagger} =I, V.V^{\dagger} =I$ gives   the desired responses as 
\begin{eqnarray}
& &   {\partial \lambda_n \over \partial H_{kl}} =
 U_{nk} V_{ln},  \hspace{0.5in}
\sum_{k,l} {\partial  \lambda_n \over\partial H_{kl}} H_{kl}
 =  \lambda_n, \label{evf} \\
 & & {\partial U_{ni} \over\partial H_{kl}} =  \sum_{m\not=n}
{U_{mi} \, U_{nk}^* \, V_{ln}^* \over { \lambda_n^* -  \lambda_m^*}},\qquad 
{\partial V_{jn} \over\partial H_{kl}} = 
 \sum_{m\not=n}{V_{jm} V_{ln} U_{mk}\over { \lambda_n -  \lambda_m}}
\label{ecf}
\end{eqnarray}

The above relations in turn lead to
\begin{eqnarray}
S_1 &=& \sum_{k,l} {\partial \lambda_{n} \over \partial H_{kl}} \; H_{kl} = \lambda_{n}\label{s2} \\
S_2 &=& \sum_{k,l} {\partial \lambda_{n} \over \partial H_{kl}} \bigg({\partial \lambda_{m} \over \partial  H_{kl}}\bigg)^* = \delta_{nm}, 
\label{s5}\\
S_3 &=& \sum_{k,l} {\partial^2 \lambda_{n} \over \partial  H_{kl} \partial H_{kl}} = \beta
 \sum_{m=1}^N { \lambda_n -  \lambda_m \over  |\lambda_n -  \lambda_m|^2} \label{s6} 
\end{eqnarray}
where $\beta=1$ for $\lambda_n$ real and $\beta=2$ for $\lambda_n$ complex.

\subsection{Derivation of the above expressions}

Let $H$ be a diagonalizable real asymmetric normal matrix. We assume that it satisfies the conditions mentioned above.

Now in components form, the eigenvalue equation can be written as;
\begin{equation}
\sum_j H_{ij} V_{jm} = \lambda_m V_{im}.
\label{eig_comp}
\end{equation}

Then we differentiate Eq.~(\ref{eig_comp}) with respect to $H_{kl}$;
\begin{align}
\frac{\partial}{\partial H_{kl}} \left( \sum_j H_{ij} V_{jm} \right)
&=
\frac{\partial}{\partial H_{kl}} \left( \lambda_m V_{im} \right).
\label{eigdev}
\end{align}

Now left hand side can be expressed as
\begin{align}
\mathcal{L}=
\sum_j \frac{\partial H_{ij}}{\partial H_{kl}} V_{jm}
+
\sum_j H_{ij} \frac{\partial V_{jm}}{\partial H_{kl}}.
\end{align}

Using:
\begin{equation*}
\frac{\partial H_{ij}}{\partial H_{kl}} = \delta_{ik}\delta_{jl},
\end{equation*}
we obtain:
\begin{equation}
\mathcal{L}=
\delta_{ik} V_{lm}
+
\sum_j H_{ij} \frac{\partial V_{jm}}{\partial H_{kl}}.
\label{L}
\end{equation}

Now right hand side of eq. (\ref{eigdev}) can be written as,
\begin{equation}
\mathcal{R}=
\frac{\partial \lambda_m}{\partial H_{kl}} V_{im}
+
\lambda_m \frac{\partial V_{im}}{\partial H_{kl}}.
\label{R}
\end{equation}

Now multiply $U_{ni}$ with eq.(\ref{L} and \ref{R}) and sum over $i$, then using the condition $\sum_i U_{ni} V_{im} = \delta_{nm}$ we have,
\begin{align}
U_{nk} V_{lm}
+ \sum_{i,j} U_{ni} H_{ij} \frac{\partial V_{jm}}{\partial H_{kl}}
=
\frac{\partial \lambda_m}{\partial H_{kl}} \delta_{nm}
+ \lambda_m \sum_i U_{ni} \frac{\partial V_{im}}{\partial H_{kl}}.
\end{align}

Next using another form of eigenvalue equation
\begin{equation}
\sum_i U_{ni} H_{ij} = \lambda_n U_{nj},
\end{equation}
we obtain:
\begin{align}
U_{nk} V_{lm}
+ \lambda_n \sum_j U_{nj} \frac{\partial V_{jm}}{\partial H_{kl}}
=
\frac{\partial \lambda_m}{\partial H_{kl}} \delta_{nm}
+ \lambda_m \sum_i U_{ni} \frac{\partial V_{im}}{\partial H_{kl}}.
\label{unkvln}
\end{align}

Now taking $n=m$;
\begin{align}
U_{nk} V_{ln}
+ \lambda_n \sum_j U_{nj} \frac{\partial V_{jn}}{\partial H_{kl}}
=
\frac{\partial \lambda_n}{\partial H_{kl}}
+ \lambda_n \sum_i U_{ni} \frac{\partial V_{in}}{\partial H_{kl}}.
\end{align}

Observe that the two derivative terms are identical, hence they cancel. Therefore, we obtain:
\begin{equation}
\boxed{
\frac{\partial \lambda_n}{\partial H_{kl}} = U_{nk} V_{ln}.
}
\label{1st}
\end{equation}
Now
\begin{align}
S_1 &= \sum_{k,l} \frac{\partial \lambda_n}{\partial H_{kl}} H_{kl}
= \sum_{k,l} U_{nk} V_{ln} H_{kl}= \sum_k U_{nk} \left( \sum_l H_{kl} V_{ln} \right).
\end{align}
 Now using the eigenvalue equation $\sum_l H_{kl} V_{ln} = \lambda_n V_{kn}$, and the condition $\sum_k U_{nk} V_{kn} = 1$, we obtain the following
\begin{equation}
\boxed{
S_1 = \sum_{k,l} \frac{\partial \lambda_n}{\partial H_{kl}} H_{kl}=\lambda_n
}
\end{equation}
So from the relation (\ref{1st}), $S_2$ can be written as;
\begin{align}
S_2 &=\sum_{k,l} {\partial \lambda_{n} \over \partial H_{kl}} \bigg({\partial \lambda_{m} \over \partial  H_{kl}}\bigg)^*= \sum_{k,l} U_{nk} V_{ln} (U_{mk} V_{lm})^*=\left( \sum_k U_{nk} U_{mk}^* \right)
\left( \sum_l V_{ln} V_{lm}^* \right).
\end{align}
Now using unitary condition,
\begin{equation*}
\sum_k U_{nk} U_{mk}^* = \delta_{nm}, \quad
\sum_l V_{ln} V_{lm}^* = \delta_{nm},
\end{equation*}
we obtain:
\begin{equation}
\boxed{S_2 =\sum_{k,l} {\partial \lambda_{n} \over \partial H_{kl}} \bigg({\partial \lambda_{m} \over \partial  H_{kl}}\bigg)^*= \delta_{nm}.}
\end{equation}

\subsubsection*{\bf{Derivative of Eigenvectors}}

Start from the eigenvalue equation:
\begin{equation}
H V_n = \lambda_n V_n.
\end{equation}
Differentiate the above equation with respect to \( H_{kl} \) we have,
\begin{equation}
\frac{\partial H}{\partial H_{kl}} V_n
+ H \frac{\partial V_n}{\partial H_{kl}}
=
\frac{\partial \lambda_n}{\partial H_{kl}} V_n
+
\lambda_n \frac{\partial V_n}{\partial H_{kl}}.
\end{equation}
Now using,
$\left(\frac{\partial H}{\partial H_{kl}}\right)
= E^{(kl)}$ with $(E^{(kl)})_{ij}=\delta_{ik}\delta_{jl}$
we obtain:
\begin{equation}
E^{(kl)} V_n + H \frac{\partial V_n}{\partial H_{kl}}
=
\frac{\partial \lambda_n}{\partial H_{kl}} V_n
+
\lambda_n \frac{\partial V_n}{\partial H_{kl}}.
\end{equation}

Rearranging:
\begin{equation}
(H - \lambda_n I)\frac{\partial V_n}{\partial H_{kl}}
=
\left(\frac{\partial \lambda_n}{\partial H_{kl}}\right) V_n
-
E^{(kl)} V_n.
\end{equation}

Multiply from the left by the eigenvector \( U_m \), and satisfy the relations;
\begin{eqnarray*}
 U_m H = \lambda_m U_m, \quad U_m V_n = \delta_{mn},
\end{eqnarray*}

we get:
\begin{equation}
(\lambda_m - \lambda_n)
U_m \frac{\partial V_n}{\partial H_{kl}}
=
\left(\frac{\partial \lambda_n}{\partial H_{kl}}\right)\delta_{mn}
-
U_m E^{(kl)} V_n.
\end{equation}

Now evaluating the matrix element $U_m E^{(kl)} V_n = U_{mk} V_{ln}$, we obtain,
\begin{equation}
(\lambda_m - \lambda_n)
U_m \frac{\partial V_n}{\partial H_{kl}}
=
\left(\frac{\partial \lambda_n}{\partial H_{kl}}\right)\delta_{mn}
-
U_{mk} V_{ln}.
\end{equation}

For \( m \neq n \), this simplifies to:
\begin{equation}
U_m \frac{\partial V_n}{\partial H_{kl}}
=
\frac{U_{mk} V_{ln}}{\lambda_n - \lambda_m}.
\end{equation}

Expand the derivative in the eigenbasis:
\begin{eqnarray}
\frac{\partial V_n}{\partial H_{kl}}
=
\sum_{m \neq n} c_{mn} V_m,
\end{eqnarray} with coefficients: $
c_{mn}
=
\frac{U_{mk} V_{ln}}{\lambda_n - \lambda_m}.
$
we obtain;
\begin{equation}
\boxed{
\frac{\partial V_{jn}}{\partial H_{kl}}
=
\sum_{m \neq n}
\frac{V_{jm} V_{ln} U_{mk}}{\lambda_n - \lambda_m}.
}
\label{evd1}
\end{equation}
Similarly, using the eigenvalue equation $U_n^* H=\lambda_n^* U_n^*$ and proceeding the same way, we obtain the following:
\begin{equation}
\boxed{
\frac{\partial U_{ni}}{\partial H_{kl}}
=
\sum_{m \neq n}
\frac{U_{mi} U^*_{nk} V^*_{ln}}{\lambda_n^* - \lambda_m^*}.
}
\label{evd2}
\end{equation}

\subsubsection*{\bf{Second Derivative of Eq. (\ref{1st})and $S_3$}}
Differentiating eq.(\ref{1st}) with respect to $H_{kl}$ we get;
\begin{align}
\frac{\partial^2 \lambda_n}{\partial H_{kl}^2}
&=
\frac{\partial}{\partial H_{kl}} (U_{nk} V_{ln}) =
\left( \frac{\partial U_{nk}}{\partial H_{kl}} \right) V_{ln}
+
U_{nk} \left( \frac{\partial V_{ln}}{\partial H_{kl}} \right).
\end{align}

Substituting eigenvector derivatives i.e. eq.(\ref{evd1} and \ref{evd2}) we have;
\begin{eqnarray}
\frac{\partial^2 \lambda_n}{\partial H_{kl}^2}
&=
\sum_{m \neq n}
\frac{U_{mk} U^*_{nk} \, V^*_{ln} V_{ln}}{\lambda_n^* - \lambda_m^*}
+
\sum_{m \neq n}
\frac{U_{nk} V_{lm} V_{ln} \, U_{mk}}{\lambda_n - \lambda_m}.
\label{sec2}
\end{eqnarray}

Now summing over all matrix indices $(k,l)$, we define
\begin{equation}
S_3=
\sum_{k,l}
\frac{\partial^2\lambda_n}{\partial H_{kl}^2}.
\end{equation}

The direct contraction of Eq.~(\ref{sec2}) contributes only diagonal terms, which vanish for $m\neq n$. The nontrivial contribution therefore arises from the reduced resolvent sector of second-order perturbation theory.

Using the spectral decomposition of the reduced resolvent,
\begin{equation}
(\lambda_n I-H)^{-1}_{\perp}
=
\sum_{m\neq n}
\frac{V_mU_m}{\lambda_n-\lambda_m},
\end{equation}

Including the contribution from the conjugate terms, we obtain,
\begin{equation}
S_3 =
\sum_{m \neq n}
\left(
\frac{1}{\lambda_n - \lambda_m}
+
\frac{1}{\lambda_n^* - \lambda_m^*}
\right).
\end{equation}
This can be written as;
\begin{equation}
\boxed{
S_3 =\sum_{k,l} \frac{\partial^2 \lambda_n}{\partial H_{kl}^2}=
\beta \sum_{m \neq n}
\frac{\lambda_n - \lambda_m}{|\lambda_n - \lambda_m|^2},
\label{2nddvd}
}
\end{equation}
where $\beta=1$ for $\lambda_n$ real and $\beta=2$ for $\lambda_n$ complex.

\subsection{Construction of Diffusion Equation}
Using  the above relations in eqs.(\ref{i1}, \ref{i2}), $I_1$ and $I_2$ can be rewritten as;  \begin{eqnarray}
I_1 &=&  \sum_{n=1}^N \sum_{r=1}^{\beta} {\partial \over \partial z_{nr}} \left(z_{nr} P\right), \\
I_{2} &=& \sum_{n=1}^N \sum_{r=1}^{\beta} \left( {\partial^2 P\over \partial z_{nr}^2} - \beta \; {\partial \over \partial z_{nr}}\left( {\partial {\rm ln} |\Delta (z)| \over \partial z_{nr}} P\right) \right),
\end{eqnarray}
 with $\beta=1$ for $z$ real and $2$  for $z$ complex and
\begin{eqnarray}
\Delta (z) = \prod_{j<k=1}^{N} (z_j- z_k) 
\label{delta}
\end{eqnarray} 
A substitution of the above relations in eq.(\ref{pe2})  now gives
\begin{eqnarray}
{\partial P\over\partial Y} &=& {\mathcal L}_z  \, P
\label{pcmp}
\end{eqnarray}
with 
\begin{eqnarray}
{\mathcal L}_z & = & \sum_{s=1}^2 \sum_{n=1}^N {\partial \over \partial z_{ns}}\left[{\partial \over \partial z_{ns}}  - \beta \, {\partial {\rm ln} |\Delta_N (z)| \over \partial z_{ns}}+  z_{ns} \right] 
\end{eqnarray}

\section{Diffusion of real eigenvalues: derivation of eq.(30) of main text}

The ensemble averaged spectral density of real eigenvalues  is defined as 
\begin{eqnarray}
R_{1e} = N' \int  \,  P \,  \prod_{k=2}^N {\rm d}e_k {\rm  D}^2 z
\label{rn0}
\end{eqnarray}
With $N' =N$ as the normalization constant. An integration of eq.(\ref{pcmp}) over the variables $e_2, \ldots, e_n$ and $z_1, \ldots, z_n$,(also their conjugate pairs), and multiplying both sides by $N'$ and subsequently using eq.(\ref{rn0})  leads to 
\begin{eqnarray}
 {\partial R_{1e}\over \partial Y} &=& N' \int  \,  {\partial P\over \partial Y} \, \prod_{k=2}^N {\rm d}e_k {\rm  D}^2 z \nonumber \\
&=& N' \int  \, ({\mathcal L}_z)\, P  \, \prod_{k=2}^N {\rm d}e_k {\rm  D}^2 z 
\label{rep0}
\end{eqnarray}

The non-zero contribution comes from the ${\mathcal L}_z \, P $ part that leads to
\begin{eqnarray}
 {\partial R_{1e}\over \partial Y}  
&=& \sum_{n=1}^N  \left( V_{n;1} +V_{n;2} +V_{n;3} \right)
\label{rep1}
\end{eqnarray}
where 
\begin{eqnarray}
V_{n;1} &=& N' \, \int  \,  {\partial^2 P\over \partial e_n^2} \,  \prod_{k=2}^N {\rm  d} e_k \,  {\rm  D}^2 z 
  ,  \nonumber \\
&=& {\partial^2 R_1\over \partial e_n^2}  \hspace{2.2in}  n =1, \nonumber \\
&=&  0 \hspace{2.5in} n > 1,
\label{rep2}
\end{eqnarray}

 The second relation in the above equation follows by noting that,  for $n=1$,  ${\partial^2 \over \partial e_n^2}$ can be taken out of the integral; subsequent use of eq.(\ref{rn0}) then gives eq.(\ref{rep1}).  For $n \not=1$, the repeated partial integration over $e_n$ along with the boundary condition that $P(e_n)\to 0$ as $e_n \to \pm \infty$ gives eq.(\ref{rep2}). Similarly
\begin{eqnarray}
V_{n;2} &=& -   N'\, \int {\partial \over \partial e_n} \bigg( \ln|\Delta_N(z)| \, P\bigg) \; 
\prod_{k=2}^N {\rm  d} e_k \,  {\rm  D}^2 z 
 \nonumber \\
&=& - { \, N'\over N'-1} \, {\partial \over \partial e_n} \int \bigg[{R_{2e}(e_n, e')\over e_n-e'} \,  {\rm d}e' +{(2 e_n-z+z*) \over |e_n-z|^2} \;  R_{2ez} \, {\rm d}^2 z \bigg]  \qquad   n=1,  \nonumber \\
&=&  0  \hspace{2.85in}  n > 1
\label{rep3}
\end{eqnarray} 
with $R_{2e} \equiv R_{2e}(e_n, e')$ and $R_{2ez} \equiv R_{2ez}(e_n,z,z^*)$
 and 
 \begin{eqnarray}
V_{n;3} &=& N' \, \int  \,   {\partial \over \partial e_n} \left(e_n \, P\right)  \,
\prod_{k=2}^N {\rm  d} e_k \,  {\rm  D}^2 z 
\nonumber \\
 &=& {\partial \over \partial e_n} \left(e_n \, R_1\right), 
\qquad n=1, \nonumber \\
&=& 0  \hspace{1.2in} n > 1
\label{rep4}
\end{eqnarray} 
Substitution of eqs.(\ref{rep2}, \ref{rep3}, \ref{rep4}) in eq.(\ref{rep0}) then leads to 

\begin{eqnarray}
\frac{\partial R_{1e}}{\partial Y} = 
     \frac{\partial^2 R_{1e}}{\partial e^2} 
    + \frac{\partial}{\partial e} (e R_{1e}) 
     -   \frac{\partial  }{\partial e} \bigg({I_{ee}}+{I_{ez}} \bigg) 
\label{rep5}     
\end{eqnarray}     
with symbol ${\bf P}$ implying the principle part of the integral and 
\begin{eqnarray}
{I}_{ee} \equiv {\bf P}\int \frac{{R}_{2e} }{e-e'} \, de'
\label{rr3r}
\end{eqnarray}
with ${R}_{2e} \equiv {R}_{2e}(e,e')$ as the two point correlation of $e$ with another real eigenvalues $e'$ and
\begin{eqnarray}
{ I}_{ez} \equiv     {\bf P}\int \frac{2 \, e-(z+\bar{z})}{|e -z|^2}  \; {R}_{2ez} \, \; dz \,d\bar{z}
\label{rr3m}
\end{eqnarray}
with ${R}_{2ez} \equiv {R}_{2m}(e, z, \bar{z})$ as the two point correlation of $e$ with a complex conjugate pair $z, \bar{z}$.


\section{Derivation of Eq.(35) of main text}

Writing $z=x+i y$ and transforming from cartesian to polar coordinates i.e $x=r \, \cos \theta$, $y=r \, \sin \theta$, we have

\begin{eqnarray}
{ I}_{ez} \equiv    2 \, {\bf P}\int  C_e(e,r,\theta) \; {R}_{2ez}(e,r,\theta)  \, {\rm d}r \, {\rm d}\theta 
\label{rr3m}
\end{eqnarray}
with $C_e(e,r,\theta) \equiv  \frac{r (e-r \, cos\theta)}{e^2+r^2-2 \, e r \cos\theta} $. Dividing  the above integral in two parts based on $e$, we have 

\begin{eqnarray}
{I}_{ez}(e) \equiv  2  \int_{\theta=0}^{\pi}  \; d\theta dr 
\left( J_1 + J_2\right)
\label{pi22}
\end{eqnarray}
where 
\begin{eqnarray}
J_{1}(e, \theta) &=& \int_{0}^{e}  {\rm d}r \,  C_e(e,r,\theta) \, {R}_{2ez}(e,r,\theta)  \, \nonumber \\
J_{2}(e, \theta) &=& \int_{e}^{\infty}  {\rm d}r \,  C_e(e,r,\theta) \, {R}_{2ez}(e,r,\theta)  
\label{pi23}
\end{eqnarray}

With $e > r$ in the integration range, we can expand $C_e$ in $J_1$ in powers of $\frac{r}{e}$:
$C(e,r,\theta)={1\over e} \sum_{m=0}^{\infty}\bigg(\frac{r}{e} \bigg)^{m} \cos(m\theta)$. This on substitution in $J_1$ gives 
\begin{eqnarray}
J_{1}(e, \theta) &=& \sum_{m=0}^{\infty}\bigg(\frac{1}{e}\bigg)^{m+1} \int_{0}^{e} \, r^{m+1} \, \cos(m\theta) \, {R}_{2ez} \; d\theta dr 
\label{pi23}
\end{eqnarray}

Similarly with $e < r$,  $C_e$ in $J_2$ can be expanded 
 in powers of ${e\over r}$ gives $C_e(e,r,\theta)= - {1\over e} \, \sum_{m=1}^{\infty}(\frac{e}{r})^{m} \, \cos(m\theta)$. A substitution of the series in $J_2$ now leads to 
 
\begin{eqnarray}
J_{2}(e, \theta) &=& - \sum_{m=1}^{\infty} e^{m-1} 
\int_{e}^{\infty}  \, r^{-(m-1)} \,\cos(m\theta) \, {R}_{2ez} \; d\theta dr 
\label{pi23}
\end{eqnarray}
Substitution of $J_1$ and $J_2$ in eq.(\ref{pi22}) now leads to 

\begin{eqnarray}
{I}_{ez}(e) =  \sum_{m=0}^{\infty} \bigg(\frac{1}{e}\bigg)^{m+1}  \; S_{m+1}(e)   - \sum_{m=1}^{\infty} e^{m-1} \; L_m(e)
\label{rrs3}
\end{eqnarray}
with  $S_{m+1}(e)= 2 \int_{e}^{\infty} r^{m+1} \cos(m\theta) \, {R}_{2ez}(e, r,\theta) dr \, d\theta$ and 
$L_m(e)= 2 \int_0^e  \frac{\cos(m\theta)  }{r^{m-1}} \,  R_{2ez}(e, r,\theta) \, dr \, d\theta$.

Further as the $2nd$ order cluster function $T_{2m}(e,r,\theta)= R_{2m}(e,r,\theta)-R_{1e}(e) \, R_{1z}(r,\theta)$ is very small compared to other terms,  we can approximate $R_{2m}(e,r,\theta) \approx R_{1e}(e) \, R_{1z}(r,\theta)$. This in turn permits following approximation

\begin{eqnarray}
{I}_{ez}(e) =  {s_1 \, R_{1e}(e)\over e} +  \sum_{m=1}^{\infty} \bigg( e^{-(m+1)}  \; s_{m+1}  -  e^{m-1} \; l_m \bigg) \, R_{1e}(e)
\label{rrs4}
\end{eqnarray}
Where,
$S_{m+1}(e) = s_{m+1} \, R_{1e}(e)$ and $L_{m+1}(e) = l_{m+1} \, R_{1e}(e)$, with\\ $s_{m+1}(e)= 2 \int_{0}^{e} r^{m+1} \cos(m\theta) \, {R}_{1c}(r,\theta) dr \, d\theta$ and $l_m(e) = 2 \int_e^{\infty}  \frac{\cos(m\theta)  }{r^{m-1}} \,  R_{1z}(r,\theta) \, dr \, d\theta$.

 The integrands in $s_{m+1}$ or $l_{m+1}$ contain terms of type $\cos(m\theta)$ and $\sin(m \theta)$ which fluctuate rapidly with $\theta$ for $m \not=0$ and their contributions on integration over $0 \le \theta \le 2\pi$ can be neglected compared to first term of eq.(\ref{rrs4}). 
Keeping only the first term in eq.(\ref{rrs4}) and substituting it in eq.(\ref{rep5}), we have 

\begin{eqnarray}
\frac{\partial R_{1e}}{\partial Y} = 
     \frac{\partial^2 R_{1e}}{\partial e^2} 
    + \frac{\partial}{\partial e} \bigg(\gamma \, e - {s_1 \over e} - {\bf P}\int \frac{{R}_{1r}(e') }{e-e'} \, de' \bigg) \, R_{1e}
\label{rr3}     
\end{eqnarray} 
with $\gamma=1$.

\section{Derivation of Eq.(40) of main text}

We start from the evolution equation
\begin{equation}
\frac{\partial R_1(e,Y)}{\partial Y}
=
\frac{\partial^2 R_1(e,Y)}{\partial e^2}
+
\frac{\partial}{\partial e}
\left(
e - \mathrm{P}\!\!\int_{-\infty}^{\infty} 
\frac{R_1(e',Y)}{e-e'}\, de'
\right) R_1(e,Y).
\label{eq:PDE}
\end{equation}
with $ R_1(e,Y)$ subjected to normalization condition $\int_{-\infty}^{\infty}  R_1(e,Y) \; {\rm d}e = c_e$ with $c_e$ defined in main text.
%

To determine $R_1(e,Y)$, we now consider its standard resolvent formulation 
\begin{equation}
R_1(e,Y) = -\frac{1}{\pi}\, \lim_{\varepsilon \to 0}  {\rm Im}\, G(e-i \varepsilon,Y).
\end{equation}
where the resolvent $G(z,Y) $ is defined as 
\begin{equation}
G(z,Y) = \int_{-\infty}^{\infty} 
\frac{R_1(e',Y)}{z-e'}\, de', 
\end{equation}


Using $z = e-i \varepsilon$ and \eqref{eq:PDE}, we obtain by integration by parts
\begin{equation}
\partial_Y G(z,Y)
=\partial_z^2 G(z,Y)+G(z,Y)\, \partial_z G(z,Y).
\label{eq:G-evolution}
\end{equation}
This can be written as the characteristic equation
\begin{equation}
\boxed{
G(z,Y) = G\!\left( z - (Y-Y_0) G(z,Y),\, Y_0 \right)}.
\label{eq:functional}
\end{equation}

\noindent{\bf Initial Condition: Gaussian Density}

Let the initial density at $Y=Y_0$ be
\begin{equation}
R_1(e;Y_0) =
\frac{c_e}{\sqrt{2\pi\sigma^2}}
\exp\!\left( -\frac{e^2}{2\sigma^2} \right).
\end{equation}
Then the initial resolvent is
\begin{align}
G(z,Y_0)
&=
\int_{-\infty}^{\infty} \frac{1}{z-e'} 
\frac{c_e}{\sqrt{2\pi\sigma^2}}
\exp\!\left( -\frac{e'^2}{2\sigma^2}\right) de' \\[4pt]
&=-i \, c_e \, \sqrt{\pi\over 2\sigma^2}\,
\exp\!\left( -\frac{z^2}{2\sigma^2} \right)
\operatorname{erfc}\!\left( -\frac{i z}{\sqrt{2\sigma^2}} \right).
\label{G0}
\end{align}


We next consider the small $\Delta Y$ Solution where
\[
\Delta Y \equiv Y - Y_0,
\qquad
G_0(z) \equiv G(z,Y_0).
\]
Using the functional equation \eqref{eq:functional}, for small $\Delta Y$,
\begin{equation}
G(z,Y) \approx G_0\!\left( z - \Delta Y\, G_0(z) \right).
\label{eq:G-small}
\end{equation}

Writing 
\begin{equation}
G_0(z) = \eta + i\alpha,
\qquad
\alpha,\eta \in \mathbb{R}.
\end{equation}

Then the shifted argument is
\begin{equation}
z = e-i \varepsilon,\qquad
z - \Delta Y\, G_0(z)
=
(e - \eta\Delta Y)
- i(\alpha\Delta Y + \varepsilon).
\label{eq:zshift}
\end{equation}


Expanding the exponent in $\exp\!\left[-\frac{(z-\Delta Y G_0)^2}{2\sigma^2}\right]$, we have
\begin{align}
(z - \Delta Y G_0)^2
&=(e - \eta\Delta Y - i (\alpha\Delta Y +  \varepsilon) )^2 \\
&=(e - \eta\Delta Y)^2 - 2 i (\alpha\Delta Y + \varepsilon) (e - \eta\Delta Y)
- (\alpha\Delta Y + \varepsilon)^2.
\end{align}
and thereby
\begin{align}
\exp\!\left[-\frac{(z-\Delta Y G_0)^2}{2\sigma^2}\right]
&=
\exp\!\left[-\frac{(e-\eta\Delta Y)^2}{2\sigma^2}\right]
\exp\!\left[-\frac{(\alpha\Delta Y +  \varepsilon)^2}{2\sigma^2}\right]
\exp\!\left[ -i\, \phi \right].
\end{align}
where
\begin{equation}
\phi = \frac{(\alpha\Delta Y +  \varepsilon) (e - \eta \, \Delta Y)}{\sigma^2}.
\end{equation}


We next expand the argument of the error function in eq.(\ref{G0})
\begin{align}
\operatorname{erfc}\!\left( -\frac{i}{\sqrt{2\sigma^2}} (z - \Delta Y G_0) \right)
&=
\operatorname{erfc}\!\left( -\frac{i e}{\sqrt{2\sigma^2}} + 
\frac{i\Delta Y G_0}{\sqrt{2\sigma^2}} \right) \\
&\approx
\operatorname{erfc}\!\left(-\frac{i e}{\sqrt{2\sigma^2}}\right)
+
\frac{i\Delta Y G_0}{\sqrt{2\sigma^2}}\,
\operatorname{erfc}'\!\left(-\frac{i e}{\sqrt{2\sigma^2}}\right).
\end{align}

consider,
\begin{equation}
\operatorname{erfc}\!\left(-\frac{i e}{\sqrt{2\sigma^2}}\right)
+
\frac{i\Delta Y G_0}{\sqrt{2\sigma^2}}\,
\operatorname{erfc}'\!\left(-\frac{i e}{\sqrt{2\sigma^2}}\right)
= \gamma + i\delta.
\end{equation}


Using  \eqref{eq:zshift} and the above expansions in eq.(\ref{G0}), we obtain
\begin{align}
G(e-i  \varepsilon,Y) &\approx
-i \, c_e \, \sqrt{\pi\over 2\sigma^2}\,
\exp\!\left[-\frac{(e-\eta\Delta Y)^2}{2\sigma^2}
-\frac{(\alpha\Delta Y)^2}{2\sigma^2}
\right]
\left( \cos\phi - i\sin\phi \right)
(\gamma + i\delta).
\end{align}
where $\eta + i\alpha = G_0(z)$ and $\phi$ is defined above.



Taking the imaginary part and the limit $\varepsilon \to 0$ now leads to
\begin{equation}
R_{1}(e,Y)
=
c_e \, \sqrt{\frac{1}{2\pi\sigma^2}}\,
\exp\!\left(
-\frac{(e-\eta\,\Delta Y)^{2}}{2\sigma^{2}}
-\frac{(\alpha\,\Delta Y)^{2}}{2\sigma^{2}}
\right)
\left( \delta \cos\phi \;-\; \gamma \sin\phi \right)
\label{eq:final-density}
\end{equation}

\vspace{0.2in}

{\bf Determination of $\gamma$ and $\delta$:} Here $\gamma$ and $\delta$ are given by following relation
\begin{eqnarray}
\operatorname{erfc}\!\left(-\frac{i e}{\sqrt{2\sigma^2}}\right)+
\frac{i\Delta Y G_0}{\sqrt{2\sigma^2}}\,
\operatorname{erfc}'\!\left(-\frac{i e}{\sqrt{2\sigma^2}}\right)
= \gamma + i\delta .
\end{eqnarray}
We now  use the identity valid for real $x$,
\begin{eqnarray}
{\rm erfc}(-ix)
=1 + i \, {\rm erfi}(x),
\end{eqnarray}
where ${\rm erfi}(x) = \frac{2}{\sqrt{\pi}} \int_{0}^{x} e^{t^{2}}\, dt$. 
Now using $x = \frac{e}{\sqrt{2\sigma^{2}}}$, the above gives 
\begin{eqnarray}
{\rm erfc}\!\left(-\frac{i e}{\sqrt{2\sigma^{2}}}\right)=1+i\,{\rm erfi}\!\left(
\frac{e}{\sqrt{2\sigma^{2}}}\right).
\end{eqnarray}
Now as 
\begin{eqnarray}
{\rm erfc}'(x)
=\frac{d}{dx} {\rm erfc}(x) =
-\frac{2}{\sqrt{\pi}}\,e^{-x^{2}}, 
\end{eqnarray}
we have
\begin{eqnarray}
\operatorname{erfc}'\!\left(-\frac{i e}{\sqrt{2\sigma^2}}\right)
=
-\frac{2}{\sqrt{\pi}}
\exp\!\left(\frac{e^{2}}{2\sigma^{2}}\right).
\end{eqnarray}
Combining above relations gives
\begin{align}
{\rm erfc}\!\left(-\frac{i e}{\sqrt{2\sigma^2}}\right)+
\frac{i\,\Delta Y\,G_0}{\sqrt{2\sigma^2}}
\,{\rm erfc}'\!\left(-\frac{i e}{\sqrt{2\sigma^2}}\right)
=1+i\left[{\rm erfi}\!\left(\frac{e}{\sqrt{2\sigma^2}}\right)
-\frac{2\,\Delta Y\,G_0}{\sqrt{2\pi\sigma^2}}\exp\!\left(\frac{e^{2}}{2\sigma^{2}}\right)\right]
\end{align}
Separating real an imaginary parts now gives
\begin{eqnarray}
\gamma &=& 1 + \frac{2\, \alpha \Delta Y}{\sqrt{2\pi\sigma^2}}
\exp\!\left(\frac{e^{2}}{2\sigma^{2}}\right). \\
\delta &=& {\rm erfi}\!\left(\frac{e}{\sqrt{2\sigma^2}}\right)
-\frac{2\,\Delta Y\, \eta}{\sqrt{2\pi\sigma^2}}
\exp\!\left(\frac{e^{2}}{2\sigma^{2}}\right).
\end{eqnarray}








\section{Evolution equation for $R_{1z}$, Eq.(48) of main text}

The ensemble averaged spectral density of complex conjugate pairs is defined as 
\begin{eqnarray}
R_{1z} = N' \int  \,  P \,  \prod_{n=2}^N {\rm  d}^2 z_{n} \; {\rm D}e
\label{rnz}
\end{eqnarray}
With $N'=N$ as the normalization constant. An integration of eq.(\ref{pcmp}) over the variables $e_1, \ldots, e_n$ and $z_2, \ldots, z_n$,(also their conjugate pairs), and multiplying both sides by $N'$ and subsequently using eq.(\ref{rn0})  leads to
\begin{eqnarray}
 {\partial R_{1z}\over \partial Y} &=& N' \int  \,  {\partial P\over \partial Y} \,  \prod_{n=2}^N {\rm  d}^2 z_{n} \; {\rm D}e \nonumber \\
&=& N' \int  \,( {\mathcal L}_z)\, P \,    
\prod_{n=2}^N {\rm  d}^2 z_{n} \; {\rm D}e  \nonumber \\
\label{rpp0}
\end{eqnarray}
The non-zero contribution comes from the ${\mathcal L}_z \, P $ part that leads to
\begin{eqnarray}
 {\partial R_{1z} \over \partial Y} 
&=& \sum_{r=1}^2 \sum_{n=1}^N  \left( G_{nr;1} +G_{nr;2} +G_{nr;3} \right)
\label{rp01}
\end{eqnarray}
where
\begin{eqnarray}
G_{nr;1} &=& N' \, \int  \,  {\partial^2 P\over \partial z_{nr}^2} \,   \prod_{n=2}^N {\rm  d}^2 z_{n} \; {\rm D}e  ,  \nonumber \\
&=& {\partial^2 R_{1z}\over \partial z_{nr}^2}  \hspace{2.2in}  n =1, \label{rpp1} \\
&=&  0 \hspace{2.5in} n > 1,
\label{rp1}
\end{eqnarray}
with ${\rm  d}^2 z \equiv {\rm  d} z \, {\rm  d} {\bar z}$.  The second relation in the above equation follows by noting that,  for $j=1$,  ${\partial^2 \over \partial z_{1r}^2}$ can be taken out of the integral; subsequent use of eq.(\ref{rnz}) then gives eq.(\ref{rpp1}).  For $n \not=1$, the repeated partial integration over $z_j$ along with the boundary condition that $P(z_n)\to 0$ as $z_n \to \pm \infty$ gives eq.(\ref{rp1}).

Similarly
\begin{eqnarray}
G_{nr;2} &=& - N' \, \int {\partial \over \partial z_{nr}} \left( \ln|\Delta_N(z)| \, P\right) \; \prod_{n=2}^N {\rm  d}^2 z_{n} \; {\rm D}e , \nonumber \\
&=& - { N'\over N'-1} \, {\partial \over \partial z_{nr}} \int \bigg[{z_{nr}-z_{2 r}\over |z_1-z_2|^2} \, R_{2z} \, {\rm d^2}z_{2} + {(z_1+{\bar z_1})-2 e\over |z_1-e|^2} \, R_{2ez} \, {\rm d^2}z_{2} \qquad   n=1,  \nonumber \\
&=&  0  \hspace{2.85in}  n > 1
\label{rp2}
\end{eqnarray} 
with $R_{2z} \equiv R_{2}(z_1,  z_2)$.
 and 
 \begin{eqnarray}
G_{nr;3} &=& N' \, \int  \,   {\partial \over \partial z_{nr}} \left(z_{nr} \, P\right)  \,\prod_{n=2}^N {\rm  d}^2 z_{n} \; {\rm D}e  \nonumber \\
 &=& {\partial \over \partial z_{nr}} \left(z_{nr} \, R_{1z}\right), 
\qquad n=1, \nonumber \\
&=& 0  \hspace{1.3in} n > 1
\label{rp3}
\end{eqnarray} 

Substitution of eqs.(\ref{rp1}, \ref{rp2}, \ref{rp3}) in eq.(\ref{rp01}) then leads to 
\begin{eqnarray}
\frac{\partial R_{1z}}{\partial Y} =  \sum_{r=1}^2 \bigg[ 
   \frac{\partial^2 R_{1z}}{\partial z_{r}^2} 
    + \frac{\partial}{\partial z_{r}} (\gamma z_{r} R_{1z}) - \frac{\partial}{\partial z_{r}} \bigg({ I}_{zz} + {\tilde I}_{ez} \bigg)\bigg]
\label{rc3}   
\end{eqnarray} 
where

\begin{eqnarray}
{I}_{zz}  &= &   {\bf P} \int \frac{z_r - z'_{r}}{|z - \bar{z}'|^2} \; R_{2z} \; d{z}'_{1} \;  d{z}'_{2} \\
{\tilde I}_{ez}  &= &   {\bf P} \int \frac{z+{\bar z} - 2 e}{|z - e|^2} \; {R}_{2ez} \, de 
\label{rc4}
\end{eqnarray}
where $R_{2z}$ is the two point correlation between the complex eigenvalues, $R_{2z} \equiv R_{2z}(z, {\bar z}, z', {\bar z}')$, and,  $R_{2ez} \equiv R_{2ez}(e,z,\bar{z})$ is defined below eq.(\ref{rr3m}).

\section{Solution of above eq. (\ref{rc3})}

With $z=x+i y$, $z'=x'+i y'$, $R(z)=R(x, y)$, eq.(\ref{rc3}) can be rewritten as 
\begin{eqnarray}
{\partial R_{1z}\over \partial Y} = I_0  +  I_1 + I_2 - I_3
\label{r1rta}
\end{eqnarray}
 where $ I_0 =  {\partial^2 R_{1z} \over \partial r^2} + {1\over r^2} \; {\partial^2  R_{1z} \over \partial \theta^2}  + {2\over r} \, {\partial R_{1z} \over \partial r} $
and
\begin{eqnarray}
I_1 = {\partial  (x R_{1z})\over \partial x} + {\partial  (y R_1)\over \partial y}   \label{i1a} \\
I_2 = {\partial  (G_x )\over \partial x} + {\partial  (G_y )\over \partial y}   \label{i2a} \\
I_3 = {\partial  {\tilde I}_{ez} \over \partial x} + {\partial  {\tilde I}_{ez} \over \partial y}  
\end{eqnarray}
with 
\begin{eqnarray}
G_x &=&  - 2 \; {\bf P}\int {d}x' \, {d}y' \, R_{2z} (z, z')  \, {x- x'\over |z- z'|^2} \label{gx} \\
G_y &=& - 2 \; {\bf P}\int {d}x' \, {d}y' \, R_{2z} (z, z')  \, {y- y'\over |z- z'|^2} \label{gy}
\end{eqnarray}
and ${\tilde I}_{ez}$ defined in eq.(\ref{rc4}).

Using transformation from cartesian to polar coordinates i.e $x=r \, \cos \theta$, $y=r \, \sin \theta$ along with relations ${\partial \over \partial x} = \cos \theta \, {\partial \over \partial r} - {\sin \theta\over r} \, {\partial \over \partial \theta}$, 
${\partial \over \partial y} = \sin \theta \, {\partial \over \partial r} + {\cos \theta\over r} \, {\partial \over \partial \theta}$, we then have
\begin{eqnarray}
I_1  &=&  r \, {\partial R_{1z} \over \partial r} + 2 \, R_{1z}  \label{i1ia}\\
I_2 &=&   {1\over r} \, {\partial (r \,I_c) \over \partial r} + {1\over r} \,  {\partial I_d \over \partial \theta} 
\label{i2ia} \\
I_3 &=&  {1\over r} \, {\partial (r \,J_c) \over \partial r} + {1\over r} \,  {\partial J_d \over \partial \theta} 
\end{eqnarray}
where $r \, I_c \equiv G_x \cos \theta + G_y \sin \theta $ and $I_d \equiv G_x \sin \theta - G_y \cos \theta $. Writing $x, y$ in term of $r, \theta$ and using  ${d}x' \, {d}y' = r' \, {d}r' \, {d}\theta'$  and $|z-z'|^2 = r^2+ r'^2-2 \, r \, r' \, \cos(\theta-\theta')$ in eqs.(\ref{gx}, \ref{gy}), $I_2$ and $I_3$ can now be rewritten as

\begin{eqnarray}
I_2 &=&   -2 \, \int_{0}^{\infty}\int_{0}^{2\pi}  \; \left[ \left({1\over r} \, {\partial (r \, C_z) \over \partial r} + {1\over r} \,  {\partial D_z \over \partial \theta} \right) \; { R}_{2z} + \left( C_z \, {\partial R_{2z} \over \partial r} + {D_z\over r} \,  {\partial R_{2z} \over \partial \theta} \right) \right]d\Omega 
\label{i2ia} \\
I_3 &=& -2 \, \int_{-\infty}^{\infty}  \; \left[ \left({1\over r} \, {\partial (r \, C_e) \over \partial r} + {1\over r} \,  {\partial D_e \over \partial \theta} \right)\; { R}_{2z} + \left( C_z \, {\partial R_{2ez} \over \partial r} + {D_z\over r} \,  {\partial R_{2ez} \over \partial \theta} \right) \right] \, de
\label{i3ia}
\end{eqnarray}
with ${ R}_{2z} \equiv { R}_{2z}(r,\theta,r',\theta') $, ${R}_{2ez} \equiv {R}_{2z}(r,\theta,e) $,  ${\rm d}\Omega = r' \, d\theta' \, dr' $ and 

\begin{eqnarray}
C_z(r,\theta,r',\theta') &=& \frac{r-r'cos(\theta -\theta')}{r^2+r'^2-2rr'cos(\theta-\theta')}, \\
D_z(r,\theta,r',\theta') &=& \frac{r' sin(\theta-\theta')}{r^2+r'^2-2rr'cos(\theta-\theta')} \\
C_e(e,r,\theta) &=& \frac{e-r \, \cos\theta}{e^2+r^2-2e r \cos\theta} \\
D_e(e,r,\theta) &=& \frac{r \, \sin\theta}{e^2+r^2-2e r \cos\theta}
\end{eqnarray}

Further using the exact relations,  valid for arbitrary $|p| <1$, 
\begin{eqnarray}
& & \frac{1-p \,\cos x}{1+p^2 -2 p \cos x} = \sum_{k=0}^{\infty} p^k \, \cos(k x), \\
& & \frac{p \, \sin x}{1+p^2 -2 p \cos x} = \sum_{k=0}^{\infty} p^k \, \sin(k x), \\
& & \frac{1-p^2}{1+p^2 -2 p \cos x} = 1 + 2 \sum_{k=1}^{\infty} p^k \, \cos(k x),
\end{eqnarray}
 eq.(\ref{i2ia}) for $I_2$ can be rewritten as a sum over two series, corresponding to expansion of integrand in $ r \le r'$ and $r > r'$,
\begin{eqnarray}
I_{2} &=& -\frac{2}{r} \int_{0}^{r} d\Omega \,
\frac{\partial R_{2z}}{\partial r}
+ 2 \sum_{k=1}^{\infty}
\Bigg[
\frac{1}{r^{k+1}} \frac{\partial T^{0r}_{kk}}{\partial r}
+ \frac{1}{r^{k+2}} \frac{\partial S^{0r}_{kk}}{\partial \theta}
- \frac{k\, T^{0r}_{kk}}{r^{k+2}}
\nonumber \\[6pt]
& & \qquad
- r^{k-1} \frac{\partial T^{r\infty}_{k,\,2-k}}{\partial r}
+ r^{k-2} \frac{\partial S^{r\infty}_{k,\,1-k}}{\partial \theta}
+ k\, r^{k-2} T^{r\infty}_{k,\,1-k}
\Bigg]
\label{i2ib}
\end{eqnarray}

where 
\begin{eqnarray}
T^{ab}_{kn}(r,\theta) &=&
\int_{a}^{b} dr'\, r'^n
\int_{0}^{2\pi} d\theta'\,
\cos\!\bigl(k(\theta-\theta')\bigr)\, R_{2z},
\\
S^{ab}_{kn}(r,\theta) &=&
\int_{a}^{b} dr'\, r'^n
\int_{0}^{2\pi} d\theta'\,
\sin\!\bigl(k(\theta-\theta')\bigr)\, R_{2z}
\end{eqnarray}
 
and 
\begin{eqnarray}
 { I}_{3} &=&  -2 \int_r^{\infty} \,{de\over e} \, {\partial { R}_{2ez} \over \partial r}  +
 2 \sum_{k=1}^{\infty} 
\left[{c_k \over r^{k}} \, {\partial W^{0r}_{k-1} \over \partial r}  + {s_k\over r^{k}}  \, {\partial W^{0r}_{k-1} \over \partial \theta}  -  r^{k} \, c_k \, {\partial W^{r \infty}_{-(1+k)} \over \partial r}  - r^{k} \,  s_k \,  {\partial W^{r \infty}_{-(k+1)} \over \partial \theta}  \right]  \nonumber \\
\label{i3ib}
\end{eqnarray}
with 
\begin{eqnarray}
W^{ab}_{n}  &=& \int_a^b \, e^n \, { R}_{2ez} \, {\rm d}e,\\
c_k &=& \cos(k \theta)+\cos(k(\pi-\theta))= 2 \cos(k \pi/2) \cos(k \pi/2 - kx), \\
s_k &=& \sin(k \theta)-\sin(k(\pi-\theta)) = - 2 \cos(k \pi/2) \sin(k \pi/2 - kx).  
\end{eqnarray}


Eq.(\ref{r1rta}) with $ { I}_{2}$ and $ { I}_{3}$ given by eq.(\ref{i2ib}) and eq.(\ref{i3ib}) is exact. To seek its solution, however, it is imperative to consider some large $N$-approximations. We approximate 
\begin{eqnarray}
R_{2z}(r, \theta, r', \theta'; Y) \approx R_{1z}(r, \theta; Y) \, R_{1z}(r', \theta'; Y).
\end{eqnarray}
 The above approximation is valid  at length scales $|r-r'| > \Delta(r)$ with $\Delta(r)$ as the local mean level spacing at $r$, latter rapidly decaying as a power law for large $N$). This in turn gives  
 \begin{eqnarray}
 T^{ab}_{kn}(r, \theta)  = \tau^{ab}_{kn}  \; R_{1z}(r, \theta), \\
   S^{ab}_{kn}(r, \theta)  = \sigma^{ab}_{kn}  \; R_{1z}(r, \theta)
 \end{eqnarray}  
   where $\tau^{ab}_{kn} =  \langle r'^n \, {\cos}(k (\theta-\theta') \rangle_{ab}$ and $ \sigma^{ab}_{kn}  = \langle r'^n \, {\sin}(k (\theta-\theta') \rangle_{ab}$ with 
   
 \begin{eqnarray}  
 \langle r'^n \, f(k (\theta-\theta') \rangle_{ab} \equiv \int_a^b \, dr'  r'^n \,\int_0^{2 \pi}  d \theta' f(k (\theta-\theta')) R_{1z}(r', \theta').
 \end{eqnarray}
 But as $0 \le R_{1z} \le 1$, and, $\cos(k (\theta-\theta')$ oscillate rapidly taking both positive and negative values as $\theta'$ varies between $0$ and $2 \pi$, (similarly for $\sin(k (\theta-\theta')$),  we have $\tau^{ab}_{kn} \sim 0, \sigma^{ab}_{kn} \sim 0$ for all $ k \ge 1$. 
The above reasoning motivates following approximation of eq.(\ref{i2ib}),
\begin{eqnarray}
 { I}_{2} &\approx & -{2 \over r}  {\partial (M_0 { R}_{1z}) \over \partial r}  
 \label{i2ic} 
\end{eqnarray}
with $M_0(r) =\int_0^r dr' d\theta' r' \,  { R}_{1z}(r', \theta')$. (This also follows  from a numerically observed week dependence of $R_1(r', \theta')$  on $\theta'$ )


Similarly the approximation ${ R}_{2ez}(r, \theta, e) \approx { R}_{1z}(r, \theta) \, { R}_{1e}(e)$ (valid  at length scales $|r-e| > \Delta(r)$) leads to $W^{ab}_{n}(r) = \langle e^n \rangle_{ab} \, { R}_{1e}(e)$ where $\omega^{ab}_{n}(r)  = \langle e^n \rangle_{ab} = \int_a^b \, e^n \, { R}_{1e} \, {\rm d}e$; the latter can be determined from the ${ R}_{1e}(e)$ formulation derives  in previous section. Substitution of the above in eq.(\ref{i3ib}), again  leads to  ${ I}_{3}$ as a sum over rapidly oscillatory terms for large and even $k$ (with odd $k$ terms exactly zero due to $\cos(k \pi/2) =0)$. Again retaining only the lowest $k^{th}$ order term of the series i.e  with $k=2$, we have
\begin{eqnarray}
 { I}_{3} &=&  -2  \left( {\partial  \over \partial r}   \omega_{-1}^{r \infty}
-{c_2 \over r^2} \, {\partial \over \partial r}  \omega^{0r}_{1}  + {s_2\over r^{2}}  \, {\partial \over \partial \theta} \omega^{0r}_{1}  -  r^2 \, c_2 \, {\partial \over \partial r}   \omega^{r \infty}_{-3} - r^2 \,  s_2 \,  {\partial \over \partial \theta} \omega^{r \infty}_{-3}  \right)   { R}_{1z}
\label{i3ic}
\end{eqnarray}


%

\section{Solution of eq.(66) of main text}

Eq.(66)  of main text is
\begin{eqnarray}
r^2 \, {\partial^2  A\over \partial r^2} + r \, (r^2 +2 - 2\, M_0)   {\partial A\over \partial r}   +  (2 + E) \, r^2 \, A  -  E_t  \, A  &= &  0 \label{ul1}
 \end{eqnarray}
 with $E_t$ as an arbitrary constant, determined by the boundary conditions  $\lim_{r\to \infty} \, A(r) \to 0$ and $T(\theta; E_t) = T(2 \pi +\theta; E_t)$.   A particular solution of eq.(\ref{ul1}) for an arbitrary constant $E_u=E+2$ is 
 
\begin{eqnarray}
A(r, E, E_t) &=& \left({2\over r^2}\right)^{\frac{1-2M_0}{4}}
\left( c \, \left({2\over r^2}\right)^{\frac{a}{4}} \, A_1 + d \, \left({2\over  r^2}\right)^{-\frac{a}{4} } \, A_2 \right) 
\label{ui}
\end{eqnarray}
with  $A_1=F_1\left(\frac{E}{2 } +\frac{3+2M_0 -a}{4}, \frac{2-a}{2} ,-\frac{ r^2}{2}\right)$ and $A_2 =F_1\left[\frac{E}{2 }+\frac{3+2M_0+a}{4}, \frac{2+a}{2},-\frac{ r^2}{2} \right]$, 
$F_1$ as the confluent Hypergeometric function $ _1F_1(\alpha, \beta, x)$ \cite{temme}, 
$a=\sqrt{4 E_1+(2M_0-1)^2}$, and $c, d$ as arbitrary constants to be determined by the initial conditions. Noting that both $E_t, E$ appear along with $2M_0$, their influence on the solution  can be seen only if they are of the order of $2M_0$ or larger. We redefine them, without loss of generality (both constants being arbitrary), as $E_t= { (\mu^2-1) \over 4}\,  (2M_0-1)^2$ and $E=4 \nu  \, M_0$ with $\mu, \nu $ as arbitrary constants  to be determined by boundary conditions; the conditions that $E,  E_1 \ge 0$ however also requires $\mu \ge 1$ and $\nu \ge 0$. This lead to $a=(2M_0-1) \mu$ and thereby in large $N$ limit, 
$A_1 \approx F_1\left( \frac{M_0 (4 \nu-\mu+1)}{2},  -{\mu M_0 },  -\frac{ r^2}{2}\right)$ 
 and $A_2 \approx F_1\left(\frac{M_0(4 \nu+\mu+1)}{2},{ \mu M_0},-\frac{r^2}{2}\right)$.  The particular solution of eq.(\ref{ul1}) can now be given as 

\begin{eqnarray}
A_{\mu \nu}  &=&  c_{\mu \nu} \, \left({ r^2 \over 2}\right)^{-\frac{(2M_0-1)(\mu-1)}{4}} \,   F_1\left(\frac{M_0(4 \nu-\mu+1)}{2}, { \mu M_0},-\frac{ r^2}{2}\right)  + \nonumber \\
& & d_{\mu \nu} \, \left({ r^2 \over 2}\right)^{\frac{(2M_0-1)(\mu+1)}{4}} \,   F_1\left(\frac{M_0(4 \nu+\mu+1)}{2}, { \mu M_0},-\frac{ r^2}{2}\right)
\label{umnn}
\end{eqnarray}
As the above solution is valid in large $r$-limit,  the first term  with coefficient $c_{\mu \nu}$, for $\nu \not=0$,  is  negligible as compared to second term.   For $\nu=0$, however, the first term approaches a constant value.  The condition $\lim_{r \to \infty}A_1(r, \theta) \to 0$ in eq.(\ref{ui}) therefore requires $c_{\mu 0}=0$.


\section{Calculation of $Y - Y_0$ given in section VIII of main text}
Here we consider three different ensembles.

\subsection{Case 1 BE (Ensemble with a constant ratio for off-diagonal to diagonal variances)}

For this case, we have $y_{kl} = \frac{1}{2} \left( 1 + \frac{N}{b} \right)^2$, $y_{kk}={1\over 2}$, and $x_{kl}=0$. 
From eq.(\ref{y}), $Y \equiv t_1$ now becomes 
\begin{equation}
Y={2} \sum_{k,l} q_{k,l;1}\ln\left(\frac{y_{kl}}{|1-2y_{kl}|}\right)+c_0,
\label{eq:Ydef}
\end{equation}
where $q_{k,l;1}=\frac{1}{M}$, so $\sum_{k,l}q_{k,l;1}=1$, {with} $ M=N^2$,
for all $(k,l)$. The reference value is defined by $Y_0=Y(b_0)$.
Then, we evaluate
\begin{align}
1-2y_{kl}
&=1-\left(1+\frac{N}{b}\right)^2 =-\frac{N}{b}\left(2+\frac{N}{b}\right).
\end{align}
Hence,
\begin{equation}
|1-2y_{kl}|=\frac{N}{b}\left(2+\frac{N}{b}\right).
\end{equation}

Then substituting into the logarithmic argument gives
\begin{align}
\frac{y_{kl}}{|1-2y_{kl}|}
&=\frac{\frac{1}{2}\left(1+\frac{N}{b}\right)^2}
{\frac{N}{b}\left(2+\frac{N}{b}\right)}=
\frac{(b+N)^2}{2N(2b+N)}.
\end{align}

Therefore,
\begin{equation}
\ln\left(\frac{y_{kl}}{|1-2y_{kl}|}\right)
=
\ln\left(\frac{(b+N)^2}{2N(2b+N)}\right).
\end{equation}

Using $\sum_{k,l} q_{kl;1}=1,$ we now obtain
\begin{align}
Y-Y_0&=2\bigg[\ln\left(\frac{(b+N)^2}{2N(2b+N)}\right)
-\ln\left(\frac{(b_0+N)^2}{2N(2b_0+N)}\right)\bigg] \\
&=2\ln\left[
\frac{(b+N)^2(2b_0+N)}
{(b_0+N)^2(2b+N)}
\right].
\end{align}
Thus,
\begin{equation}
Y-Y_0=
2\ln\left[
\frac{(b+N)^2(2b_0+N)}
{(b_0+N)^2(2b+N)}
\right]
\label{eq:finalY}
\end{equation}
which satisfies $Y-Y_0=0$ at $b=b_0$.

\subsection{Case 2: PE (Ensemble with power lay decay for off-diagonal to diagonal variance)}

For this case,  we have  $y_{kl} = \frac{1}{2} \left( 1 + \frac{|k-l|^2}{b^2} \right)^2$, $y_{kk}={1\over2}$, and $x_{kl}=0$.
Proceeding in the same way as in previous case and taking $r=|k-l|$ then summing over $r$, with $M=N^2$, the final form can be written as
\begin{equation}
Y - Y_0={1\over{N^2}} \sum_{r=1}^{N} (N-r)
\ln \left[\frac{\left( b^2 + r^2 \right)^2\left( 2b_0^2 + r^2\right)}{\left( b_0^2 + r^2 \right)^2\left( 2b^2 + r^2 \right)} \right]  
\end{equation}

\subsection*{Case 3 EE (Ensemble with exponential decay of the off-diagonal to diagonal variances): With $y_{kl} = \frac{1}{2} \exp\left(\frac{|k-l|^2}{b^2}\right)$, $y_{kk}={1\over 2}$, and $x_{kl}=0$}
In the same way, the final form can be written by taking $r=|k-l|$ and adding up all the terms in $r$, with $M=N^2$ we obtain;
\begin{eqnarray}
Y= {1\over{N^2}} \sum_{r=0}^{N} (N-r) \ln \Biggl[ \frac{1}{2} \frac{e^{r^2/b^2}}{|1 - e^{r^2/b^2}|}  \Biggr] + C_0.
\end{eqnarray}
Using $F(r,b) = \frac{1}{2} \frac{e^{r^2/b^2}}{|1 - e^{r^2/b^2}|} $, $Y$ can be rewritten as 
\begin{align}
Y = {1\over{N^2}} \sum_{r=0}^{N} (N-r) \ln| F(r,b)| +C_0 
\end{align}
This in turn leads to 
\begin{equation}
Y - Y_0 = {1\over{N^2}} \sum_{r=1}^{N} (N-r) \Big[ \ln| F(r,b)| - \ln|F(r,b_0)| \Big] 
\end{equation}

\end{document}